\documentclass[journal]{IEEEtran}
\usepackage{ifpdf}
\ifCLASSINFOpdf
\else
\fi
\usepackage{amsmath}
\usepackage{url}
\usepackage{subfig}
\usepackage{longtable}
\usepackage{array}
\usepackage{booktabs}
\usepackage{multirow}
\usepackage{graphicx}
\usepackage{cite}
\usepackage{subcaption}
\usepackage{paralist}
\usepackage{makecell}
\usepackage{enumitem}
\setlist{leftmargin=*}
\usepackage[para]{threeparttable}
\usepackage{mathtools}
\usepackage{wasysym}
\usepackage{orcidlink}
\usepackage[utf8]{inputenc}
\usepackage{caption}
\DeclareCaptionLabelSeparator{periodspace}{.\quad}    % “.␣␣” 
\usepackage{stfloats}   

\usepackage{etoolbox}

\usepackage{xcolor}

\usepackage[normalem]{ulem}

\usepackage[markup=underlined]{changes}

\definecolor{addcolorone}{RGB}{0,102,204}     % Reviewer 1 (same as current addcolor)
\definecolor{addcolorthree}{RGB}{0,102,0} % Reviewer 3 Darker green 
\definecolor{addcolorcommon}{RGB}{255,140,0} % Reviewer 1, 3 (Orange)
\definecolor{commentcolor}{RGB}{0,0,150}   % soft purple
\definecolor{default}{RGB}{0,0,0}   % soft purple

\newcommand{\addedone}[1]{\textcolor{default}{#1}}
\newcommand{\addedthree}[1]{\textcolor{default}{#1}}
\newcommand{\addedcommon}[1]{\textcolor{default}{#1}}

\definecolor{delcolor}{RGB}{204,0,0}

\setaddedmarkup{\addedone{#1}}
\setaddedmarkup{\addedthree{#1}}
\setaddedmarkup{\addedcommon{#1}}
\setdeletedmarkup{\textcolor{delcolor}{\sout{#1}}}

\begin{document}
\title{Review of Power Electronic Solutions for Dielectric Barrier Discharge Applications}

\author{Hyeongmeen Baik\orcidlink{0009-0007-1060-8916},~\IEEEmembership{Graduate Student Member,~IEEE,}
        Jinia Roy\orcidlink{0000-0002-1446-0417},~\IEEEmembership{Senior Member,~IEEE,}
        
        % <-this % stops a space
\thanks{The authors are with the Department
of Electrical and Computer Engineering, University of Wisconsin-Madison, Madison, WI 53706 USA e-mail: (hyeongmeen.baik@wisc.edu).}% <-this % stops a space
}


\markboth{IEEE TRANSACTIONS ON POWER ELECTRONICS, DECEMBER 2025}%
{Shell \MakeLowercase{\textit{et al.}}: Bare Demo of IEEEtran.cls for IEEE Journals}



\maketitle

\begin{abstract}
\addedthree{This paper presents a comprehensive review of dielectric barrier discharge (DBD) power supply topologies, aiming to bridge the gap between DBD applications and power electronics design. Two key aspects are examined: the dependence of the DBD electrical model on reactor geometry, and application-driven requirements for injected waveform characteristics, including shapes, voltage amplitude, frequency, and modulation techniques. On this basis, the paper systematically reviews two major categories of power supplies: sinusoidal types comprising transformerless and transformer-based resonant inverters, and pulsed power supplies (PPSs). The review summarizes performance trade-offs, highlights untested topologies and emerging applications, and offers guidance for advancing high-performance DBD power supply design for next-generation systems.}
\end{abstract}


% \fi
\IEEEpeerreviewmaketitle

\section{Introduction}
%%%%%%%%%%%%%%%%%%%%%%%%%%%%%%%%%%%%%%%%%%%%%%%%%%%%%%%%%%%%%%%%%%%%%%%%%%%%%%%%%%%%%%%%%%%%%%%%%%

\IEEEPARstart{D}{ielectric} barrier discharge (DBD) generates stable, low-temperature plasma under atmospheric conditions by incorporating a dielectric layer on metal electrodes~\cite{DBD_basics1, DBD_basic2, DBD_basic3}. DBD systems typically consist of two electrodes separated by a discharge gap and insulated by dielectric materials such as glass, ceramics, polymers, or metal oxides~\cite{DBD_material1, DBD_material2, DBD_material3}. This configuration prevents arc formation and supports a uniform electric field distribution.

Due to its capability to generate non-thermal plasma under atmospheric pressure, DBD technology is recognized as an effective solution for plasma-based processing applications. Unlike high-temperature plasma systems, DBD produces reactive species such as radicals and ions without the need for expensive vacuum infrastructure or the risk of thermal damage to temperature-sensitive materials, while achieving higher energy efficiency~\cite{DBD_temp1, DBD_temp3, DBD_temp4}. These characteristics make DBD particularly effective for applications involving polymers, biological tissues, and other delicate substrates~\cite{DBD_polymer2}. With the advent of DBD structures featuring flexible geometries, the range of applications has expanded significantly, offering potential solutions for biomedical applications, including wound healing~\cite{DBD_wound1,DBD_wound2,DBD_wound3}. Furthermore, the ability of DBD to operate under atmospheric conditions broadens its applicability across various industries, including food sterilization~\cite{DBD_food_review, DBD_food_review2, DBD_food1}, and catalytic reactions~\cite{DBD_catalyst_review,DBD_catalyst_review2, DBD_catalyst_review3}.

% DBD_food_review3

These diverse and expanding applications demand precise and reliable plasma generation, which is fundamentally governed by the operation of the power supply. As a critical component of DBD systems, the power supply directly influences key parameters such as discharge uniformity and plasma dynamics. To support the wide range of DBD applications, power supply designs must address the unique challenges posed by the capacitive and dynamic load behavior of DBD systems while meeting application-specific requirements.

The increasing scope of DBD applications has introduced diverse operational requirements, such as high-frequency (HF), high-voltage (HV), and pulsed- or AC-excitation modes. A variety of power supplies have been proposed in the literature, over the years to drive DBDs with unique characteristics. However, a comprehensive review of the state-of-the-art in power supply technologies for DBD systems has not yet been presented. It is essential to summarize recent advancements, address challenges such as energy losses, reactor geometry compatibility, and cost constraints, and finally identify opportunities for further research.

This paper presents an extensive review of the power electronics solutions developed for various DBD applications. It firstly identifies different DBD applications/reactor geometries and corresponding power supply requirements and presents a comprehensive review of the existing power supplies for DBD systems. Such a review would provide a systematic foundation, offering insights into the design and selection of power supplies for DBD systems. Furthermore, it would highlight performance trade-offs, optimization strategies, and underexplored areas, thereby advancing the development of robust and efficient DBD systems for a broad range of industrial and research applications.

The remainder of this paper is structured as follows: Section~\ref{sec:2} provides an overview of DBD reactor geometries and modeling approaches. To address the waveform requirements of DBD applications and guide the development of suitable power supplies, Section~\ref{sec:3} and Section~\ref{sec:4} discuss the unique characteristics of DBD systems, focusing on waveform shapes, frequency considerations, and power supply requirements. Building on this foundation, Sections~\ref{sec:5} and~\ref{sec:6} present a detailed examination of sinusoidal supply and pulsed power supply (PPS) topologies, along with their operational characteristics. Sections~\ref{sec:7} and~\ref{sec:8} highlight remaining challenges and future research directions. Finally, Section~\ref{sec:9} presents the conclusions.

\section{Overview of DBD Applications} \label{sec:2}
%%%%%%%%%%%%%%%%%%%%%%%%%%%%%%%%%%%%%%%%%%%%%%%%%%%%%%%%%%%%%%%%%%%%%%%%%%%%%%%%%%%%%%%%%%%%%%%%%%
An important advantage of DBD technology lies in its energy efficiency and flexibility in controlling plasma chemistry. %By adjusting gas composition and operational parameters, DBD systems can generate reactive species suitable for processes such as surface modification, pollutant degradation, and coating deposition.
This section presents the equivalent impedance model of DBD systems followed by different DBD geometries and their equivalent model considerations. There can be various configurations of DBD systems, such as Volume DBD (VDBD), Surface DBD (SDBD), Floating-Electrode DBD (FE-DBD), and Flexible DBD (FX-DBD), each optimized for specific applications. The power supply requirements for these systems are strongly influenced by the reactor design and the intended application. Consequently, a review of the DBD setup is essential to ensure proper alignment of power supply specifications with reactor configurations and operational demands.

\vspace{-10pt}
\subsection{The Classical Equivalent Model of DBD}

\begin{figure}[!t]
    \centering    
    \subfloat[][]{\label{fig:DBDequi1}
    \centering
    \includegraphics[height=3.8cm]{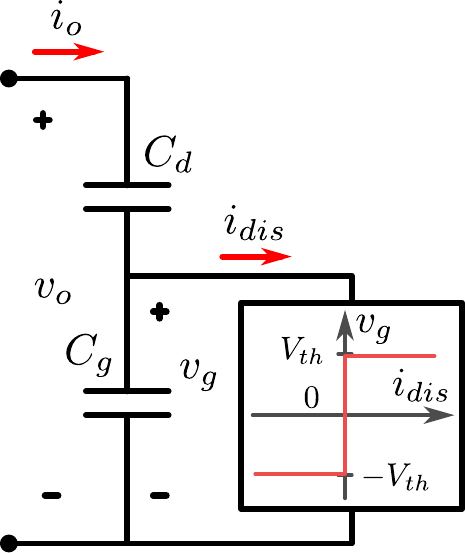}
    % \caption{Closed Loop Control}
    }
    \subfloat[][]{\label{fig:DBDequi2}
    \centering
    \includegraphics[height=3.8cm]{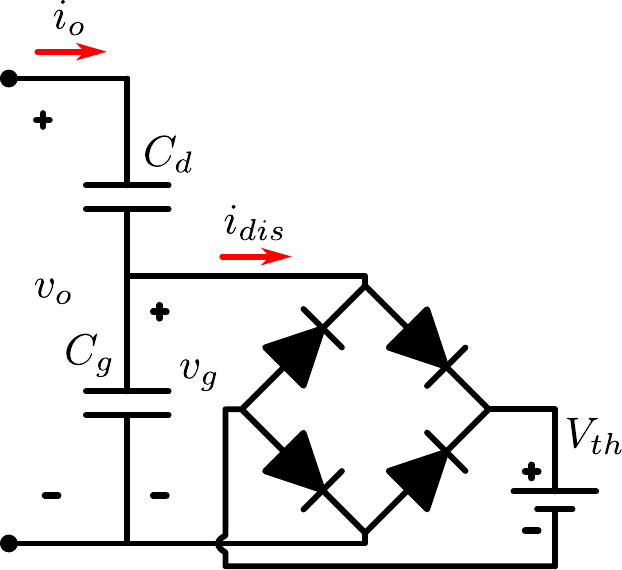}

    }
    \caption{DBD equivalent circuits. (a) Threshold-based. (b) Rectifier-based.}
    \label{fig:DBD_equi}

    \centering
    \includegraphics[width=0.6\linewidth]{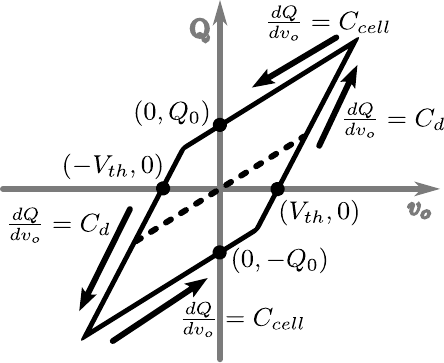}
    \caption{Ideal Q-V plot.}
    \label{fig:lissajous_ideal}
    \vspace{-10pt}
\end{figure}

Fig.~\ref{fig:DBD_equi} shows the classical equivalent circuit model for DBD. This model consists of the dielectric capacitance ($C_d$) and gas gap capacitance ($C_g$) arranged in series, with the plasma represented as a resistive element during active discharge phases~\cite{mis_equi}. This resistive element is often replaced with a threshold component, as shown in Fig.~\ref{fig:DBD_equi}(a), or a constant voltage source with a full-bridge rectified configuration, as depicted in Fig.~\ref{fig:DBD_equi}(b). 

In the passive phase, when the applied voltage remains below the threshold voltage, $V_{th}$, the entire circuit—referred to as the "cell"—operates predominantly as a capacitive network. The total impedance of the cell is determined by the combined effects of its components, including the two capacitances ($C_d$ and $C_g$) and the threshold component. In contrast, during the active phase, when the applied voltage across the gas material, $v_g$, surpasses the threshold voltage, $V_{th}$, plasma conduction alters the circuit dynamics, which are then dominated by the dielectric capacitance in conjunction with the voltage source. Achieving $v_g > V_{th}$, typically in the range of several kV to 100 kV, is critical because surpassing $V_{th}$ initiates gas ionization. At this point, the observed current corresponds to the discharge process, marking the active plasma generation phase. 

This model is often analyzed using charge-voltage ($Q{-}V$) characteristics, visualized as Lissajous plots, as shown in Fig.~\ref{fig:lissajous_ideal}. These plots represent the energy dissipated per cycle and enable back-calculation of parameters such as $C_d$ and $C_g$ for DBD modeling~\cite{DBD_impedance}.

Overall, the DBD system can be simplified into two distinct operating modes with different capacitances: the passive mode, governed by the capacitive network of $C_d$ and $C_g$, and the active mode, dominated by the dielectric capacitance and plasma conduction. While the classical model is effective for basic diagnostics and power analysis, it assumes uniform discharge characteristics, neglecting complexities such as filamentary behavior, non-uniform charge distributions, and parasitic effects. For more detailed investigations and applications involving intricate discharge dynamics, extensions of the classical model are necessary to accurately capture these phenomena and adapt to diverse geometrical configurations~\cite{dbd_equi_non_ideal1, dbd_equi_non_ideal2, dbd_equi_non_ideal3, DBD_review1}.

\vspace{-10pt}
\subsection{Volume DBD}

\begin{figure}[!t]
    \centering    
    \subfloat[][]{\label{fig:VDBD11}
    \centering
    \includegraphics[width=0.9\linewidth]{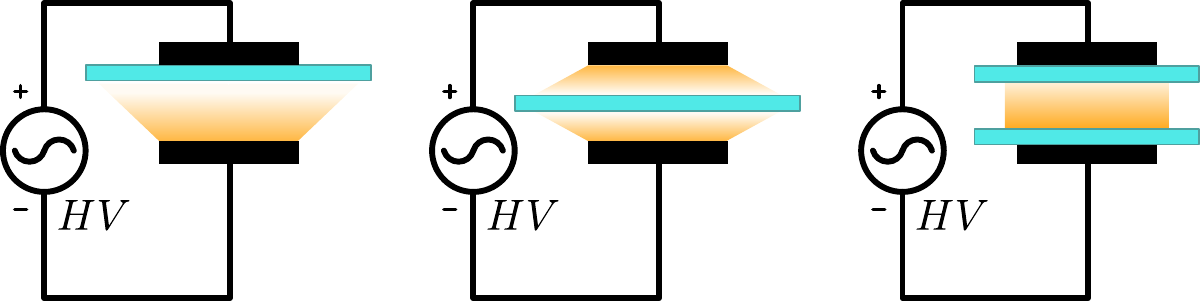}
    % \caption{Closed Loop Control}
    }
    \\
    \subfloat[][]{\label{fig:VDBD12}
    \centering
    \includegraphics[width=0.9\linewidth]{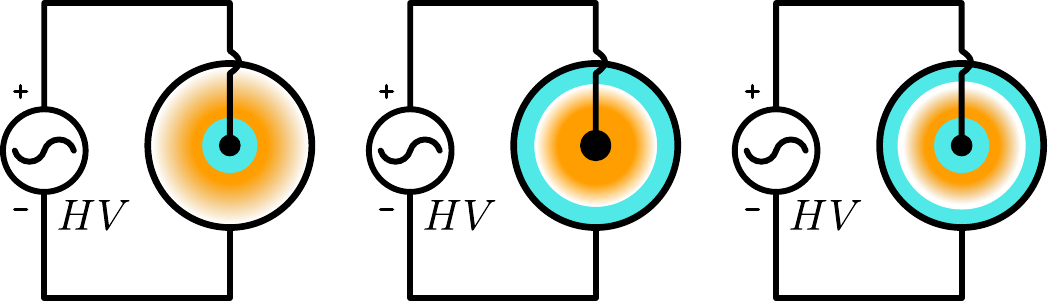}

    }
    \\
    \subfloat[][]{\label{fig:VDBD13}
    \centering
    \includegraphics[height=2.3cm]{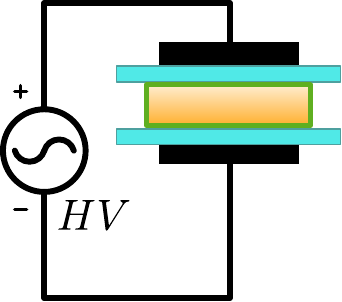}

    }
    \subfloat[][]{\label{fig:VDBD14}
    \centering
    \includegraphics[height=2.3cm]{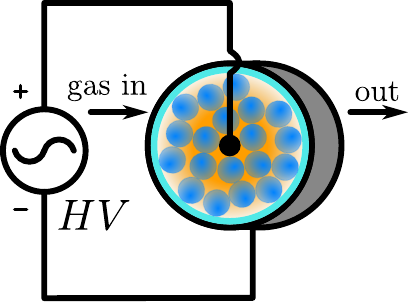}

    }
    \subfloat[][]{\label{fig:VDBD15}
    \centering
    \includegraphics[height=2.3cm]{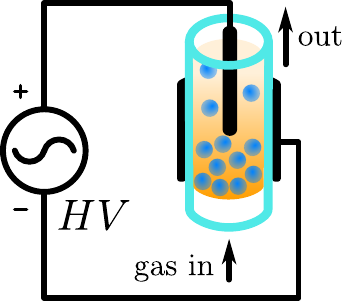}

    }
    \caption{VDBD reactor geometries. Cyan: dielectric material. Orange: generated plasma. Black: electrical connection including electrodes. Green: package. Blue: packed material. (a) Plate-to-plate. (b) Cylindrical. (c) In-package. (d) Packed-bed. (e) Fluidized-bed.}
    \label{fig:VDBD1}
    \vspace{-10pt}
\end{figure}

VDBD systems, as shown in Fig.~\ref{fig:VDBD1}, represent a foundational configuration of DBD technology, characterized by plasma generation within a confined volume between two dielectric-covered electrodes. This configuration is widely used in gas-phase applications, such as catalysis and thin-film treatments, where precise plasma generation within a controlled environment is essential~\cite{DBD_catalyst}.

However, the limited discharge volume inherent to VDBD systems presents challenges in treating large or irregularly shaped samples, restricting their applicability in certain industrial processes. To address these limitations, advanced adaptations such as Packed-Bed DBD (PB-DBD) and Fluidized-Bed DBD (FB-DBD) have been developed, significantly enhancing functionality and extending the range of applications.

Fig.~\ref{fig:VDBD1}(d) illustrates an example of PB-DBD, where dielectric or catalytic materials are introduced into the discharge zone. This configuration enhances plasma-catalyst interactions by increasing the available surface area and redistributing the electric field~\cite{DBD_PB2}.

Building on the principles of PB-DBD, FB-DBD (see Fig.~\ref{fig:VDBD1}(e)) incorporates fluidized particles within the plasma discharge zone to overcome challenges associated with static packed beds such as clogging and poor gas flow, especially due to the accumulation of solid byproducts like carbon nanofibers, and non-uniform heat and mass transfer~\cite{DBD_FB1,DBD_FB2,DBD_FB3}. 
FB-DBD not only improves plasma penetration but also minimizes localized arcing, creating a consistent and stable reaction environment.

\begin{figure}[!t]
    \centering
    \begin{minipage}{.28\textwidth}
        \centering
        \includegraphics[width=1\linewidth]{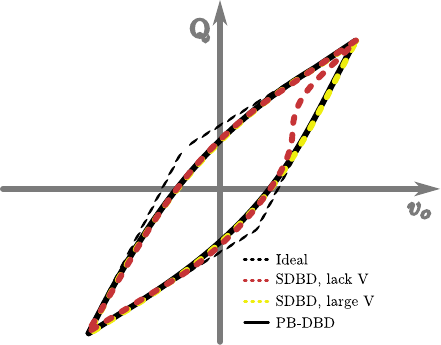}
        \caption{Non-ideal Lissajous plot.}
        \label{fig:lisa_nonideal}
    \end{minipage}%
    \begin{minipage}{0.2\textwidth}
        \centering
        \includegraphics[width=0.7\linewidth]{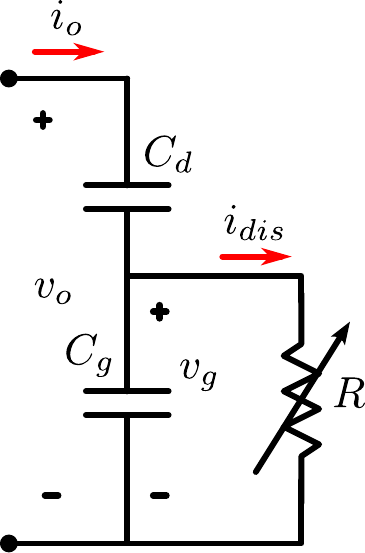}
        \caption{Augmented DBD equivalent circuit.}
        \label{fig:DBD_equi_non_ideal}
    \end{minipage}
    \vspace{-10pt}
\end{figure}

It should be noted that PB-DBD and FB-DBD systems deviate from ideal VDBD behaviors, as evident in their Lissajous plots. Fig.~\ref{fig:lisa_nonideal} shows a lens-shaped trajectory for PB-DBD, diverging from the ideal parallelogram associated with symmetrical VDBD designs~\cite{DBD_PB_lissa1}. This deviation underscores the classical equivalent circuit model's limitations in capturing PB-DBD's time-varying characteristics~\cite{DBD_equivalent1}. In FB-DBD, dynamic capacitance changes due to varying gas velocities and fluidized geometries further complicate the modeling. For reactors with non-parallelogram Lissajous plots, equivalent circuit models incorporating variable resistors, as shown in Fig.~\ref{fig:DBD_equi_non_ideal}, provide more accurate representations but add complexity to the analysis. This is important to note, as it guides the control design of the power converter driving FB-DBD.

\begin{figure}[!t]
    \centering    
    \subfloat[][]{\label{fig:SDBD1}
    \centering
    \includegraphics[height=2cm]{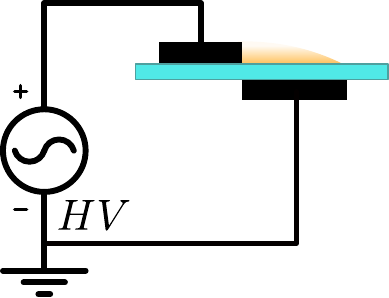}
    % \caption{Closed Loop Control}
    }
    \subfloat[][]{\label{fig:SDBD_in_package}
    \centering
    \includegraphics[height=2cm]{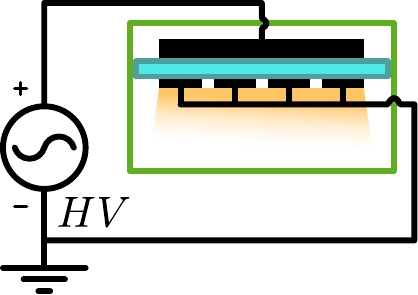}

    }
    \subfloat[][]{\label{fig:FEDBD}
    \centering
    \includegraphics[height=2cm]{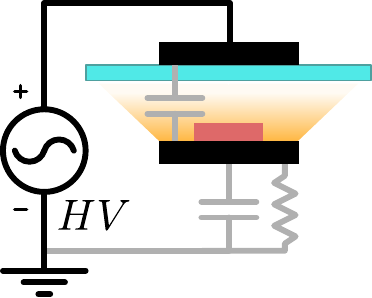}

    }

    \caption{Exemplary reactor geometries. Red: treatment target. (a) SDBD. (b) In-package SDBD. (c) FE-DBD.}
    \label{fig:SDBD}
    \vspace{-10pt}
\end{figure}

\vspace{-10pt}
\subsection{Surface DBD}

SDBD addresses the spatial limitations of VDBD by enabling plasma generation over larger surface areas. Fig.~\ref{fig:SDBD} shows typical geometries of SDBD including in-package geometry. By embedding electrodes within a dielectric barrier, SDBD creates a planar discharge area ideal for extensive surface treatments, such as sterilization and wound healing. The co-planar electrode design reduces the gap between electrodes, lowering voltage requirements and enhancing energy efficiency for surface-based plasma applications.

FX-DBD, on the other hand, extends this concept by incorporating flexible materials, enabling plasma treatment of non-planar or irregular surfaces. It functions as both an SDBD and VDBD alternative, leveraging flexible electronics to conform to complex geometries for uniform plasma exposure~\cite{Food7}. However, like PB-DBD, SDBD exhibits lens-shaped Lissajous plots, as shown in Fig.~\ref{fig:lisa_nonideal}, which can be modeled using a variable resistor~\cite{DBD_SDBD_lissa1,DBD_SDBD_lissa2}. Additionally, the asymmetrical geometry of SDBD causes Q-V trajectory discrepancies between cycles at low voltages, emphasizing the importance of maintaining sufficient voltage levels for symmetric reactor operation.

\begin{figure}[!t]
    \centering    
    \includegraphics[height=4cm]{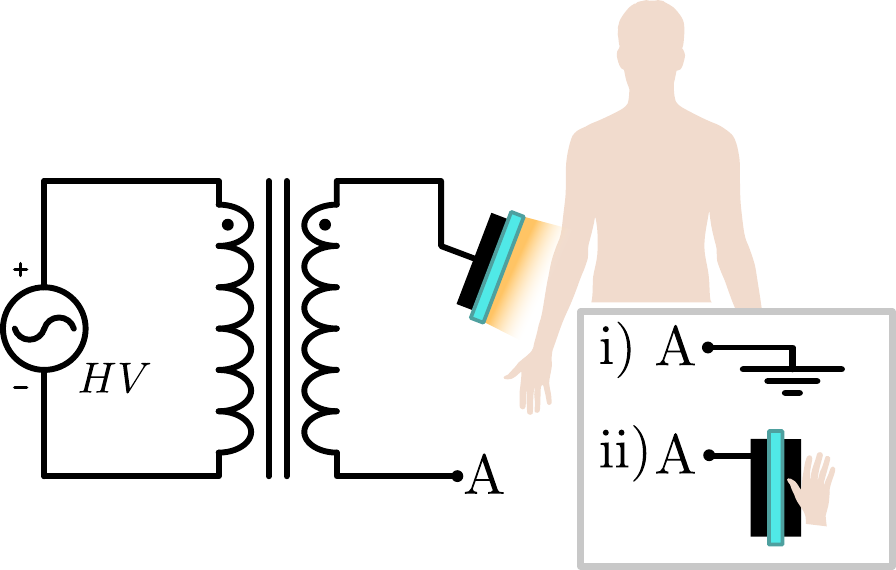}
    \caption{FE-DBD system for human skin treatment.}
    \label{fig:FE_DBD_human}
    \vspace{-10pt}
\end{figure}

\vspace{-10pt}
\subsection{Floating-Electrode DBD}

FE-DBD enhances the classical DBD geometry by introducing a floating electrode concept, where the treated sample or a floating electrode forms capacitive coupling with the other electrode, as shown in Fig.~\ref{fig:SDBD}(c). This eliminates the need for a direct ground connection, reducing electrical hazards and enabling localized plasma generation at the target surface. This configuration is particularly suitable for sensitive applications such as wound healing, combining simplified electrode structures with efficient plasma generation.

However, the dynamic interaction between the floating electrode and the actual ground node demands careful consideration. In applications like human skin treatment, the positioning of the ground node relative to the body significantly impacts system performance~\cite{FE_DBD_human2}. As shown in Fig.~\ref{fig:FE_DBD_human}, directly connecting the transformer's secondary winding to the body bypasses capacitive coupling, leading to unsafe current levels, while grounding the secondary node causes impedance mismatch and instability~\cite{FE_DBD_human}.

\section{Waveforms on DBD Systems}\label{sec:3}
The choice of waveform in DBD systems, both in terms of shape and frequency, is highly application-dependent, as different waveforms uniquely influence discharge behavior, energy efficiency, and process outcomes. Specific plasma characteristics, such as discharge uniformity, energy transfer efficiency, and temporal behavior can be optimized by adjusting waveform parameters like shape, amplitude, frequency, rise time and modulation. Consequently, the selection of power supply setups in DBD systems must align closely with the specific requirements of the intended application.
\vspace{-10pt}

\subsection{Effect of Waveform Shapes}
\begin{figure}[tp!]
    \centering
    \includegraphics[width=0.7\linewidth]{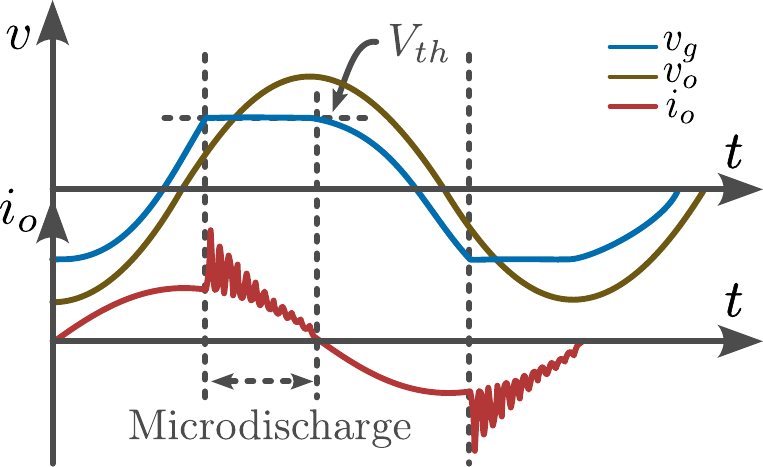}
    \caption{Filamentary discharge of sinusoidal excitation.}
    \label{fig:FDBD}
    
    \centering
    \includegraphics[width=0.7\linewidth]{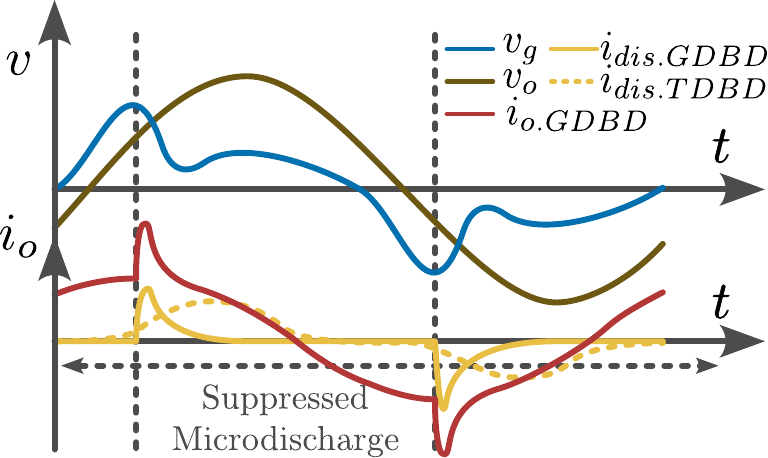}
    \caption{Example waveforms of GDBD and TDBD.}
    \label{fig:GDBD}

    \includegraphics[width=0.7\linewidth]{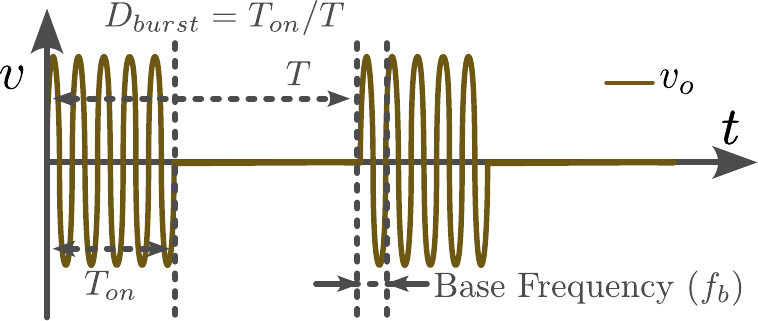}
    \caption{Typical waveform of burst-mode excitation.}
    \label{fig:burst}
    \vspace{-10pt}
\end{figure}

For DBD systems, excitation waveforms can be sinusoidal, square, or pulsed. Square and pulsed waveforms can be realized as either bipolar or unipolar excitations.
For instance, in ozone generation, which is a common intermediate reactive species determining performance of DBD applications, \textit{sinusoidal} and \textit{pulsed} voltage waveforms have been extensively studied. Pulsed waveforms are preferred for their sharp rise times, which enhance ionization and energy transfer efficiency, while sinusoidal waveforms produce more stable discharges, making them ideal for continuous ozone production setups~\cite{DBD_tendency1,DBD_tendency6}. The different characteristics of AC and short-pulse excited DBDs arise mainly from their differing discharge dynamics. In AC-excited DBDs, the discharge typically consists of multiple microdischarges (as shown in Fig.~\ref{fig:FDBD}) that occur randomly across different locations and times during both the positive and negative half-cycles of the voltage waveform~\cite{DBD_tendency11}. In contrast, short-pulse excited DBDs produce a single and simultaneous discharge across the entire area, resulting in an almost homogeneous discharge appearance~\cite{DBD_equivalent1}. 
% Thus, optimized pulsed waveforms ensure higher plasma uniformity and reduced energy consumption, providing both scalability and operational efficiency~\cite{DBD_tendency6}.

Similarly, in aerodynamic applications using DBD plasma actuators, waveform selection plays a critical role in determining thrust generation and electric wind characteristics. \textit{Square} waveforms produce the highest thrust output but at the cost of increased power consumption. Conversely, \textit{sinusoidal} waveforms yield smoother electric wind velocity profiles, which are advantageous for precise flow control applications~\cite{DBD_tendency9,DBD_tendency8,DBD_tendency5}. DBD ionization in portable mass spectrometry systems benefits from \textit{square} waveforms, which minimizes power consumption while achieving homogeneous plasma for efficient ion detection~\cite{DBD_tendency10}.

Moreover, \textit{unipolar pulses} are preferred for certain applications due to their ability to suppress reverse discharges, resulting in more uniform and stable plasma generation. In contrast, \textit{bipolar pulses} tend to introduce secondary reverse discharges, which disrupt discharge uniformity and lead to increased energy dissipation~\cite{DBD_bi_uni}. Thus, unipolar pulses are particularly advantageous for applications requiring precise energy control, such as surface treatments and thin-film deposition.

However, it is not evident which waveform shapes outperform others, as this depends strongly on the DBD operating mode, which in turn is heavily influenced by application characteristics, electrode geometry, frequency, voltage amplitude, and rise and fall times. These dependencies will be discussed in detail in the following subsections.

\begin{table*}[tp!]
\centering
\caption{Comparison of Different Discharge Modes in DBDs}
\label{table:different_modes}
\begin{threeparttable}
\scalebox{0.9}{
\begin{tabular}{c|c|c|c|c}
\toprule
\textbf{Feature} 
& \textbf{Filamentary DBD} 
& \textbf{TDBD} 
& \textbf{GDBD} 
& \textbf{RF-DBD} \\ 
\midrule

\makecell[c]{Mode of \\Ionization} 
& \makecell[c]{Localized, intense \\ ionization in discrete \\ filaments.} 
& \makecell[c]{Uniform, low-density \\ ionization via avalanches.} 
& \makecell[c]{Sustained, bulk \\ ionization near the \\ cathode.} 
& \makecell[c]{Bulk ionization \\ driven by oscillating \\ electric fields.} \\ 
\hline

\makecell[c]{Plasma \\Uniformity} 
& \makecell[c]{Non-uniform; dominated \\ by microdischarge \\ filaments.} 
& \makecell[c]{Highly uniform across \\ the discharge gap.} 
& \makecell[c]{Moderately uniform \\ with distinct plasma \\ regions.} 
& \makecell[c]{Highly uniform; \\ no microdischarges.} \\ 
\hline

\makecell*[c]{Power \\Density} 
& \makecell[c]{High} 
& \makecell[c]{Low} 
& \makecell[c]{Moderate} 
& \makecell[c]{High} \\ 
\hline

\makecell*[c]{Operating \\Frequency} 
& \makecell[c]{Low\tnote{1}} 
& \multicolumn{2}{c|}{\makecell[c]{Moderate; frequency depends on gas \\ composition and velocity.}} 
& \makecell[c]{High (\textgreater 1 MHz)} \\ 
\hline

\makecell[c]{Advantages} 
& \makecell[c]{Best for localized, \\ high-intensity discharges \\ (e.g., ozone generation).} 
& \makecell[c]{Ideal for highly \\ uniform, low-energy \\ plasma treatments.} 
& \makecell[c]{Balances uniformity \\ and power, suitable \\ for surface treatment.} 
& \makecell[c]{Achieves lowest \\ breakdown voltage \\ and high plasma density.} \\ 
\bottomrule
\end{tabular}
}

\begin{tablenotes}
\scriptsize
\addtolength{\itemindent}{10cm}
\hspace*{3cm}
\begin{minipage}{\linewidth}
\item[1] Filamentary DBD can occur at any frequency if the applied voltage is sufficiently high.
\end{minipage}
\end{tablenotes}
\end{threeparttable}
\vspace{-10pt}
\end{table*}

%In summary, the choice of waveform has distinct advantages and disadvantages, which can be preferred by each applications

\vspace{-10pt}
\subsection{Effect of Frequency}

As mentioned before, the selection of the excitation waveform frequency also plays a critical role in the operation of DBD systems. The frequency requirement can range from tens of Hz to even several MHz. And the requirements for the applications also vary due to distinct operational characteristic of DBD depending on the frequency. 

At low frequencies, typically in the range of tens of Hz to a few kHz, the discharge operates in a filamentary mode, characterized by discrete microdischarges, as shown in Fig.~\ref{fig:FDBD}. These are formed through localized streamer breakdown, resulting in filamentary discharge channels with high current density. This mode is advantageous for applications where higher ionization efficiency and reactive species production are critical~\cite{DBD_tendency1}. However, localized streamer breakdown induces thermal effects due to localized heating, which can degrade overall system performance and efficiency. 

As the frequency increases to the tens of kHz range, the number of filaments decreases due to a memory effect—the residual influence of previous discharge cycles, which stabilizes discharges in this frequency range~\cite{DBD_diffuse}. This transition leads to more homogeneous DBD types, often referred to as diffuse DBD. Depending on the reactor configuration, dielectric material, the gas used, and even gas flow rate, this can manifest as Glow DBD (GDBD), driven by a combination of Townsend ionization and secondary electron emission, or Townsend DBD (TDBD), dominated by slow electron avalanches associated with Townsend ionization~\cite{DBD_tendency_freq1}. As shown in Fig.~\ref{fig:GDBD}, the output waveforms of GDBD and TDBD show suppressed microdischarges. These diffuse modes offer improved plasma uniformity and suppressed microdischarges, making them well-suited for applications requiring consistent plasma exposure, such as thin-film deposition and surface treatment. 

At even higher frequencies, in the MHz range—referred to as Radio Frequency (RF) DBD—the dynamics of gas ionization and charge carrier behavior shows a notable shift~\cite{DBD_tendency12}. The shorter voltage cycles limit ion recombination time, allowing the oscillating electric field to maintain a higher density of free electrons. This enhances ionization efficiency and significantly reduces breakdown voltages compared to low-frequency operation~\cite{mis_gas_breakdown2}. Studies of atmospheric-pressure gas breakdown in the 1–100 MHz range confirm that higher frequencies improve ionization efficiency and lower the voltage threshold for discharge initiation~\cite{mis_gas_breakdown, DBD_tendency_RF1}. 

\begin{figure*}[!h]
    \vspace{-10pt}
    \centering    
    \subfloat[][]{\label{fig:total1}
    \kern-1em
    \centering
    \includegraphics[height=7.5cm]{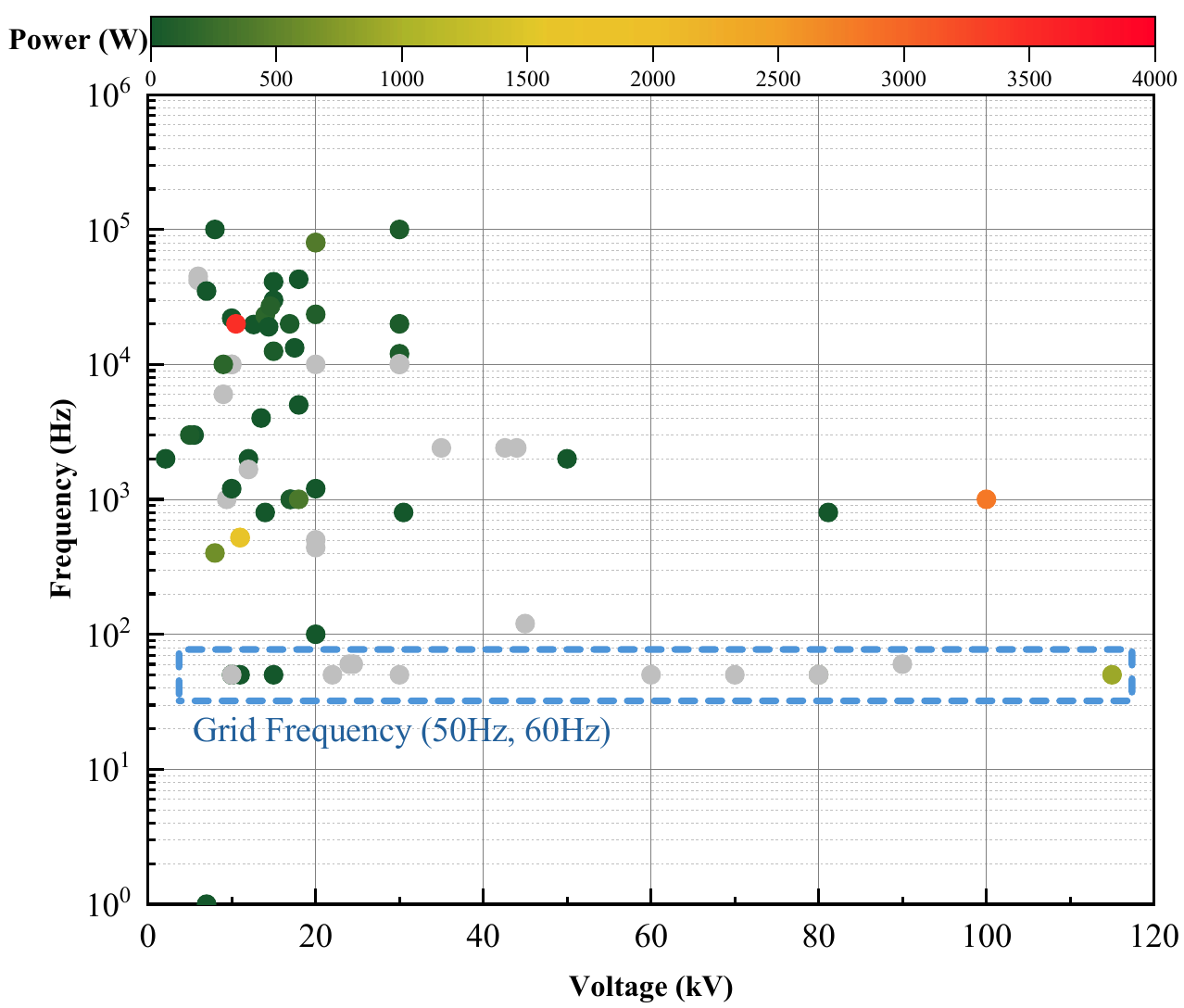}
    % \caption{Closed Loop Control}
    }
    % \hspace{-5pt}
    \subfloat[][]{\label{fig:total2}
    \kern-1em
    \centering
    \includegraphics[height=7.5cm]{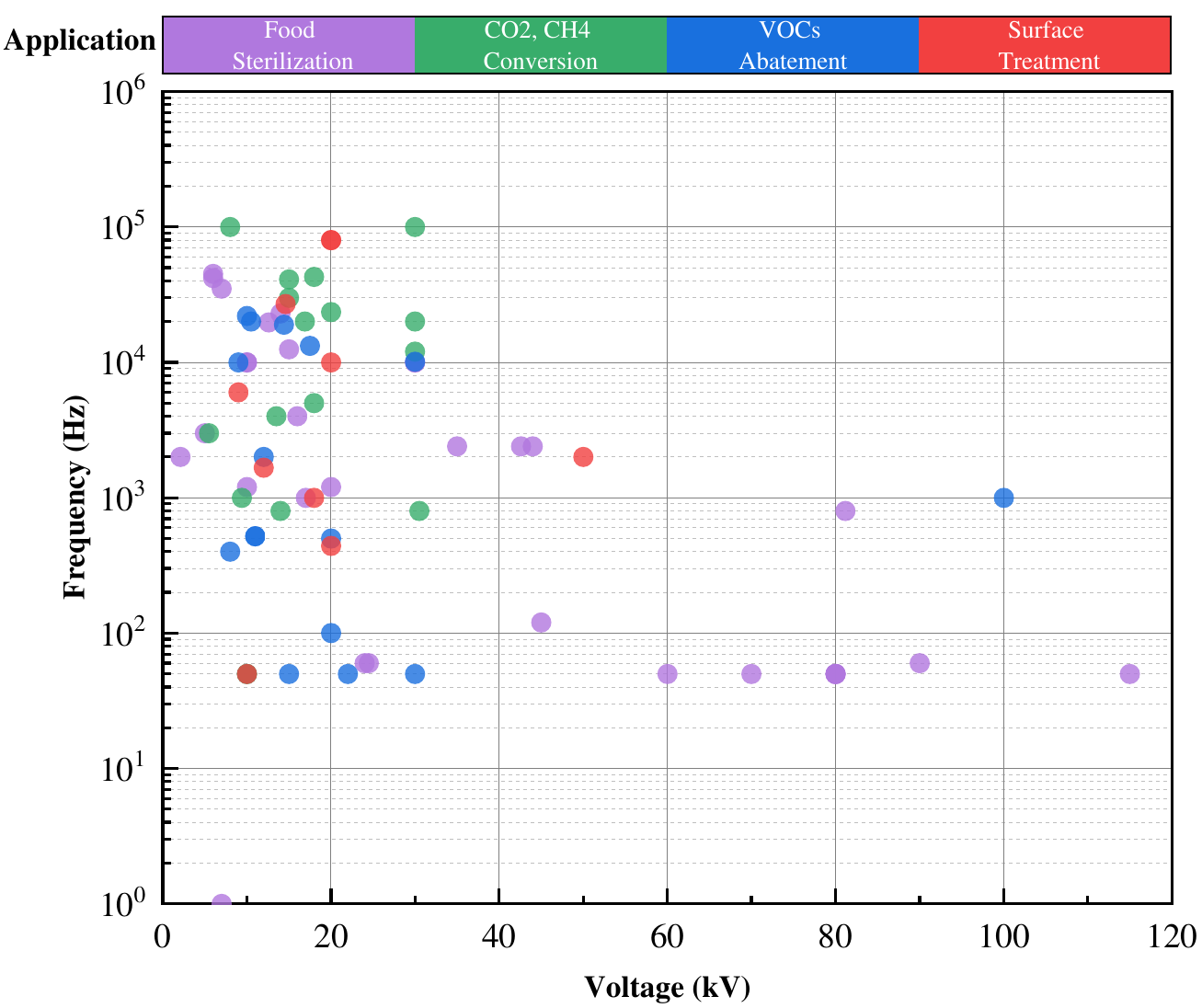}

    }
    \\ 
    \vspace{-10pt}
    \subfloat[][]{\label{fig:total3}
    \centering
    \includegraphics[width=0.48\linewidth]{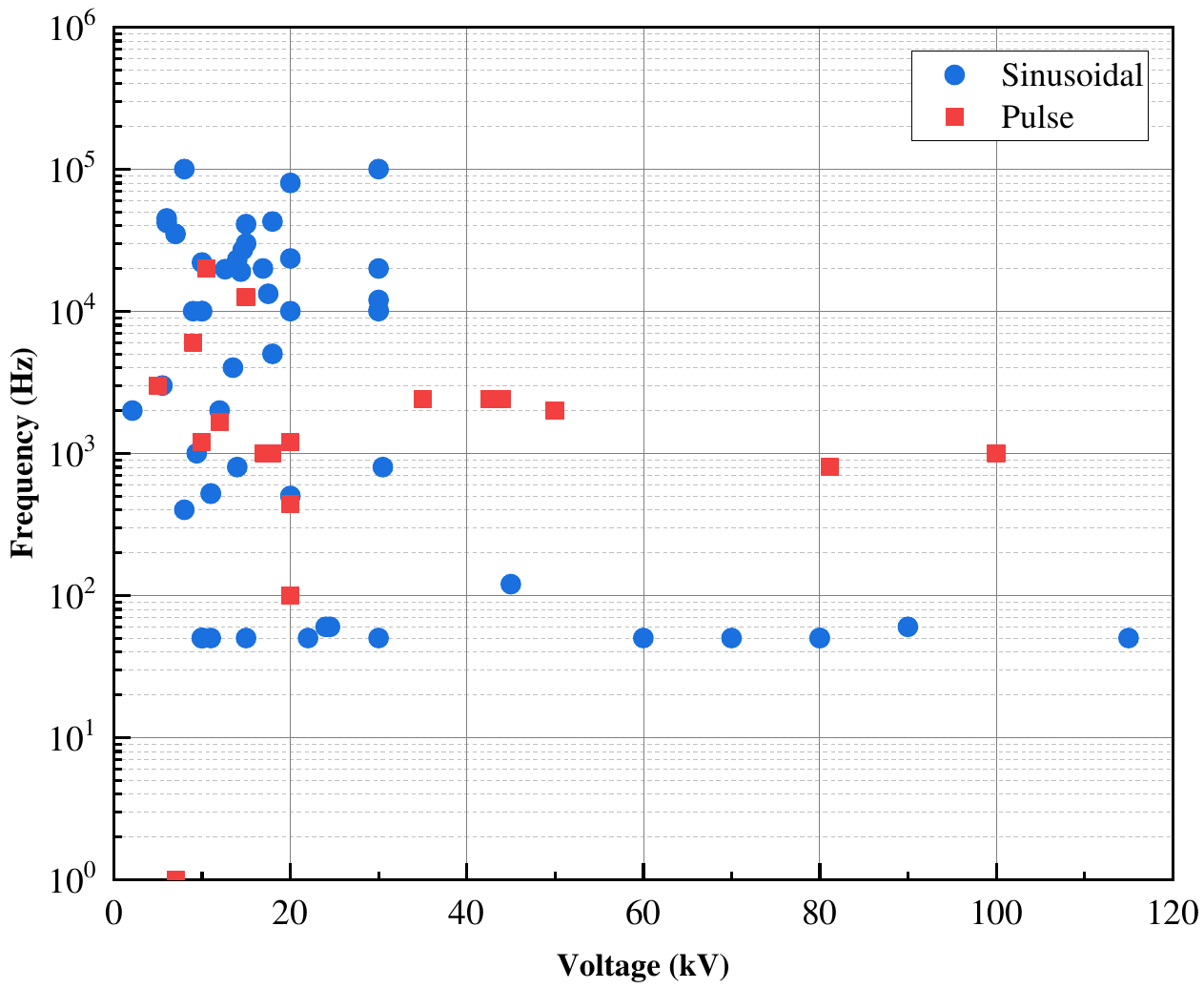}
    % \caption{Closed Loop Control}
    }
    % \hspace{-5pt}
    \subfloat[][]{\label{fig:total4}
    \centering
    \vspace{-5pt}
    \includegraphics[width=0.48\linewidth]{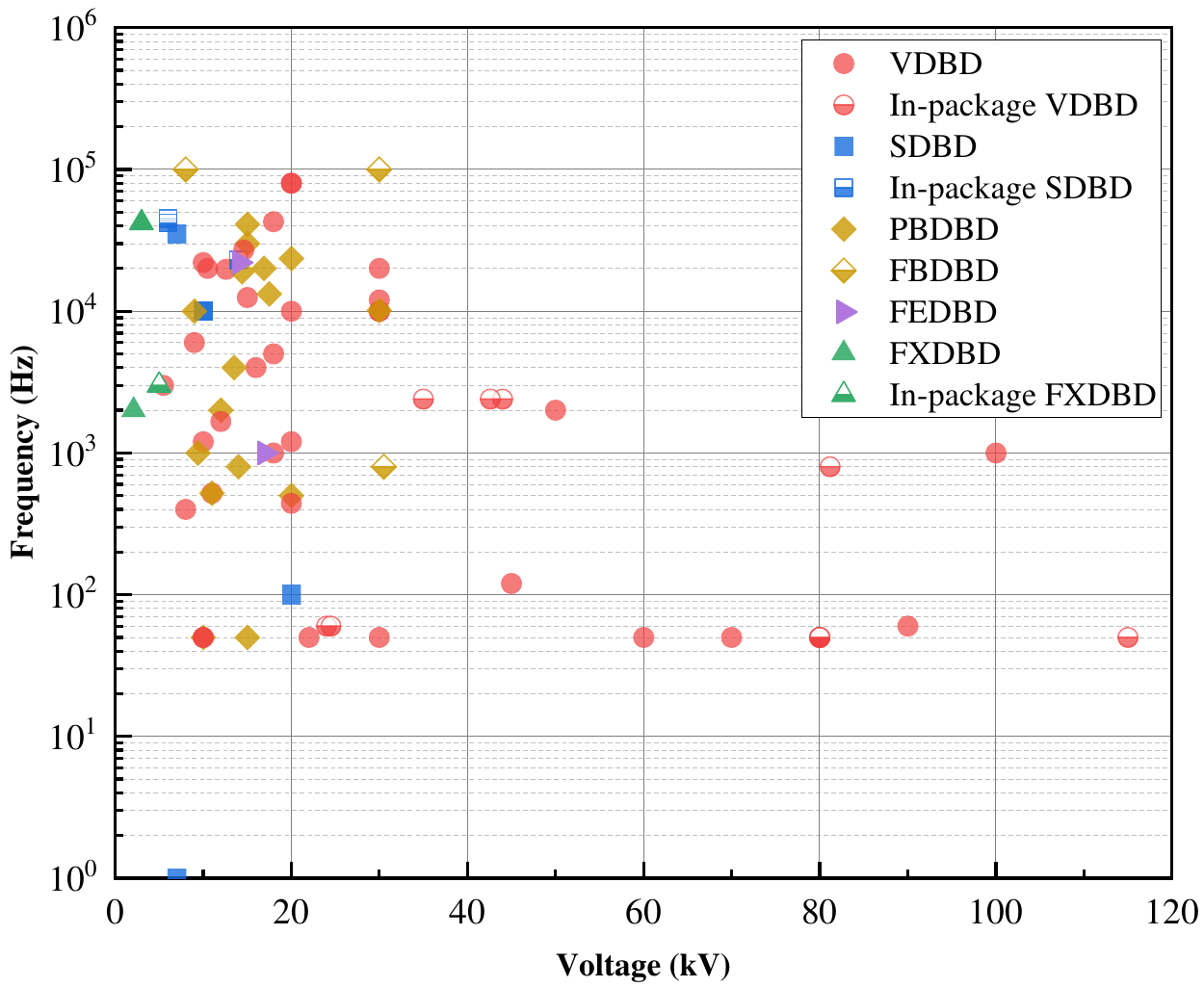}

    }
    
    % \subfloat[][]{\label{fig:VDBD13}
    % \centering
    % \includegraphics[height=2.3cm]{figs/vdbd_PB_svg-tex.pdf}
    % % \includegraphics[width=0.45\textwidth]{Figs/is_VIv_1_svg-tex.pdf}
    
    % }
    % \subfloat[][]{\label{fig:VDBD14}
    % \centering
    % \includegraphics[height=2.3cm]{figs/vdbd_FB_svg-tex.pdf}
    % % \includegraphics[width=0.45\textwidth]{Figs/is_VIv_1_svg-tex.pdf}
    
    % }
    \vspace{-5pt}
    \caption{Specifications of DBD setups from the reported cases. Each data point occupies the same position across all four subfigures to represent the same case. (a) Voltage vs. frequency with average output power levels. Gray markers indicate cases where power is not reported. (b) Voltage vs. frequency for various applications. (c) Voltage vs. frequency for sinusoidal and pulsed excitation methods. (d) Voltage vs. frequency for different DBD geometries.}
    \label{fig:total}
    \vspace{-5pt}
\end{figure*}

\begin{table*}[!h]
\centering
\caption{Power Supply Specifications from the Reported Cases}
% \vspace{-3pt}
\label{tab:power_specs}
\scalebox{0.85}{%
\begin{tabular}{c|c|cc|c|c|c}
\toprule
\textbf{Applications}
  & \textbf{DBD Geometry}
    & \multicolumn{2}{c|}{\textbf{Waveforms}} 
      & \textbf{Voltage} (kV)
        & \textbf{Power} (W) 
          & \textbf{Frequency} (Hz) \\ 
\midrule
%---------------- Food Sterilization ----------------
\multirow{10}{*}{\makecell{Food Sterilization}}
  & \multirow{2}{*}{VDBD}
    & Sinusoidal  &~\cite{Food5,Food6,Food11,Food13,Food15,Food16,Food34,Food43}
      & 11$-$90    & 6.48$-$600     & 50$-$19,700 \\ \cline{3-7}
  &  
    & Pulse       &~\cite{Food14,Food30,Food40,Food41}
      & 10$-$20    & 20$-$57        & 1,200$-$12,500 \\
\cline{2-7}
  & \multirow{2}{*}{\makecell{In-package VDBD}}
    & Sinusoidal  &~\cite{Food10,Food17,Food20,Food24,Food25}
      & 11.5$-$80  & 900          & 50$-$60 \\ \cline{3-7}
  &  
    & Pulse       &~\cite{Food4,Food22,Food27,Food28}
      & 35$-$81.2  & 33.9         & 800$-$2,400 \\
\cline{2-7}
  & \multirow{2}{*}{SDBD}
    & Sinusoidal  &~\cite{Food45,Food46}
      & 7$-$10     & 15           & 10,000$-$35,000 \\ \cline{3-7}
  &  
    & Pulse       &~\cite{Food21}
      & 7        & 13.5         & 1 \\
\cline{2-7}
  & In-package SDBD
    & Sinusoidal  &~\cite{Food8,Food9,Food44}
      & 6$-$14     & 190          & 23,000$-$45,000 \\ 
\cline{2-7}
  & FXDBD
    & Sinusoidal  &~\cite{Food37}
      & 2.1      & 10           & 2,000 \\
\cline{2-7}
  & In-package FXDBD
    & Pulse       &~\cite{Food7}
      & 5        & 45           & 30,000 \\
\cline{2-7}
  & FE-DBD
    & Pulse       &~\cite{Food2}
      & 17       & 48.4         & 1,000 \\
\cline{2-7}
  & Conveyer type
    & Sinusoidal  &~\cite{Food19}
      & 10       & –            & 10,000 \\
\hline
%---------------- CH4 & CO2 Conversion ----------------
\multirow{3}{*}{\makecell{CH$_4$ \& CO$_2$ Conversion}}
  & VDBD
    & Sinusoidal  &~\cite{CH441,CH442,CO243,CO246,CO248}& 5.5$-$30   & 6$-$80         & 50$-$42,800 \\
\cline{2-7}
  & PB-DBD
    & Sinusoidal  &~\cite{CH439,CO244,CO249,CO252,CO253}& 9.4$-$20   & 3$-$100        & 50$-$41,000 \\
\cline{2-7}
  & FB-DBD
    & Sinusoidal  &~\cite{CO254,CO255,CO256}
      & 8$-$30.5   & 4$-$60         & 800$-$100,000 \\
\hline
%---------------- VOC Abatement ----------------
\multirow{4}{*}{\makecell{VOCs Abatement}}
  & \multirow{2}{*}{VDBD}
    & Sinusoidal  &~\cite{VOC107,VOC108,VOC117}& 8–30     & 39.6$-$1,600  & 50$-$22,000 \\ \cline{3-7}
  &  
    & Pulse       &~\cite{VOC114,VOC117}
      & 10.5$-$100 & 2,830$-$3,450 & 1,000$-$20,000 \\
\cline{2-7}
  & SDBD
    & Pulse       &~\cite{VOC110}
      & 20       & 7            & 100 \\
\cline{2-7}
  & PB-DBD
    & Sinusoidal  &~\cite{VOC106,VOC109,VOC112,VOC113}& 9$-$30     & 0.96$-$1,600  & 50$-$19,000 \\
\hline
%---------------- Surface Treatment ----------------
\multirow{2}{*}{\makecell{Surface Treatment}}
  & \multirow{2}{*}{VDBD}
    & Sinusoidal  &~\cite{surface1,surface2,surface3,surface4,surface6}
      & 12$-$50    & 143$-$100,000 & 50$-$80,000 \\
      \cline{3-7}
  &  
    & Pulse       &~\cite{surface8,surface9,surface10,surface11}
      & 9$-$200    & 19.2$-$387     & 25$-$6,000 \\
\bottomrule
\end{tabular}%
} % end \scalebox
\end{table*}

\vspace{-10pt}
\subsection{Effect of Rising and Falling Times}

For pulsed excitation, the rising time and falling time significantly impact plasma dynamics. A shorter rising time enhances the ionization rate and electric field strength, resulting in higher peak currents and improved plasma uniformity. This suppresses filamentary discharges and promotes a more diffuse plasma layer~\cite{DBD_rise_Xie_2019}. Unlike filamentary discharge mode—where a trade-off typically exists between high ionization rate and plasma uniformity—pulsed excitation with a fast rising time enables both to be achieved simultaneously. Furthermore, shorter rising times intensify pressure waves by increasing their amplitude and propagation distance, making them highly effective for aerodynamic applications such as flow control and plasma mixing~\cite{DBD_rise_Liu_2001}. 

While less influential than the rising time, the falling time governs the discharge termination behavior and energy dissipation. A shorter falling time allows for faster quenching of the discharge, reducing recombination processes and minimizing energy losses. This contributes to maintaining plasma uniformity and ensuring efficient energy deposition. Additionally, shorter falling times sharpen the wavefront of induced pressure waves, enhancing their aerodynamic effects~\cite{DBD_rise_benard_2012}. 
However, shorter rising and falling times can also increase stress on the electrodes due to abrupt changes in current, necessitating careful consideration of trade-offs between performance and component longevity~\cite{DBD_rise_Zhang_2019}.

\subsection{Effect of Modulation Techniques}
Modulation techniques enhance the flexibility and controllability of DBD operation by enabling the use of custom waveform profiles and tunable discharge characteristics. Among them, \textit{burst-mode excitation} has been studied in excimer lamps and ozone generation. At HF operation, the memory effect generally improves discharge uniformity. However, over time,particularly in filamentary discharge regimes or without periodic zero-voltage intervals, non-uniform charge accumulation occurs. This eventually leads to localized overcharging, reinforcing filamentary discharges, and degradation in spatial uniformity. Consequently, burst mode, a hybrid waveform between continuous sinusoidal and short-pulse excitation methods (shown in Fig.~\ref{fig:burst}), provides zero-voltage (or low-frequency) idle intervals. These intervals facilitate ion recombination on barrier surfaces, lateral drift in surface electric fields, and charge diffusion driven by gradients, improving overall discharge performance.

Burst mode is typically realized using HF AC pulses separated by zero-voltage idle intervals (see Fig.~\ref{fig:burst}) or via amplitude modulation with a HF carrier. By adjusting parameters such as the burst duty ratio ($D_{burst}$) and base frequency ($f_b$), a reset operation is enabled, allowing efficient distribution of DBD power. For example, in Xenon DBD excimer lamps, burst-mode excitation achieves improved vacuum-ultraviolet efficiency and uniform, filament-free discharge~\cite{mod_excimer1}. In ozone generation, periodic idle intervals reduce gas temperature and the rate of ozone decomposition, improving energy efficiency and enabling stable, tunable ozone concentrations~\cite{mod_ozone1, mod_ozone2, mod_ozone3}.

\subsection{Discussion}
These findings emphasize that power supply configurations for DBD systems cannot be universal but must be designed to the specific requirements of each application. Table~\ref{table:different_modes} summarizes the generalized characteristics of different DBD operating modes, each offering distinct advantages suited to particular applications. Notably, filamentary discharges, often occurring at excessively high voltage levels, may bypass intermediate diffuse modes (i.e., TDBD and GDBD). Thus, the careful selection of waveforms, operating frequencies, and voltage levels is essential to optimize energy efficiency, discharge uniformity, and system stability.

\section{Power Supply Considerations for DBD Applications} \label{sec:4}
%%%%%%%%%%%%%%%%%%%%%%%%%%%%%%%%%%%%%%%%%%%%%%%%%%%%%%%%%%%%%%%%%%%%%%%%%%%%%%%%%%%%%%%%%%%%%%%%%%
\subsection{Clarifying DBD-Enabled Processes and Applications}
Conventionally, ozone generation and excimer lamp emission are treated as applications of DBD technology. However, a more accurate description considers these as underlying physical processes enabled by DBD. In contrast, activities such as food sterilization, CO\textsubscript{2}/CH\textsubscript{4} conversion, volatile organic compound (VOC) abatement, and surface treatment are best categorized as end-use applications of DBD-enabled processes.

Since DBD produces a non-thermal plasma environment, it enables various plasma-chemical and plasma-photonic processes, including ozone generation (O\textsubscript{2} $\rightarrow$ O\textsubscript{3}), excimer formation (rare-gas dimers emitting UV light), radical generation, and molecular dissociation. These effects can be used, for example, through direct ozone treatment~\cite{clarification_food_ozone}, plasma-activated water, or UV irradiation from DBD-excimer lamps~\cite{clarification_food_excimer} to inactivate bacteria and viruses. Such plasma processes are therefore central to applications ranging from sterilization, chemical conversion, and pollution abatement to surface modification. This clarification helps align the terminology with the perspective of power electronics engineers.

\vspace{-10pt}
\subsection{Power Supply Requirements: Voltage, Frequency, Power}

Accordingly, Table~\ref{tab:power_specs} summarizes the specifications of the power supplies in various applications, including ongoing research in food sterilization, CH\textsubscript{4} and CO\textsubscript{2} conversion, VOC abatement, and surface treatment. To meet the diverse requirements discussed in Sec.~\ref{sec:3}, DBD power supplies are categorized into sinusoidal AC sources and pulsed bipolar/unipolar sources. The table highlights excitation waveform shapes, frequencies, and voltage and power requirements for different DBD applications. As shown, \textit{voltage} requirements range from several kV to over 100 kV, while \textit{operating frequencies} span from a few Hz to several MHz.

\textit{Power demand} is primarily influenced by the reactor's load configuration, including electrode and reactor geometry. Larger systems, such as VDBD setups, require higher power levels due to increased electrode separation and larger reaction volumes, which necessitate higher energy input for sustaining effective plasma generation. In contrast, SDBD systems typically operate at only a few watts because of their smaller discharge areas and closer electrode spacing.

Fig.~\ref{fig:total} shows the specifications of various DBD systems. It shows that high power levels---ranging from kW to MW---are rarely reported in the literature~\cite{ozone_kw_1, ozone_kw_2, ozone_kw_3, ozone_kw_4}. However, 100 kW-level power for surface treatment and several MW-level power for industrial ozone generation were documented as early as the 2000s~\cite{DBD_application_review}. Additionally, several commercial systems with power ratings in the tens of kW have been reported~\cite{ozone_ozcon, ozone_zenith, ozone_zonosistem}. As research and industrial interest in DBD applications continues to expand, the power requirements for these systems are anticipated to increase significantly.

\vspace{-10pt}
\subsection{Voltage-Fed vs. Current-Fed Power Supplies}\label{subsec:VsvsCs}
DBD power supplies can also be classified based on their power delivery method: voltage-fed or current-fed. Voltage-fed power supplies have long been the primary choice for DBD applications due to their simplicity, scalability, and efficiency in delivering HV excitation. Their straightforward design, often incorporating a step-up transformer, enables reliable operation with minimal complexity. Furthermore, voltage-fed supplies are cost-effective for high-power applications, making them a practical choice for industries requiring robust and scalable plasma systems. Unlike current-fed systems, voltage-fed converters do not rely on input inductors or a front-end current regulator, reducing their size and complexity—an advantage for cost-sensitive or space-constrained applications. Furthermore, implementing soft-switching schemes in voltage-fed converters enables a more compact design by minimizing switching losses and enhancing efficiency~\cite{HB_inverter}.

However, voltage-controlled systems face challenges in maintaining precise current control, particularly with the varying capacitive loads characteristic of DBD systems. For uncontrolled or discontinuous voltage waveforms—such as square-wave voltages with undefined slew rates or arbitrary non-sinusoidal waveforms—it becomes difficult to predict \( i_{dis} \) given the discharge current of a DBD system is the derivative of the applied voltage as shown in \eqref{eqn:Vs_current}.
\begin{align}\label{eqn:Vs_current}
    i_{o} = i_{dis} = C_{d} \frac{d(v_{o}-V_{th})}{dt}
\end{align} 

In contrast, current-controlled converters enable well-regulated \( i_{dis} \), making it possible to accurately predict the energy dissipation of the DBD load, while the voltage across the electrodes can be well predicted even for arbitrary current waveforms, as described by \eqref{eqn:Cs_voltage}.
\begin{align}\label{eqn:Cs_voltage}
    v_{o} = V_{th} + \frac{1}{C_{d}} \int i_{o} \, dt
\end{align}

This precise current control allows current-fed power supplies to offer significant advantages in applications where stability and uniformity of plasma discharge are critical. By directly regulating the injected current, these systems ensure consistent plasma generation and improved discharge uniformity. Such capability is particularly vital in applications like excimer lamps and biomedical treatments, where precise plasma characteristics are essential to achieving optimal results. A key benefit of current-fed designs is their ability to regulate power delivery independently of the capacitive load's varying properties. By employing a constant current source, the energy delivered to the DBD reactor is accurately controlled. 

Moreover, the capacitive load characteristics of DBD applications pose significant challenges. For instance, under square-voltage excitation with an uncontrolled slew rate, the voltage retained by the DBD load can create a significant voltage difference relative to the applied voltage. This leads to large $dV/dt$, resulting in uncontrolled inrush currents and potential short-circuit behavior. To address this, the converter must accommodate the capacitive characteristic of the DBD load and incorporate short-circuit-tolerant properties—a feature inherently supported by current-fed converters.

Beyond general capacitive effects, a particularly critical challenge arises from the threshold-triggered, nonlinear behavior of DBD loads—especially during filamentary discharge. In such cases, the abrupt collapse of load impedance upon plasma initiation can induce severe current surges in voltage-fed systems, placing significant stress on switching devices, transformers, and resonant components. In contrast, current-fed inverters---typically incorporating an input choke inductor---naturally limit these transient current spikes by maintaining a quasi-constant or regulated input current, thereby enhancing system reliability under abrupt discharge events.

This quasi-constant or constant input current behavior often leads to reduced harmonic distortion and better input power factor, compared to voltage-fed systems. Although comprehensive input power quality analyses specifically targeting current-fed DBD power supplies remain scarce~\cite{recent_current_fed}, numerous studies on general current-fed converter topologies demonstrate their inherent capability to shape input current waveforms and achieve near-unity power factor~\cite{current_fed_pf1,current_fed_pf2,current_fed_pf3,current_fed_pf4,current_fed_pf5,current_fed_pf6,current_fed_pf7}. A DBD-specific implementation employing a passive current-fed rectifier has also reported improved input power quality compared to conventional diode-rectifier systems~\cite{CS_push_pull}.

\subsection{Voltage Step-Up: Non-Isolated vs. Isolated Power Supplies}

The HV demand of DBD loads necessitates the inclusion of a voltage-boosting stage to address the limitations of input power supply voltage. This can be accomplished using either transformer-based (isolated) and transformerless (non-isolated) configurations. 

In transformerless designs, HV amplifier structures, such as those based on operational amplifiers and driven by function generators, can be utilized for DBD applications~\cite{dbd_function_gen}. However, such designs are constrained by limited power output, making them unsuitable for high-power applications. To address this limitation, sinusoidal or pulse-generating stages with HV gain can be implemented, thereby meeting the threshold voltage levels for DBD applications. This stage typically comprises resonant pulse generators for pulse excitation (e.g., half-bridge or full-bridge configurations) or resonant inverters for sinusoidal excitation~\cite{XFMR_FB, XFMR_FB_LCL, XFMR_single_switch, XFMR_thy}. 

A primary design challenge is the precise tuning of the resonant circuit, as DBD loads exhibit complex and nonlinear characteristics. Specifically, DBD loads are not only capacitive but also possess a threshold voltage. This threshold-dependent behavior causes the load's characteristics to shift dramatically once the applied voltage exceeds the threshold, as shown in Fig.~\ref{fig:DBD_equi}. Moreover, the capacitive nature of the load contributes to a low output power factor, presenting additional challenges for system efficiency~\cite{mis_low_pf}. Furthermore, in transformerless configurations, switching devices are subjected to the same HV levels as the amplified output, necessitating components with higher blocking voltage ratings.

Beyond HV requirements, power specifications introduce additional design complexities. The power rating is determined by several factors, including application type, reactor geometry, electrode design, and material properties. For high-power applications exceeding tens of kW—such as large ozone generators with power densities ranging from 1 to 10 kW/m$^2$ or surface treatment systems requiring over 50 kW~\cite{DBD_power_review}—transformerless designs are not practical. These applications demand robust isolation, typically achieved by incorporating transformers to decouple the primary power supply from the secondary load. This approach ensures compliance with operational safety standards and supports reliable performance in high-power systems. Additionally, as switching solid-state devices are typically positioned on the primary side with lower voltage levels, the blocking voltage requirements can be significantly reduced.

However, one challenge with the HV transformer is the need for \textit{isolation} in the design process. Thick dielectric materials are commonly used between windings and between the windings and the core to ensure isolation. Achieving sufficient isolation often requires increasing the spacing between windings or enlarging winding areas to meet voltage breakdown limits. While thicker insulation reduces direct capacitance, the use of high-permittivity materials to meet isolation standards can inadvertently increase parasitic capacitance. Given that the DBD load is typically modeled as a capacitive element, typically ranging in tens of pF, the interaction between the transformer's parasitic capacitance and the load capacitance creates a current divider effect, thereby reducing overall power efficiency~\cite{DBD_cap}.

Furthermore, transformer designs must account for the specific waveform requirements of the application, with the secondary output potentially being sinusoidal, bipolar, or unipolar square pulses. Unlike sinusoidal and bipolar pulse excitations, achieving unipolar operation with a pulse transformer presents additional design challenges. A reset circuit is necessary to prevent core saturation during unipolar operation. While this approach ensures reliable transformer functionality, it also introduces additional components, increases design complexity, and imposes limitations on operational frequency.

In addition to these challenges, the leakage inductance of HV transformers imposes additional challenges. Specifically, the leakage inductance restricts the minimum achievable pulse width, complicating the implementation of transformers in DBD applications requiring higher pulse repetition frequencies (PRFs). Furthermore, leakage inductance can result in higher voltage spikes across the switches during commutation, necessitating a carefully designed transformer to mitigate these effects and ensure reliable operation.

\section{Sinusoidal Power Supplies}\label{sec:5}
%%%%%%%%%%%%%%%%%%%%%%%%%%%%%%%%%%%%%%%%%%%%%%%%%%%%%
%%%%%%%%%%%%%%%%%%%%%%%%%%%%%%%%%%%%%%%%%%%%%%%%%%%%%
%%%%%%%%%%%%%%%%%%%%%%%%%%%%%%%%%%%%%%%%%%%%%%%%%%%%%%%%%%%%%%%%%%%%%%%%%%%%%%%%%%%%%%%%%%%%%%%%%%

% \subsection{Resonant Inverter}

While conventional bridge-based voltage-source inverters (VSIs) are commonly used for inductive loads such as motors, these are not well-suited for generating sinusoidal output voltage waveforms when driving DBD with sinusoidal excitation. Multi-level modular voltage inverters have been explored as a potential solution to mitigate the high $dV/dt$ associated with such loads~\cite{modular1, modular2, modular3}. However, these inverters still suffer from uncontrolled current peaks due to the capacitive nature of the DBD load, which generates HF current spikes~\cite{recent_modular}. These spikes necessitate additional circuit components, such as series inductors, to stabilize the operation. 

Alternatively, traditional transformer-based design, achieved by connecting a transformer to the power grid, allows for HV gain at grid frequencies (e.g., at 50 Hz, see Fig.~\ref{fig:total}(a)). While effective for basic setups, this method is limited by its inability to offer adjustable frequency options and variable voltage gains, reducing its suitability for advanced DBD applications requiring greater versatility.

This challenge is effectively addressed by resonant inverters, such as the class-D resonant inverter, which naturally incorporates the capacitive DBD load into its design. By leveraging resonance, these inverters provide smooth sinusoidal excitation and ensure stable operation. Both voltage-fed and current-fed resonant inverters have been successfully implemented to drive DBD systems, as elaborated in this Section.

\vspace{-10pt}
\subsection{Voltage-Fed Resonant Inverters}

\subsubsection{Transformerless Topologies}
\begin{figure}[!t]
    \centering
    \includegraphics[width=0.8\linewidth]{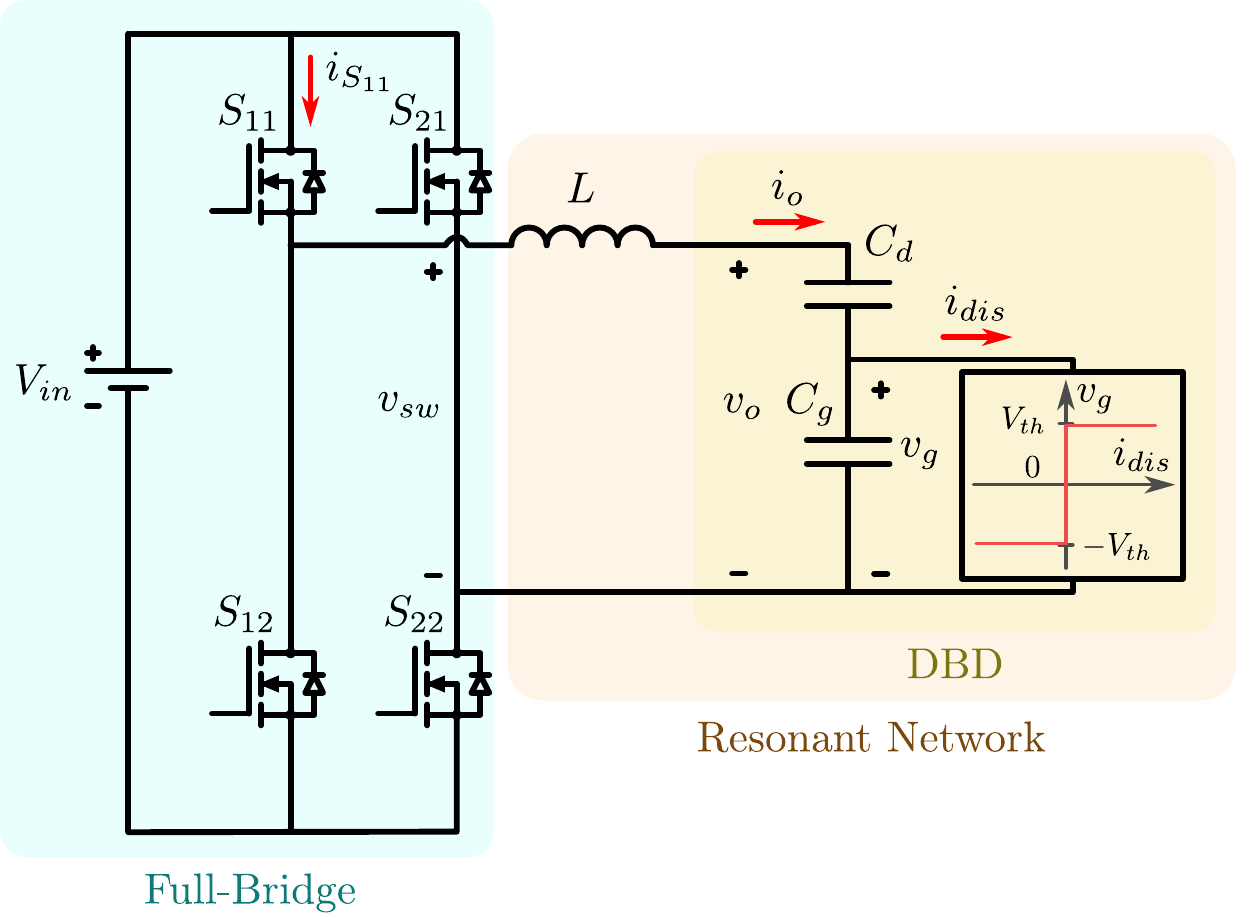}
    \caption{Transformerless voltage-fed full-bridge resonant inverter.}
    \label{fig:XFMR_FB_L}
    \vspace{-10pt}
\end{figure}

The full-bridge-based inverter with a resonant tank, as shown in Fig.~\ref{fig:XFMR_FB_L}, is a widely utilized topology for generating sinusoidal output voltages while meeting the HV requirements of DBD loads. The resonant tank facilitates voltage amplification by leveraging the intrinsic capacitances of the DBD load, including $C_g$ and $C_d$. By incorporating a single inductor, the LCC (or simply LC) resonant tank can be analytically designed to integrate the load impedance, as done in~\cite{XFMR_FB},~\cite{XFMR_thy}. The resonant tank not only facilitates voltage amplification but also provides reasonable voltage gain without increasing the blocking voltage requirements of the switching devices. Transformerless implementations of this topology further reduce the cost and complexity of the power electronics.

\begin{figure}[!t]
    \centering
    \includegraphics[width=0.6\linewidth]{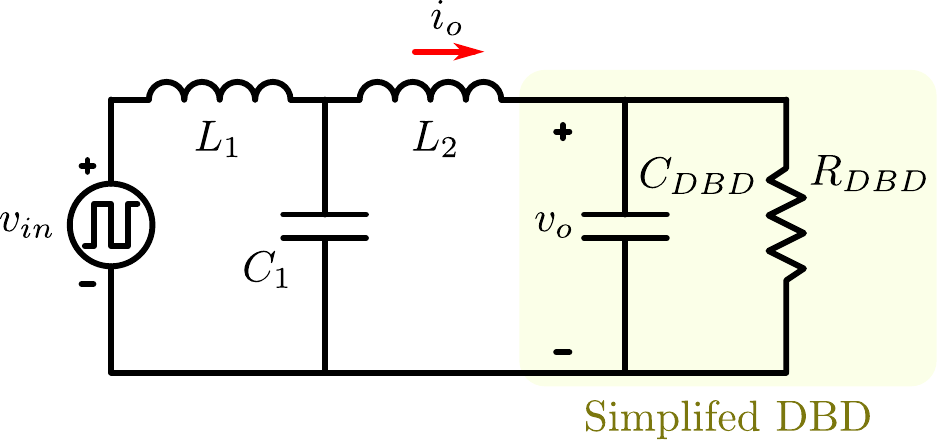}
    \caption{Equivalent circuit of LCL resonant tank with DBD load.}
    \label{fig:XFMR_FB_LCL}
    \vspace{-10pt}
\end{figure}

For applications requiring higher voltage gains, the resonant tank design can be varied. Amjad \textit{et al.} implemented an LCL resonant circuit (see Fig.~\ref{fig:XFMR_FB_LCL}), achieving greater voltage gain for ozone generation systems~\cite{XFMR_FB_LCL,recent_FB3}. However, this design operates at a significantly lower resonant frequency than LC configurations, illustrating a trade-off between operating frequency and voltage gain. This trade-off provides design flexibility, allowing the resonant tank configuration to be decided based on the specific requirements of the DBD system.

However, environmental factors, such as temperature, can cause variations in DBD capacitances, directly affecting the achievable voltage gain of the resonant tank. To mitigate the effects of capacitance drift, compensation schemes similar to those used in LC inverter topologies can be employed, as discussed in~\cite{mis_lcc_zero}. These schemes ensure consistent performance despite fluctuations in operating conditions.

\begin{figure}[!t]
    \centering
    \includegraphics[width=0.8\linewidth]{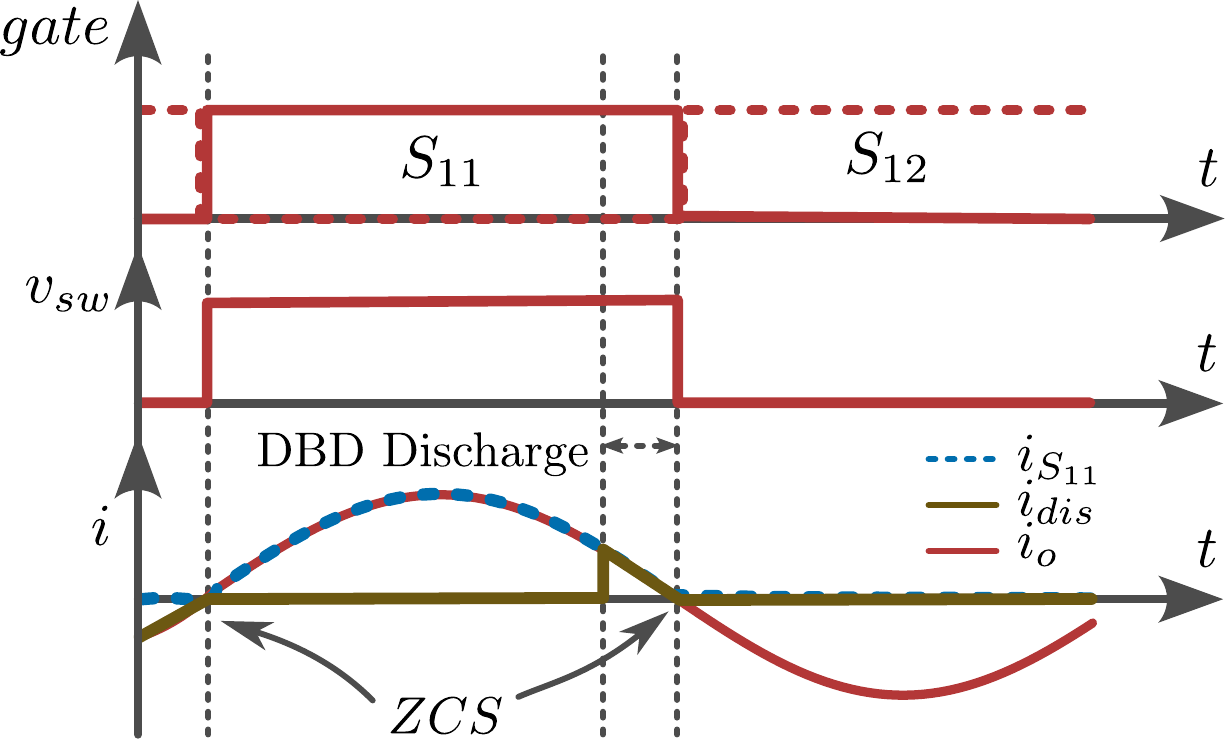}
    \caption{Exemplary waveforms of full-bridge LC inverter with DBD load in ZCS operation.}
    \label{fig:XFMR_FB_soft}
    \vspace{-10pt}
\end{figure}

Fig.~\ref{fig:XFMR_FB_soft} shows the key waveforms of the full-bridge LC inverter with a DBD load, highlighting the zero-current switching (ZCS) operation. ZCS allows for higher switching frequencies, a critical requirement for achieving efficient and stable plasma generation in DBD applications. Resonant inverters, particularly class-D and class-E configurations, are well-suited for HF operation, often reaching the MHz range. This enables DBD systems to operate in the RF-DBD mode, expanding their application potential. Among these, class-E inverters provide a cost-effective solution due to their single-switch design, higher voltage gain, and reduced component count, making them ideal for compact and high-performance power supply designs~\cite{XFMR_class_E}.

%Additionally, the sinusoidal output voltage waveform of the converter is synchronized with the switching frequency of the full-bridge operation. 
%For compact design, the inverters with single switch is possible.  
%alleviates the voltage breakdown requirement of gas in DBD reactor~\cite{mis_gas_breakdown},~\cite{mis_gas_breakdown2}. 
%In addition, LCC structure has a single dominant resonance frequency,
%However, as highlighted in~\cite{XFMR_FB}, the HV requirements of DBD applications, coupled with the voltage limitations of GaN devices (60V) and the gain constraints of the LC resonant tank, present common practical challenges. To address these limitations, a medium-voltage-level transformer is still implemented in such designs.

% \begin{figure}[!t]
%     \centering
%     \includegraphics[width=0.9\linewidth]{figs/FB_gain_LCL.png}
%     \caption{Frequency response of the full-bridge LCL inverter with DBD load \textbf{???}}
%     \label{fig:XFMR_FB_LCL_gain}
% \end{figure}

\begin{figure}[!t]
    \centering
    \includegraphics[width=0.95\linewidth]{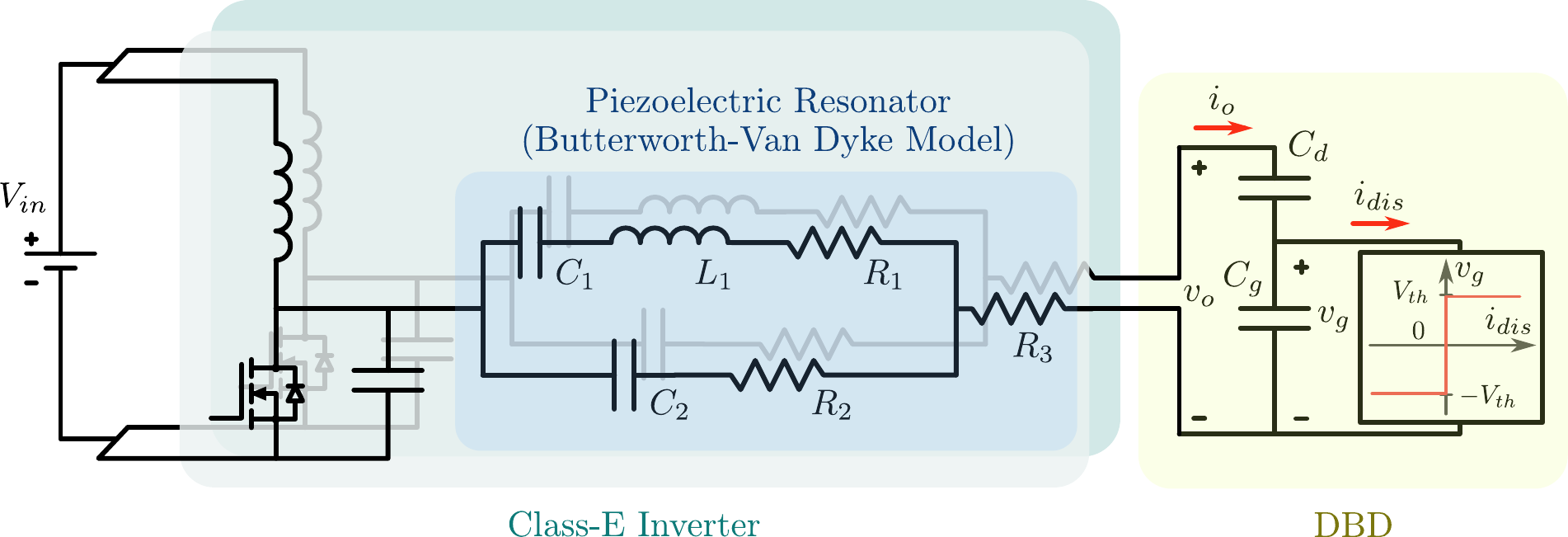}
    \caption{Piezoelectric class-E push-pull resonant inverter.}
    \label{fig:class_e_piezo_schem}
    \vspace{-10pt}
\end{figure}

% \begin{figure}[!t]
%     \centering
%     \includegraphics[width=0.8\linewidth]{figs/BVD_equi_svg-tex.pdf}
%     \caption{the BVD model of the piezoelectric resonator}
%     \label{fig:class_e_piezo_equi}
% \end{figure}

Further, the emergence of piezoelectric energy storage units represents a novel approach to integrating mechanical and electrical energy systems, leveraging the unique ability of piezoelectric materials to convert mechanical vibration into electrical resonance and vice versa. These units exhibit the high energy density and fast response of piezoelectric materials, enabling efficient energy capture, storage, and release with high quality factor compared to conventional resonant tank design. When utilized as a resonant tank, as shown in Fig.~\ref{fig:class_e_piezo_schem} using the equivalent Butterworth-Van Dyke (BVD) model, these units simplify system design, reduce losses, and support MHz range operation. A class-E inverter including piezoelectric resonator is reported in~\cite{XFMR_piezo}, achieving 6.41MHz operation with a complementary-controlled push–pull configuration to maximize voltage gain. It is noteworthy that the capacitive load affects the zero-voltage switching (ZVS) capability of the class-E inverters and the circulating current in resonant tank will reduce the overall efficiency of the converter, which requires careful tuning of circuit components.

Although the resonant tank reduces blocking-voltage stress on the switching devices and supports the HF demands of DBD excitation, the substantial voltage and power demands typical of DBD loads often necessitate the inclusion of a step-up transformer to meet application-specific requirements effectively.

%%%%%%%%%%%%%%%%%%%%%%%%%%%%%%%%%%%%%%%%%%%%%%%%%%%%%%
%%%%%%%%%%%%%%%%%%%%%%%%%%%%%%%%%%%%%%%%%%%%%%%%%%%%%%
%%%%%%%%%%%%%%%%%%%%%%%%%%%%%%%%%%%%%%%%%%%%%%%%%%%%%%
\subsubsection{Transformer-Based Topologies}

% In conclusion, while transformerless designs are feasible through optimized resonant tank implementations, a compact solution can also be achieved by substituting high-voltage transformers with smaller, medium-voltage transformers. Such adaptations maintain system efficiency while reducing the size and complexity of the power converter.

\begin{figure}[!t]
    \centering
    \includegraphics[width=0.8\linewidth]{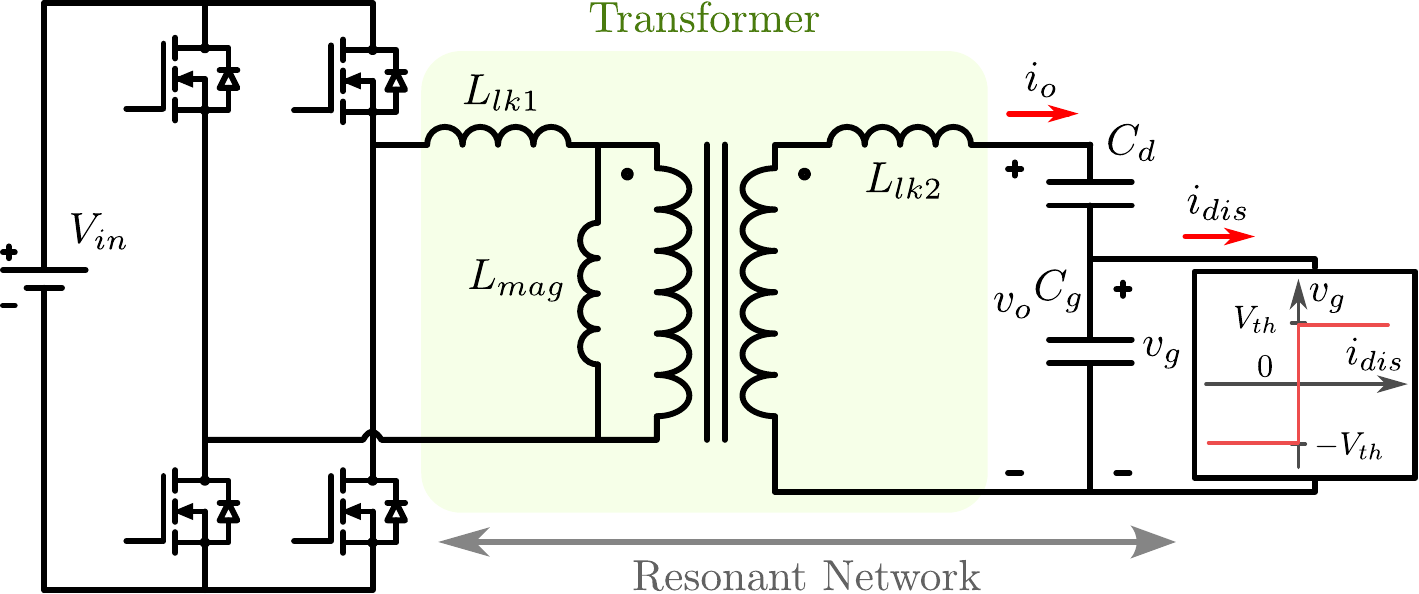}
    \caption{Transformer-based full-bridge-based topology.}
    \label{fig:VS_FB}
    \vspace{-10pt}
\end{figure}

\begin{figure}[!t]
    \centering
    \includegraphics[width=0.95\linewidth]{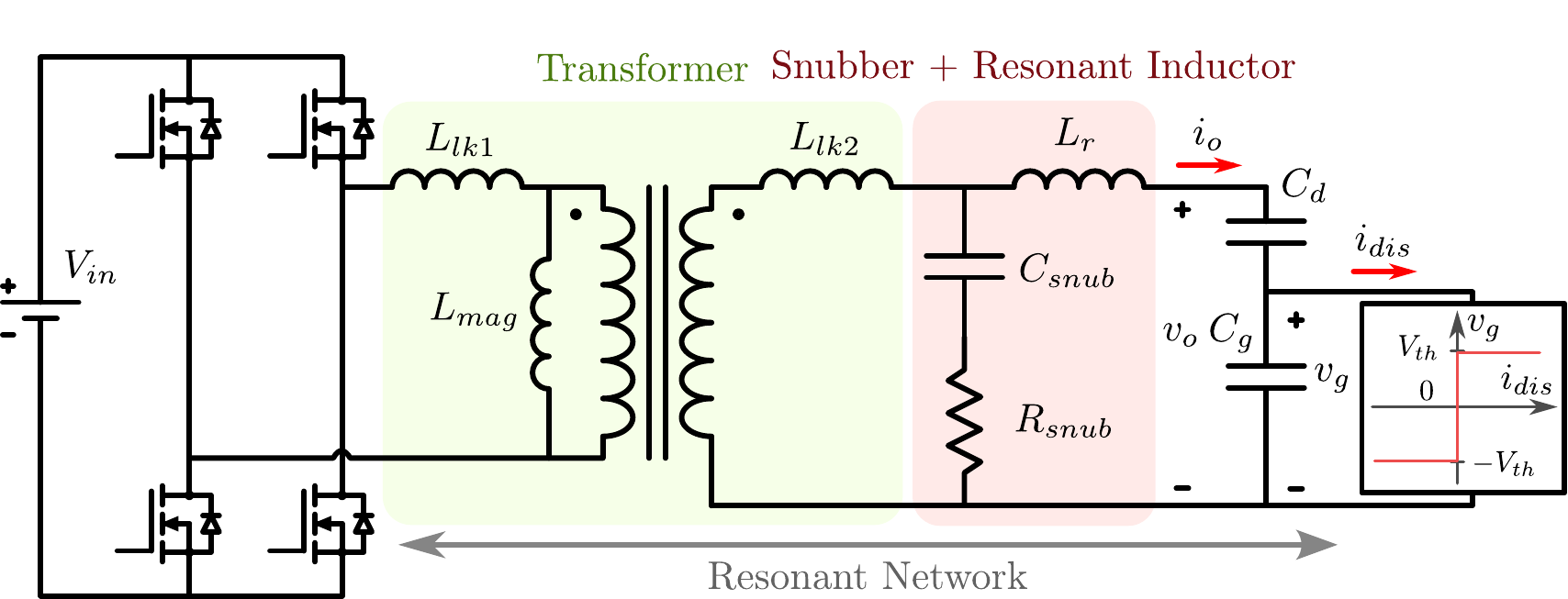}
    \caption{Transformer-based full-bridge LC resonant inverter with HF snubber.}
    \label{fig:XFMR_FB_snubber_schem}
    \vspace{-10pt}
\end{figure}

% As discussed before, HV sinusoidal voltage output in bridge-based inverters, such as bridge-based and Class-E configurations, can be achieved by incorporating a resonant tank in transformerless setups. This approach utilizes the resonance of the tank to efficiently transfer energy while ensuring soft-switching conditions for sinusoidal waveforms. The same principle can be extended to isolated configurations, as demonstrated in~\cite{XFMR_FB}, where the inclusion of a transformer facilitates sinusoidal output generation while maintaining galvanic isolation.

The bridge-based inverter structure with a transformer can be analyzed similarly to resonant DC-DC converter topologies but without a secondary rectifier. A key advantage is that soft-switching can be utilized in the primary bridge as in resonant DC-DC converters. Utilizing the inherent inductance of the transformer as a resonant component removes the need for additional inductive elements, as shown in Fig.~\ref{fig:VS_FB}. This is realized through the use of a pulse transformer, specifically designed to handle HF pulsed signals while preserving waveform integrity. 

\begin{figure}[!t]
    \centering
    \includegraphics[width=0.8\linewidth]{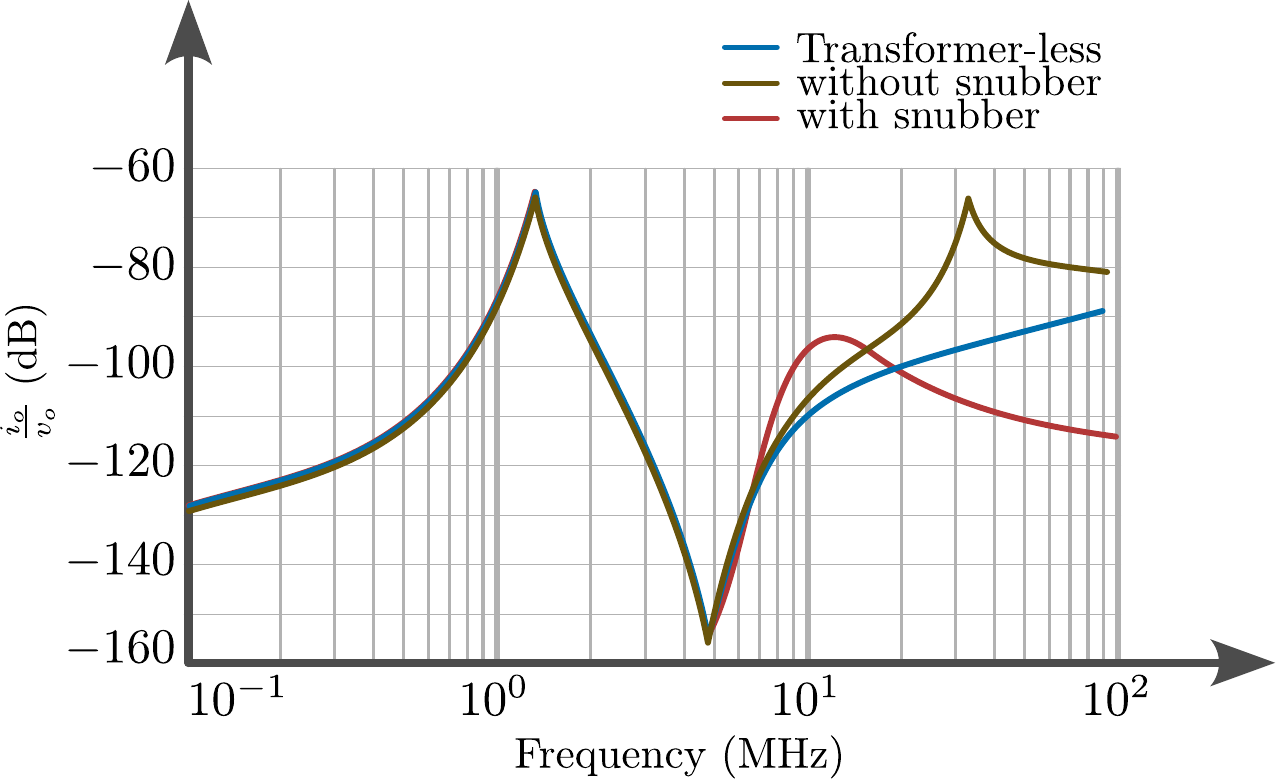}
    \caption{Exemplary frequency response of the full-bridge LC inverter with DBD load.}
    \label{fig:XFMR_FB_snubber}
    \vspace{-10pt}
\end{figure}

This approach minimizes component count and reduces overall system complexity. The combination of transformer leakage inductance and DBD capacitances forms a resonant tank that enables soft-switching operation. Phase-shifted control is employed to achieve proper ZVS, but this requires precise timing of the switching events~\cite{phase_shift_FB1,phase_shift_FB2,phase_shift_FB3, recent_FB2}. 

%On the other hand, achieving ZCS, as in transformerless LC-series designs, necessitates the inclusion of a series resonant capacitor to resonate with the leakage inductance~\cite{series_cap_FB2}. Alternatively, a series resonant inductor, $L_r$, can be added, forming a resonant network with the DBD load capacitance, as shown in Fig.~\ref{fig:XFMR_FB_snubber_schem}.

On the other hand, to achieve ZCS, as in transformerless LC-series designs, a series resonant inductor ($L_r$) can be added, forming a resonant network with the DBD load capacitance, as shown in Fig.~\ref{fig:XFMR_FB_snubber_schem}. However, it necessitates careful consideration of leakage inductance. The presence of transformer leakage inductance can introduce a secondary peak in the frequency response of the voltage gain, located near the primary resonant peak, as shown in Fig.~\ref{fig:XFMR_FB_snubber}. This secondary peak can potentially degrade the system's performance if not properly managed. To mitigate this issue, a HF snubber circuit can be employed~\cite{XFMR_FB}. The snubber circuit effectively attenuates the second peak, ensuring smoother operation and improved frequency stability. 

Moreover, the parasitic capacitance of the transformer should be managed carefully. To minimize the current flowing into the parasitic capacitance of the transformer, it is crucial to reduce the effective load impedance. This can be achieved by adding an additional inductor on the secondary side, positioned between the transformer and the DBD load. This inductive component compensates for the capacitive load, effectively minimizing the load impedance relative to the parasitic capacitance, as expressed in \eqref{eqn:L_compensation}.
\begin{align}\label{eqn:L_compensation}
    Z_{effective} = j\omega L_{res}+ Z_{DBD}
\end{align}

% \begin{figure}[!h]
%     \centering
%     \includegraphics[width=0.7\linewidth]{figs/FB_snubber.png}
%     \caption{Frequency response of the full-bridge LCC inverter with DBD load}
%     \label{fig:XFMR_FB_snubber}
% \end{figure}

It should be noted that voltage-fed resonant inverter designs for DBD applications are particularly vulnerable to the partial short-circuit behavior inherent to the threshold-like capacitive nature of DBD loads. This behavior not only compromises the system's reliability but also highlights the necessity of exploring alternative topologies. Current-fed resonant inverters emerge as a compelling solution, offering enhanced tolerance to short-circuit conditions and robust control over discharge dynamics, effectively addressing the limitations of voltage-fed designs.

% Despite these challenges, single-switch amplifiers—such as Class-E and Class-EF variations—offer compact and promising power electronic solutions for DBD applications requiring smaller converter sizes. Additionally, piezoelectric resonators can be integrated into the resonant tank, enabling innovative and efficient design approaches tailored to the unique requirements of DBD systems.
%%%%%%%%%%%%%%%%%%%%%%%%%%%%%%%%%%%%%%%%%%%%%%%%%%%%%
%%%%%%%%%%%%%%%%%%%%%%%%%%%%%%%%%%%%%%%%%%%%%%%%%%%%%

%%%%%%%%%%%%%%%%%%%%%%%%%%%%%%%%%%%%%%%%%%%%%%%%%%%%%
%%%%%%%%%%%%%%%%%%%%%%%%%%%%%%%%%%%%%%%%%%%%%%%%%%%%%
%%%%%%%%%%%%%%%%%%%%%%%%%%%%%%%%%%%%%%%%%%%%%%%%%%%%%
\vspace{-10pt}
\subsection{Current-Fed Resonant Inverters}
\subsubsection{Bridge-Based Topologies}

Beyond short-circuit tolerance, compared to voltage-fed DC/DC resonant converters, current-fed resonant converters are considered a reasonable option for applications requiring HV gain in bridge-based designs~\cite{CS_dcdc_1,CS_dcdc_2,CS_dcdc_3,CS_dcdc_4}. This is attributed not only to their achievable resonant gain but also to their reduced duty-cycle losses, the elimination of snubber circuits, and the absence of pulsating input currents, which would otherwise necessitate large filter sizes in voltage-fed designs.

\begin{figure}[!t]
    \centering
    \includegraphics[width=0.95\linewidth]{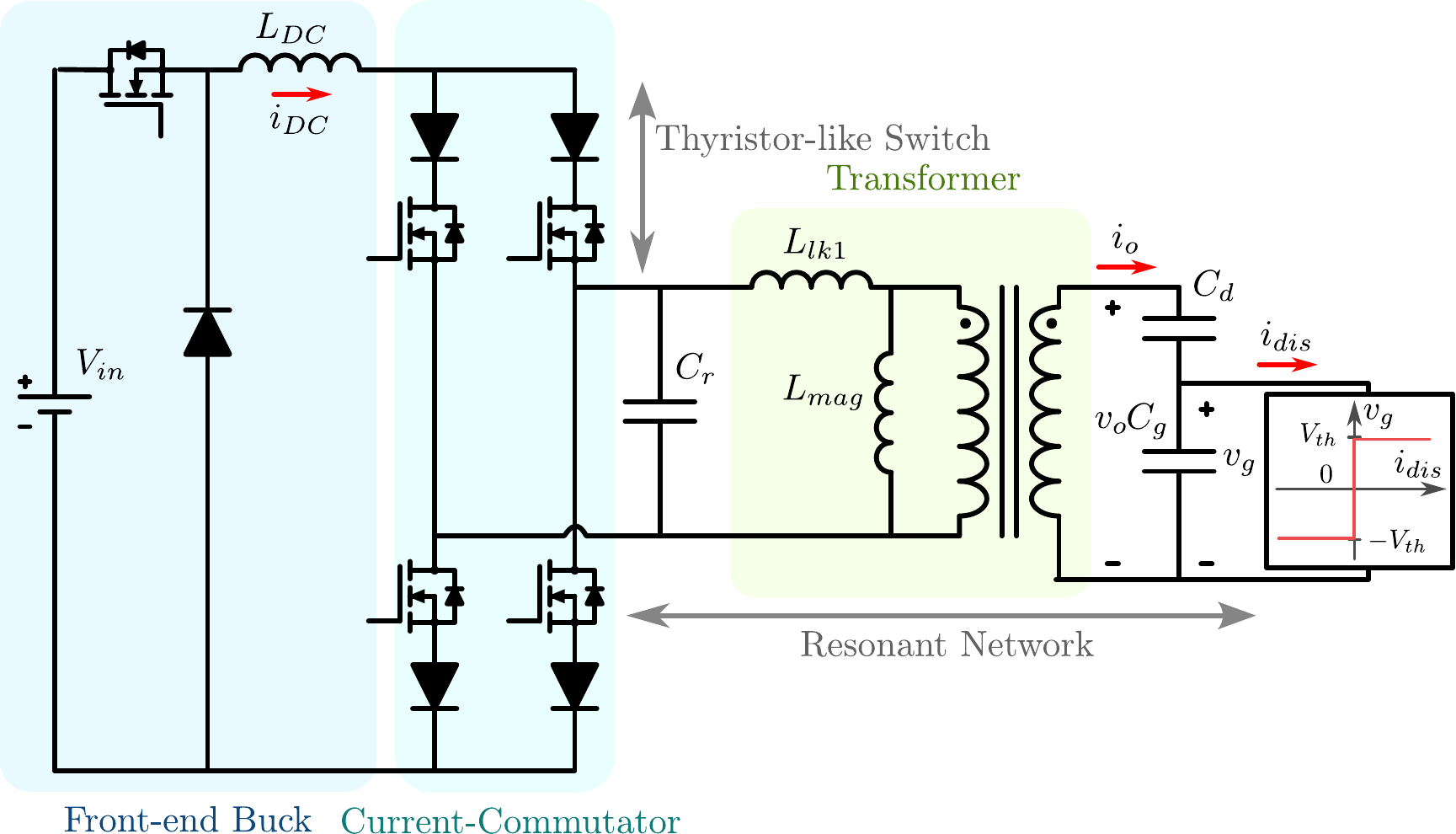}
    \caption{Current-fed CLCC resonant inverter with front-end buck.}
    \label{fig:CS_CLCC}

    \centering
    \includegraphics[width=0.8\linewidth]{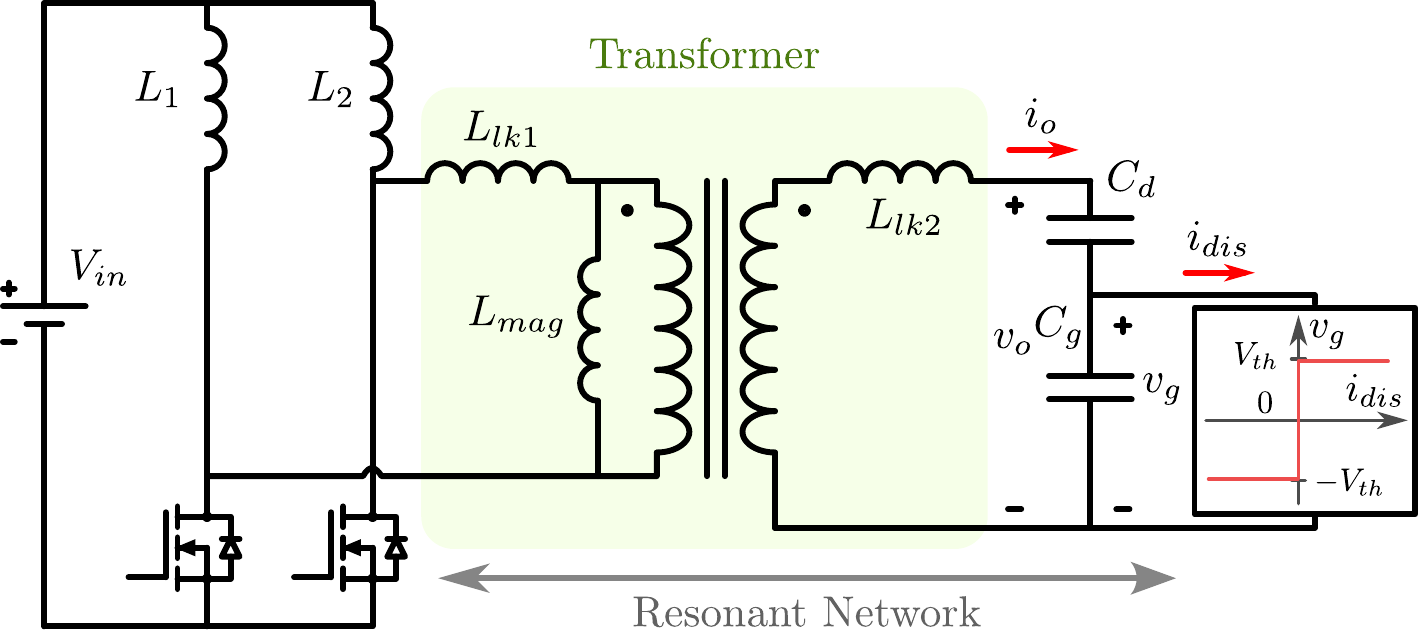}
    \caption{Current-fed parallel-resonant push-pull inverter.}
    \label{fig:CS_push_pull_inverter}
    \vspace{-10pt}
\end{figure}

In bridge-based current-fed topologies, the half-bridge or full-bridge structure typically serves as a current inverter cascaded with a front-end current regulator, such as a buck converter with constant current output~\cite{CS_third,CS_square}, as illustrated in Fig.~\ref{fig:CS_CLCC}. The primary objective of the current inverter is to generate a commutating bipolar square current waveform for the primary winding of the pulse transformer and the resonant network. The resonant design can initially be realized similarly to DC/DC resonant converters, without requiring a secondary rectifier circuit.

The output voltage waveform depends on the design of the resonant network, including the transformer and the capacitive load. For sinusoidal output waveforms, the same analysis used in the resonant tank tuning for voltage-source-fed systems can be applied; using piecewise mathematical approach~\cite{mis_math}, fundamental harmonic approximation (FHA), the rectifier compensated FHA (RCFHA), and state plane analysis, the initial tuning of the resonant tank can be performed to achieve the desired voltage waveform characteristics.

Fig.~\ref{fig:CS_CLCC} shows current-fed full-bridge CLCC resonant inverter, which incorporates the two capacitances of the DBD load within the resonant tank~\cite{CS_CLCC}. By employing a mathematical approach to tune the resonant tank, the inverter achieves high precision, accounting for the two states of the capacitive DBD load—an accuracy difficult to achieve using conventional approximation methods such as FHA.

However, this inverter design has certain limitations. The inability to utilize leakage inductance as a functional component of the resonant tank necessitates minimized leakage inductance. Additionally, soft-switching is not inherently guaranteed—a common drawback in traditional current-fed bridge-based topologies~\cite{CS_third}. To address these challenges, the current-fed full-bridge LLCC resonant inverter, proposed in~\cite{CS_LLCC}, offers a more advanced solution. By introducing three additional resonant capacitors to the lower switches of the phase legs, this design overcomes the limitations of conventional configurations. Validated using RCFHA, the asymmetric full-bridge structure achieves ZCS for the upper devices and ZVS for the lower devices, while effectively incorporating leakage inductance as a resonant component.

%The need for a high-speed front-end buck converter can be mitigated by integrating the input inductor and the transformer's leakage inductance into the resonant network. This approach achieves ZCS for the switches while maintaining a pseudo-constant input current, validated through state-plane analysis~\cite{CS_pure}. 

Unlike the aforementioned current-fed topologies, which primarily aim to control voltage output, direct control of the current is achievable due to the stiff input current regulation. In~\cite{CS_third}, it is demonstrated that the third harmonic of the injected current influences the discharge-time ratio (DTR), which represents the duration of the discharge subinterval within a single period. This enables compressed energy delivery under sinusoidal excitation. The design process leverages RCFHA. By incorporating third harmonic current injection into the reference current, the DTR becomes adjustable, providing precise control over energy compression and stretching. This capability is particularly beneficial for DBD surface treatment applications, where designed energy delivery is critical for achieving desired processing outcomes.

\addedthree{The front-end current source design is a critical part of designing current-fed resonant inverter, which must address the non-linear characteristics of DBD loads by compensating for rapid voltage transients and dynamic load behavior. Average-current-mode control, commonly implemented using current-mode buck converters or two-quadrant choppers, provides fixed switching frequency and favorable EMI characteristics for closed-loop regulation~\cite{current_fed_buck}. However, such approaches are fundamentally challenged by the fast, discontinuous current demand inherent to DBD systems. Average-based current regulation assumes a smooth and predictable relationship between input current and load behavior—an assumption that breaks down due to the threshold-dependent nature of DBD discharges. Moreover, the inherent delay introduced by the averaging process slows the control response, making it ill-suited for HF operation or rapidly changing load conditions.}

\addedthree{Conversely, hysteresis control is often favored for its fast time-domain response~\cite{CS_square}. However, its variable switching frequency complicates EMI filtering and makes the system more sensitive to load variations. When applied to capacitive and nonlinear loads such as DBDs, hysteresis bands can induce ringing or oscillations, leading to instability or degraded performance.}

As an alternative, the active front-end can be replaced with an input choke~\cite{CS_LLCC, recent_current_fed}, enabling efficient operation with improved power factor. However, such simplified current-fed configurations generally demand a large input inductance to maintain near-constant current, which can significantly increase system size and cost.

\subsubsection{Push-Pull Topologies}

A current-fed resonant inverter can also be realized using a parallel-resonant push-pull topology. Fig.~\ref{fig:CS_push_pull_inverter} illustrates the schematic of a current-fed parallel-resonant push-pull inverter. The input inductors, $L_1$ and $L_2$, establish a stiff current source for the push-pull resonant network. The primary advantage of this topology lies in its ability to achieve ZCS by ensuring the current naturally drops to zero before the switching devices turn off. The low side switching devices are implemented with series-diodes to add an extra protection to block reverse current~\cite{CS_push_pull_inverter_reverse}. However, this approach prevents the current induced by the transformer's leakage inductance from finding a proper path, leading to voltage spikes on the switching devices. To address this limitation,~\cite{CS_push_pull_inverter} proposes a design that eliminates the series diodes and incorporates the transformer's leakage inductance as part of the resonant tank. This structure can be further simplified by realizing the push–pull configuration with a center-tapped transformer~\cite{current_fed_buck}.

However, the limited practical inductance of the input choke often introduces harmonic content in the output waveform, reducing voltage gain and diminishing the expected benefits of current-fed resonant inverters.

\begin{table*}[!ht]
\centering
\caption{Specifications of Resonant Inverters for DBD applications}
\label{table:resonant}
\begin{threeparttable}
\scalebox{0.75}{
\begin{tabular}{l|c|c|c|c|c|c|c|c|c|l}
\toprule
\multirow{2}{*}{\textbf{Topology}}   &\multirow{2}{*}{\makecell[c]{\textbf{Input}\\\textbf{Mode}}}  &\multirow{2}{*}{\makecell[c]{\textbf{No. of}\\\textbf{Switches}}} & \multirow{2}{*}{\makecell[c]{\textbf{No. of}\\\textbf{Passives}\tnote{1}}} & \multirow{2}{*}{\textbf{Transformer}} & \makecell[c]{\textbf{Input}\\\textbf{Voltage}}&  \makecell[c]{\textbf{Peak-to-peak}\\\textbf{Output Voltage}} &\makecell[c]{\textbf{Output}\\\textbf{Power}} &\makecell[c]{\textbf{Efficiency}}  &\makecell[c]{\textbf{Operating}\\\textbf{Frequency}}  &\multirow{2}{*}{\makecell[c]{\textbf{Features}}}\\
&&&&&(V)&(kV)&(W)&($\%$)&(kHz)&
\\\midrule
%%%%%%%%%%%%%%%%%%%%%%%%%%%%%%%%%%%%
\makecell*[r]{\cite{XFMR_FB} Full-Bridge \\LC } & \multirow{5}{*}[-3.5em]{\rotatebox[origin=c]{90}{Voltage-Fed}}    &4  &1L, 1C &Yes   &55 &7.0  &39 &NR\tnote2  &781    &\makecell[l]{Excimer DBD lamp, HF Snubber,\\ZCS, Complementary gate signal}
\\\cline{1-1}\cline{3-11} 
\makecell*[r]{\cite{XFMR_FB_LCL} Full-Bridge \\LCLC }     &   &4  &2L, 1C   &No   &12 &3.8    &6.5    &92 &33.4   &\makecell[l]{Ozone chamber, ZCS turn-on, ZVS turn-off\\Complementary gate signal (duty 25$\%$)}
\\\cline{1-1}\cline{3-11} 
\makecell*[r]{\cite{XFMR_class_E} Class-E }     &   &1  &3L, 2C   &No   &200 &2.0    &600\tnote{3}    &91.5\tnote{3} &\makecell{12,400\textbf{,} \\15,500}   &\makecell*[l]{SDBD, ZVS,\\Frequency-selective control}
\\\cline{1-1}\cline{3-11} 
\makecell*[r]{\cite{XFMR_piezo} Class-E \\Push-Pull }     &   &2  &4L   &No   &200 &1.4    &NR    &NR &6,410   &\makecell[l]{SDBD, ZVS, piezoelectric resonator\\Complementary gate signal}
\\\cline{1-1}\cline{3-11} 
\makecell*[r]{\cite{phase_shift_FB1} Phase-Shifted \\Full-Bridge }     &&4  &1L   &Yes   &310 &6.9    &40    &87 &45$\sim$57.5   &\makecell[l]{Ozone chamber, ZVS turn-on,\\Phase-shifted control}
\\\hline
\makecell*[r]{\cite{CS_third} Full-Bridge \\$\pi$-Network }    &\multirow{6}{*}[-4em]{\rotatebox[origin=c]{90}{Current-Fed}}&5  &2L, 1C, 5D   &Yes   &310 &28.0    &350   &NR &30   &\makecell[l]{Surface treatment, pseudo-ZVS,\\Phase-locked loop}
\\\cline{1-1}\cline{3-11} 
\makecell*[r]{\cite{CS_LLCC} Asymmetric \\Full-bridge }     &   &4  &2L, 3C, 2D   &Yes   &150 &1.8    &500    &NR &32.3   &\makecell*[l]{Surface treatment,\\ZCS for top, ZVS for bottom devices,\\Phase-shifted control}
\\\cline{1-1}\cline{3-11} 
\makecell*[r]{\cite{recent_current_fed} Full-bridge \\Parallel-Resonant }     &   &4  &1L, 1C   &Yes   &100 &7.0    &551    &91 &10  &\makecell*[l]{Ozone generation, AC-DC-AC\\ with input inductor (PF: 96\%)}
\\\cline{1-1}\cline{3-11} 
\makecell*[r]{\cite{CS_push_pull_inverter_reverse} Parallel-Resonant \\Push-Pull }     &   &2  &2L, 1C, 4D  &Yes   &35 &2.1    &200    &75 &7   &\makecell[l]{Ozone chamber, ZVS,\\Complementary gate signal}
\\\cline{1-1}\cline{3-11} 
\makecell*[r]{\cite{CS_push_pull_inverter} Parallel-Resonant \\Push-Pull }     &   &2 &2L   &Yes   &12 &3.0    &7    &86.8 &50   &\makecell[l]{Cold-Cathode Fluorescent Lamp, ZCS,\\Complementary gate signal with overlap}
\\\cline{1-1}\cline{3-11} 
\makecell*[r]{\cite{current_fed_buck} Parallel-Resonant \\Push-Pull w/ Buck }     &   &2 &1L, 1C   &Yes   &325 &1.6    &60    &NR &23   &\makecell[l]{Ozone generation, ZCS, AC-DC-AC,\\ current-mode front-end buck}\\

\bottomrule
\end{tabular}}

\begin{tablenotes}
\scriptsize
\addtolength{\itemindent}{0.2cm}
\hspace*{0.5cm}
\begin{minipage}{0.8\linewidth}
\item[1] L: inductor, C: capacitor, D: diode. 
\item[2] Not Reported. 
\item[3] Tested with equivalent load.
\end{minipage}
\end{tablenotes}
\end{threeparttable}

\vspace{8pt}

\centering
\caption{Generalized Performance Comparison of Resonant Inverters for DBD Applications}
\label{table:resonant2}
\renewcommand{\arraystretch}{1.1}
\begin{threeparttable}
\scalebox{0.8}{
\begin{tabular}{c|c|c|c|c|c|c|l}
\toprule
\textbf{Topology} & \makecell{\textbf{Output}\\\textbf{Frequency}} & \makecell{\textbf{Voltage}\\\textbf{Gain}\tnote{1}} & \makecell{\textbf{Power}\\\textbf{Density}} & \makecell{\textbf{Short-Circuit}\\\textbf{Tolerance}} & \textbf{Simplicity} & \textbf{Cost\tnote{2}} & \textbf{Key Advantages and Disadvantages} \\
\midrule
\makecell{Voltage-Fed Bridge-Based\\(Class-D Variants)} & +++ & +++ & ++ & + & +++ & +++ & 
\makecell[l]{+ Only one inductor is required.\\
+ Higher voltage gain achievable with added LC network.\\
-- Trade-off: reduced frequency.} \\\hline

Voltage-Fed Class-E & ++++ & ++ & +++ & ++ & ++++ & ++++ & 
\makecell[l]{+ MHz-range operation with high power capability.\\
-- Input choke size increases with power level.} \\\hline

Current-Fed Bridge-Based & ++ & ++ & + & +++ & + & + & 
\makecell[l]{+ Short-circuit tolerant.\\
+ Injected current harmonics are controllable.\\
-- Complex design.\\
-- Requires transformer with high turn ratio.} \\\hline

Current-Fed Parallel-Resonant Push--Pull & ++ & + & + & +++ & ++ & ++ & 
\makecell[l]{+ Short-circuit tolerant.\\
+ No active front end required.\\
-- Requires large input choke.\\
-- Requires transformer with high turn ratio.} \\
\bottomrule
\end{tabular}}
\begin{tablenotes}
\scriptsize
\item[1] Normalized with transformer turns ratio.
\item[2] More plus signs (\(+\)) indicate more economical implementation.
\end{tablenotes}
\end{threeparttable}
\vspace{-10pt}
\end{table*}

\vspace{-10pt}
\subsection{Overview of Sinusoidal Power Supplies}
Table~\ref{table:resonant} summarizes the main converters discussed in this section, detailing the number of active switches and passive components, output characteristics, and soft-switching capabilities. Applications and compatible DBD geometries are also provided for clarity.

Although current-fed systems are more tolerant of the threshold behavior of DBD loads—where a conditional short circuit may occur depending on the applied waveform, voltage, and current—voltage-fed systems, including class-D variants (bridge-based) and class-E amplifiers, are often preferred for their simpler design, higher voltage gain, higher output frequency, and better efficiency enabled by inherent soft-switching.

Current-fed resonant inverters often include the requirement for a front-end current source, such as a buck-based converter or a large input inductor, a resonant network to facilitate soft-switching, and additional series diodes to mimic thyristor-like behavior in the switches. Typically, the input choke inductance for current-fed systems is in the mH range. At higher power levels required by DBD applications, this results in large component sizes that severely compromise power density. Consequently, quasi-current-fed voltage-fed systems, such as class-E resonant inverters, become preferable solutions, as they reduce the required input choke inductance, enhancing power density.

Nevertheless, topological variations remain possible because conventional voltage-fed resonant inverters—including classes D, E, EF\textsubscript{n}, E/F\textsubscript{n}, and $\Phi$\textsubscript{2}—are generally well-characterized for inductive-resistive loads~\cite{recent_res_EF}. Given the duality between voltage-driven inductive loads and current-driven capacitive loads, further analysis and novel resonant tank designs are needed to address the unique characteristics of DBD loads.

\addedthree{Regardless of whether voltage-fed or current-fed, a common limitation of resonant inverters is that soft-switching can only be maintained within a narrow range of operating conditions. As the resonant tank's soft-switching region is frequency-dependent, variations in gain or frequency can push the system out of this optimal range, reducing efficiency. Unlike DC-DC resonant converters with secondary rectifiers, the voltage waveform produced by the inverter is directly applied to the DBD load, resulting in deviations from an ideal sinusoidal shape. Consequently, pulse-frequency modulation, commonly used in DC–DC resonant converters for gain regulation, cannot be directly applied to DBD resonant inverters. This limitation makes closed-loop gain control impractical, explaining why open-loop operation is predominantly adopted, with the design instead guided by the gain characteristic, soft-switching region, and frequency dependence of the resonant tank.}

Additionally, the dynamic properties of DBD systems—such as variations in gas type, application, and operating conditions—significantly influence the load’s capacitive and resistive characteristics. These factors shift the optimal operating frequency away from the resonance frequency for which the inverter was initially designed. Furthermore, $C_d$, $C_g$, and $V_{th}$ are frequency-dependent, complicating impedance matching for the resonant network. As research on load-independent resonant inverters progresses, extending these concepts to dynamic-independent resonant inverter designs for DBD applications represents a valuable direction for future work~\cite{recent_res_load_independent}. In summary, Table~\ref{table:resonant2} presents a generalized performance comparison of reported resonant inverters for DBD applications.

%%%%%%%%%%%%%%%%%%%%%%%%%%%%%%%%%%%%%%%%%%%%%%%%%%%%%
%%%%%%%%%%%%%%%%%%%%%%%%%%%%%%%%%%%%%%%%%%%%%%%%%%%%%
%%%%%%%%%%%%%%%%%%%%%%%%%%%%%%%%%%%%%%%%%%%%%%%%%%%%%
\section{Pulsed Power Supplies}\label{sec:6}

Certain DBD applications require PPSs which can offer distinct advantages over sinusoidal excitation methods. Their capability to generate HV, HF pulses with sharp rise times establishes them as viable solutions for a diverse range of DBD applications. This section elaborates on different pulsed power supplies developed for DBD excitation. 

%This section explores various pulsed power supply topologies, focusing on their operational principles, benefits, and limitations, alongside strategies to address associated design challenges. From bridge-based converters to advanced Marx and flyback-like configurations, these topologies demonstrate their adaptability and effectiveness in meeting the demanding requirements of DBD systems.

%%%%%%%%%%%%%%%%%%%%%%%%%%%%%%%%%%%%%%%%%%%%%%%%%%%%%
%%%%%%%%%%%%%%%%%%%%%%%%%%%%%%%%%%%%%%%%%%%%%%%%%%%%%
\vspace{-10pt}
\subsection{Voltage-Fed Systems}

\subsubsection{Bridge-Based Topologies}
\begin{figure}[!t]
    \centering
    \includegraphics[width=0.7\linewidth]{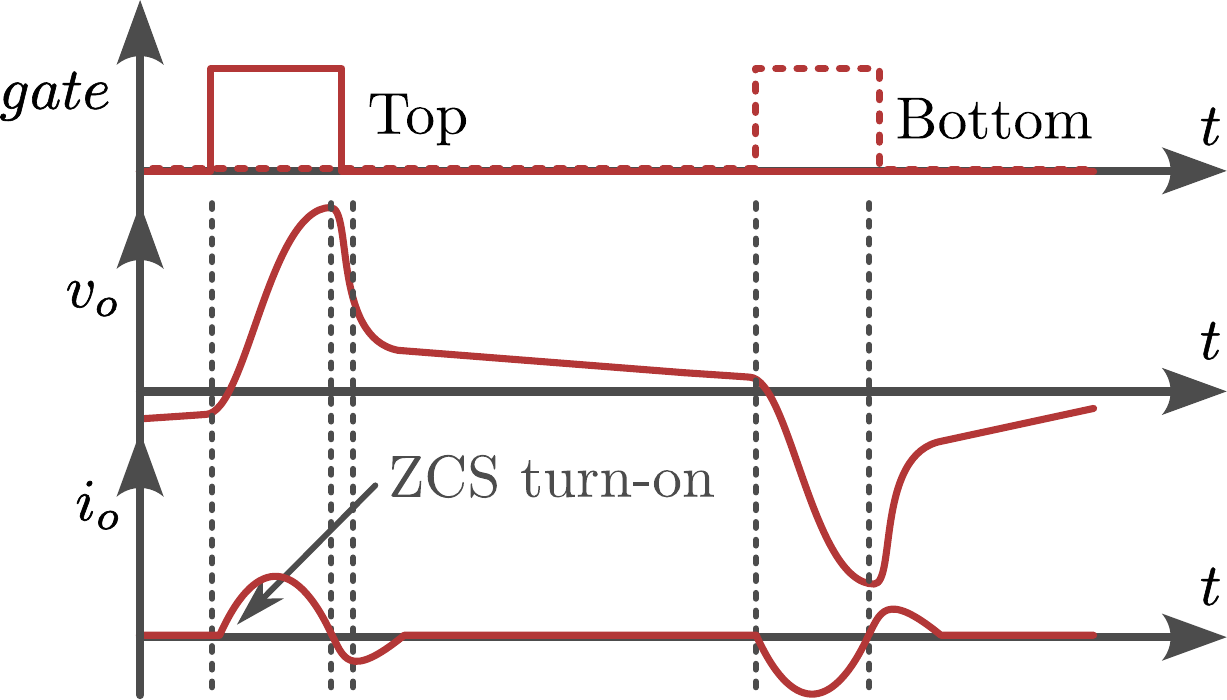}
    \caption{Waveforms of typical bridge-based PPSs with DBD load.}
    \label{fig:HB_output}
    \vspace{-15pt}
\end{figure}

A bridge-based converter without a resonant tank is capable of generating pulsed waveforms, where the secondary voltage ideally assumes a bipolar square waveform, amplified by a high turn-ratio pulse transformer~\cite{HB_inverter, new_excimer1, new_excimer2, recent_FB1}. The same topology in Fig.~\ref{fig:VS_FB} can be used. Despite its simplicity, approaches using a pulse transformer present notable limitations, including the arbitrary nature of the output voltage waveform and significant current ripple due to leakage inductance and distributed capacitance, as shown in Fig.~\ref{fig:HB_output}. These effects often require additional damping to suppress undershoot~\cite{recent_forward}.

Bridge-based topologies are frequently combined with pulse forming lines (PFLs), such as the Blumlein configuration, to achieve high PRF with nanosecond-scale pulse widths and well-defined waveform shapes~\cite{recent_pfl2}. PFLs utilize wave propagation in transmission lines to generate HV nanosecond pulses with rectangular flat-top characteristics. Owing to their stackable architecture, transformerless configurations are often feasible, enabling higher voltage gain~\cite{recent_pfl1}. For example,~\cite{recent_pfl4} reports voltage pulses with a rise time of 5.2 ns and pulse width of 10.6 ns. Although nanosecond-scale pulses are generally achievable, the required charging time limits the attainable PRF. However, by exploiting resonance, PRFs in the megahertz range can be realized~\cite{recent_pfl3}.

Nonetheless, a major limitation of PFL-based systems is the fixed pulse width imposed by their physical configuration, making it difficult to implement adjustable pulse durations. More critically, successful operation depends heavily on impedance matching between the generator and the load, an assumption that fails in the case of non-linear and time-varying DBD loads. Moreover, the ideal flat-top square waveforms produced by PFLs are often severely distorted when connected to DBD reactors~\cite{HB_BLT}. Also, physical limit of PFL often compromises power density.

%%%%%%%%%%%%%%%%%%%%%%%%%%%%%%%%%%%%%%%%%%%%%%%%%%%%%
%%%%%%%%%%%%%%%%%%%%%%%%%%%%%%%%%%%%%%%%%%%%%%%%%%%%%

\subsubsection{Marx-Based Topologies}

\begin{figure}[!t]
    \centering
    \includegraphics[width=0.8\linewidth]{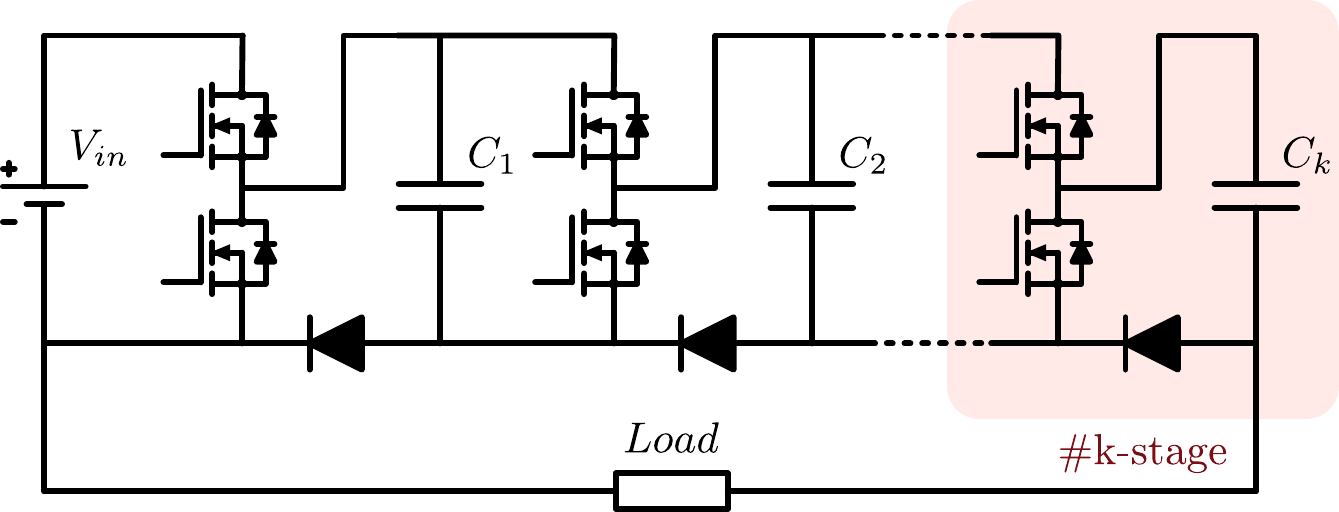}
    \caption{Schematic of typical solid-state Marx generators.}
    \label{fig:Marx_schem}
    \vspace{-10pt}
\end{figure}

\begin{figure}[!t]
    \centering
    \includegraphics[width=0.8\linewidth]{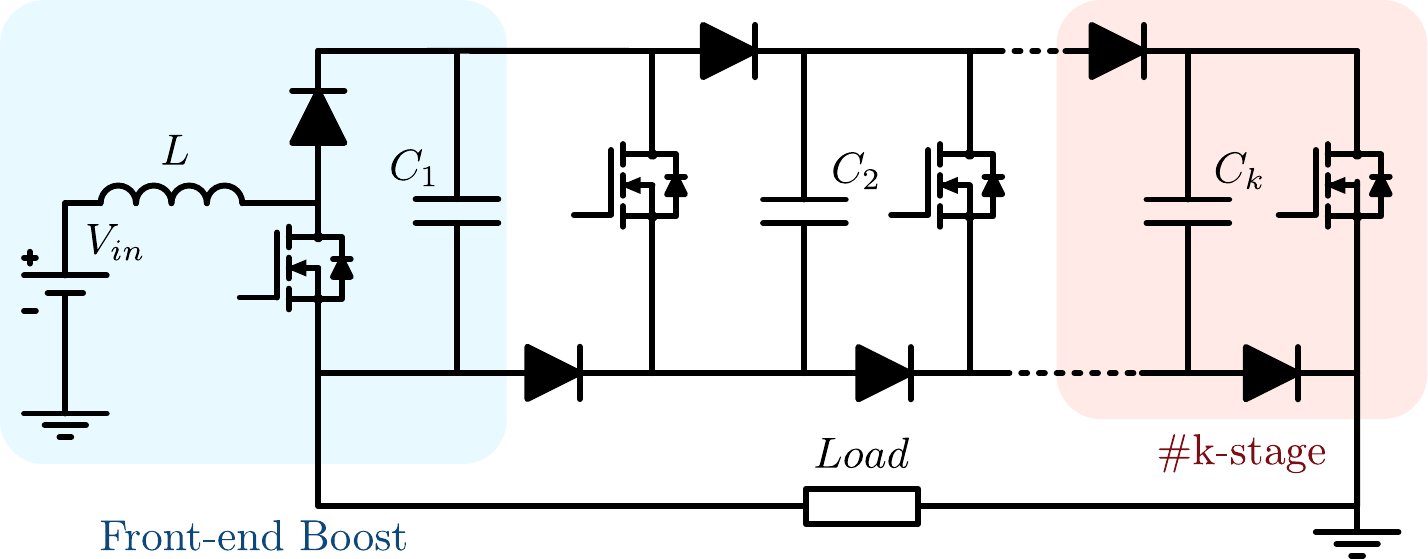}
    \caption{Marx generator with a front-end boost converter.}
    \label{fig:Marx_boost_schem}
    \vspace{-10pt}
\end{figure}

Marx generators are promising voltage-fed HF PPS for DBD applications due to their ability to produce unipolar or bipolar pulses while preventing HV stress on devices through a modular structure~\cite{recent_pulse_review}. Fig.~\ref{fig:Marx_schem} shows the basic configuration of solid-state Marx generators. By applying complementary gate signals to the half-bridge structure of each stage, the charging and discharging phases are easily controlled, generating unipolar square-wave voltage across the capacitive load. For bipolar waveforms, a full-bridge configuration for each stage can be employed, enabling full-bridge voltage commutation across the load or the use of a dual power supply.

For unipolar pulse generation, Marx generators inherently eliminate the need for a transformer in their basic operation, while isolated configurations utilizing pulse transformers may require a reset circuit to prevent core saturation. This distinction underscores the advantages of Marx generators, whose simple architecture and modular scalability make them an excellent solution for unipolar pulse generation in DBD applications, particularly when transformerless compact designs are preferred. This architecture enables nanosecond-scale pulse widths at operating frequencies of up to 1 MHz~\cite{recent_Marx2, recent_Marx4}. Zhong \textit{et al.} provide a comprehensive review of recent advancements in solid-state Marx generators~\cite{Marx_review}.

One limitation of Marx-based topologies is that the voltage gain is proportional to the number of stages, meaning the voltage gain is limited to integer values. In~\cite{Marx_boost}, adjustable voltage gain is achieved by incorporating an inductor at the input side, as illustrated in Fig.~\ref{fig:Marx_boost_schem}. Unlike conventional designs with a voltage boost converter at the front end, this approach allows boost operation during the discharging phase, resulting in a higher voltage gain compared to the combined effects of the superposition of a front-end boost converter and a standard Marx generator.

A critical issue with solid-state Marx-based topologies for capacitive loads arises during the activation of the charging switches after discharging phase. As illustrated in Fig.~\ref{fig:Marx_schem}, the capacitive load holds a significant voltage, and when the charging switches are triggered, the last capacitor in the stack could momentarily exceed its voltage rating and also have a high inrush current. This imbalance results in unequal stress on the stages, affecting both current and voltage ratings across the system.

To address this issue, introducing an inductor between the Marx generator and the DBD load is necessary to limit excessive current and mitigate EMI caused by fast transients~\cite{Marx_ind_excimer}. However, this solution comes with trade-offs. The inclusion of the inductor increases the rise time of the output voltage, which can reduce the performance of the DBD system. Additionally, the inductor forms an LCL resonant circuit with the Marx generator’s capacitors and the DBD load capacitance during the discharge phase, resulting in significant ringing voltage across the DBD load. Although the size of the inductor can be reduced for high PRF applications, it becomes a dominant factor in the size of the pulsed power supply for high-current applications.

A major limitation remains the inherently uncontrolled output voltage slew rate of Marx generators. Because the discharge current is directly proportional to the voltage slew rate ($dv/dt$), this lack of control complicates both current prediction and regulation. Consequently, careful evaluation of stray inductance in the PCB layout is necessary to suppress current overshoot or adjust the voltage rise time~\cite{recent_Marx2,recent_Marx3}. However, such optimization is often impractical in real-world implementations.

%%%%%%%%%%%%%%%%%%%%%%%%%%%%%%%%%%%%%%%%%%%%%%%%%%%%%
%%%%%%%%%%%%%%%%%%%%%%%%%%%%%%%%%%%%%%%%%%%%%%%%%%%%%

%%%%%%%%%%%%%%%%%%%%%%%%%%%%%%%%%%%%%%%%%%%%%%%%%%%%%
%%%%%%%%%%%%%%%%%%%%%%%%%%%%%%%%%%%%%%%%%%%%%%%%%%%%%
\subsubsection{Flyback-like Topologies}
\begin{figure}[!t]
    \centering
    \includegraphics[width=0.7\linewidth]{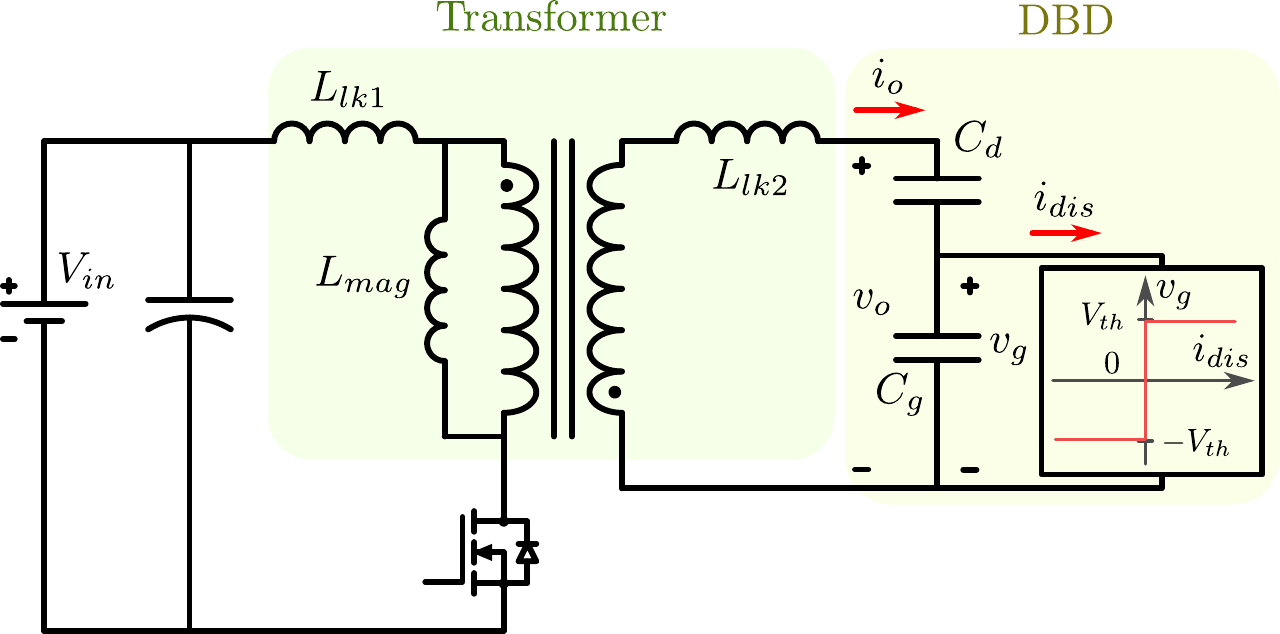}
    \caption{Basic flyback-based PPS topology.}
    \label{fig:VS_fly}
    \vspace{-10pt}
\end{figure}

\begin{figure}[!t]
    \centering
    \includegraphics[width=0.7\linewidth]{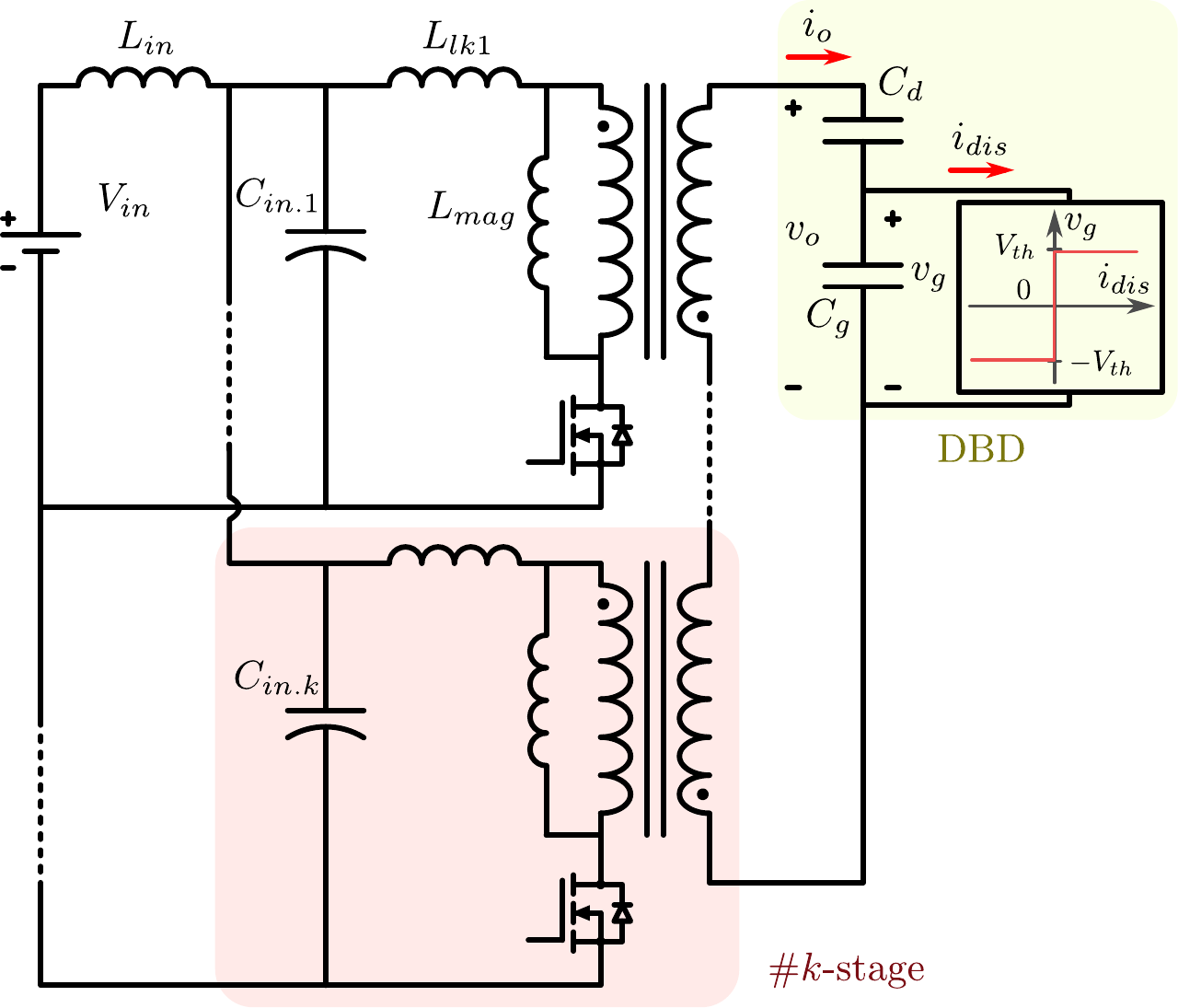}
    \caption{Flyback-based PPS topology with modular structure.}
    \label{fig:fly_stacked}
    \vspace{-10pt}
\end{figure}

The flyback-like topology in Fig.~\ref{fig:VS_fly} presents itself as a promising voltage-fed, isolated power supply for DBD applications where HV pulses are needed, offering simplicity, HV generation capability, and reliability. Central to its design is the energy storage transformer, which efficiently stores energy during the charging phase and releases it in a controlled manner to generate HV pulses~\cite{Saket}. This makes the topology particularly well-suited for applications demanding simple designs with galvanic isolation and high pulse amplitudes. It is noteworthy that the design flexibility of the transformer enables the implementation of a modular structure in flyback-like designs. A stacked configuration, where multiple parallel primary windings are connected to series-connected secondary windings, as shown in Fig.~\ref{fig:fly_stacked}, can be considered for high-power DBD applications~\cite{Fly_lc_stacked, recent_flyback2}.

\begin{figure}[!t]
    \centering
    \includegraphics[width=0.7\linewidth]{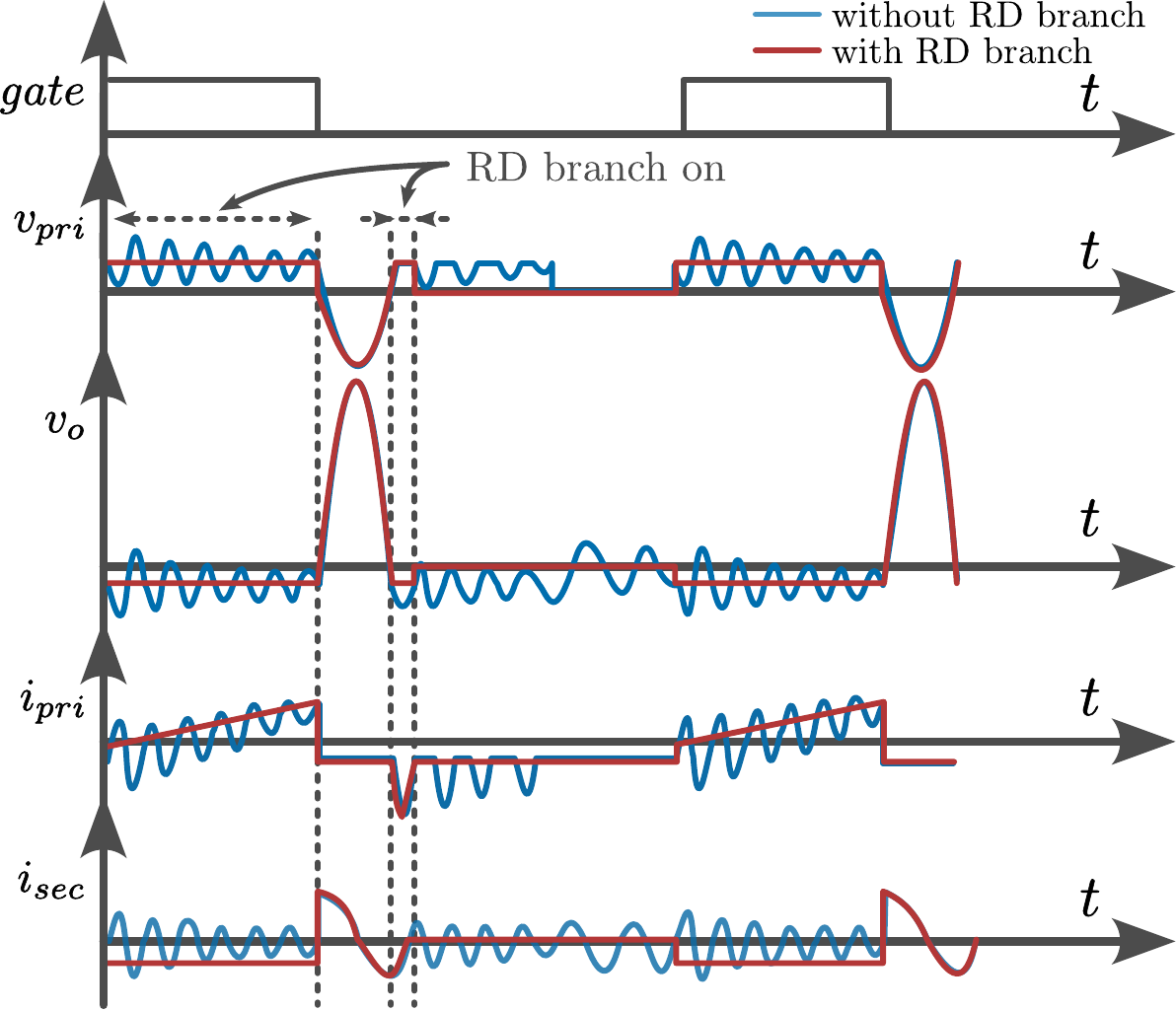}
    \caption{Output waveforms of flyback-like PPS topologies.}
    \label{fig:fly_out}
    \vspace{-10pt}
\end{figure}

However, conventional flyback-like topologies face challenges in maintaining pulse quality due to the resonances that occur between the transformer’s leakage inductance, the capacitive nature of the DBD load, and the distributed capacitance of the secondary side of the pulse transformer, as shown in Fig.~\ref{fig:fly_out}. These resonances can lead to voltage overshoots and distortions, degrading the uniformity and stability of the plasma discharge. 

\begin{figure}[!t]
    \centering
    \includegraphics[width=0.8\linewidth]{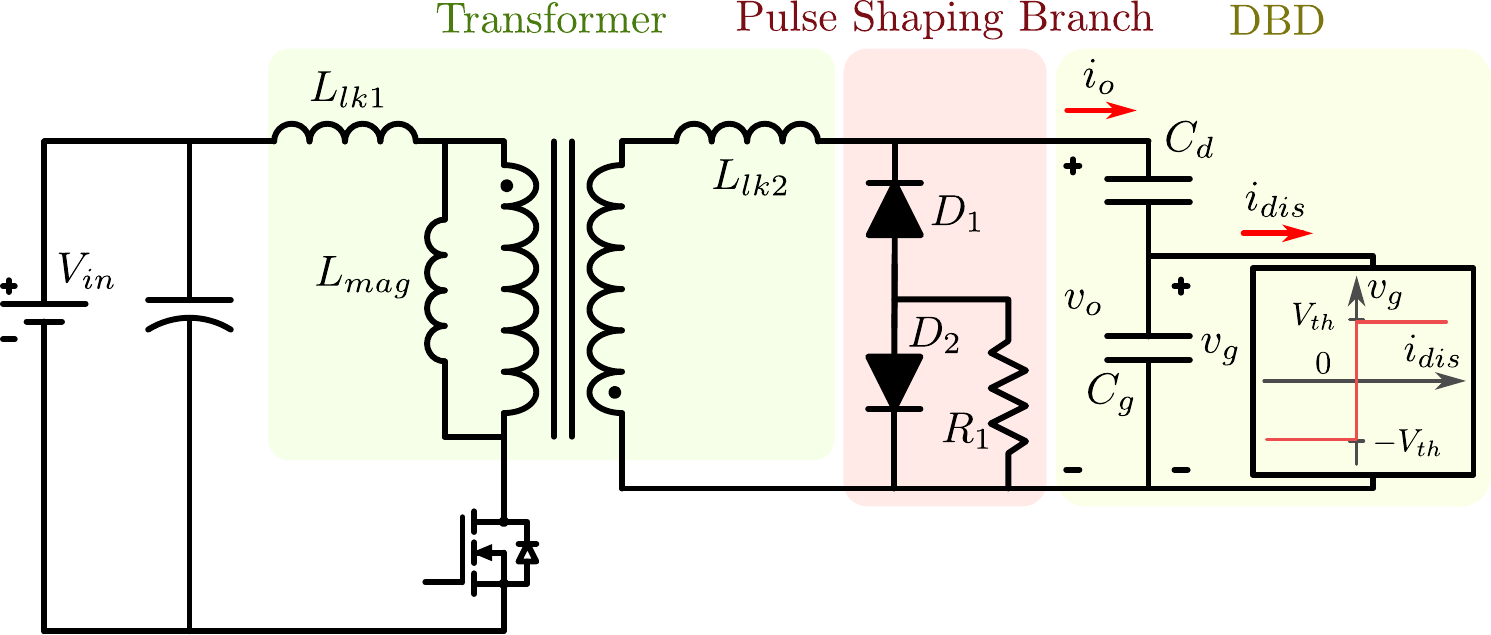}
    \caption{Flyback-based PPS topology with RDD branch.}
    \label{fig:VS_fly_RDD}
    \vspace{-10pt}
\end{figure}

To address these challenges, a pulse-shaping stage is used. Often, a parallel resistor is connected to the DBD load to stabilize operation and mitigate the impact of its capacitive characteristics~\cite{Fly_D}. However, despite stabilizing the operation, the parallel resistor allows continuous current flow, leading to efficiency losses. 

Consequently, advanced pulse-shaping methods incorporating resistor and diode configurations on the secondary side have been proposed to mitigate oscillations caused by the interaction between the transformer's leakage inductance and the capacitive load~\cite{Fly_lc_stacked}. In this resistor-diode (RD) configuration, the resistor conducts only during a sub-time interval rather than throughout the entire period, thereby minimizing resistive losses. As shown in Fig.~\ref{fig:fly_out}, the RD branch suppresses oscillations by allowing current flow during two distinct sub-time intervals. 

Building on this concept, a resistor-diode-diode (RDD) pulse-shaping branch is proposed in~\cite{Fly_RDD}. The RDD configuration, shown in Fig.~~\ref{fig:VS_fly_RDD}, not only suppresses oscillations but also incorporates an additional diode in parallel with the resistor to provide an alternative current path for over-voltage conditions. This design gradually reduces the secondary-side current during the switch’s off-state when the capacitor voltage exceeds the breakdown voltage of the main diode in the RD branch. Overall, the RD and RDD branches are effective in suppressing oscillations, preventing localized overheating at the DBD load.

% Once the current drops to zero, the pre-charged capacitor voltage enables a reverse current flow through the leakage inductance, facilitating the discharge of the capacitor voltage. This mechanism effectively addresses over-voltage issues, providing an added safety criterion for system reliability. 

\begin{figure}[!t]
    \centering
    \includegraphics[width=0.6\linewidth]{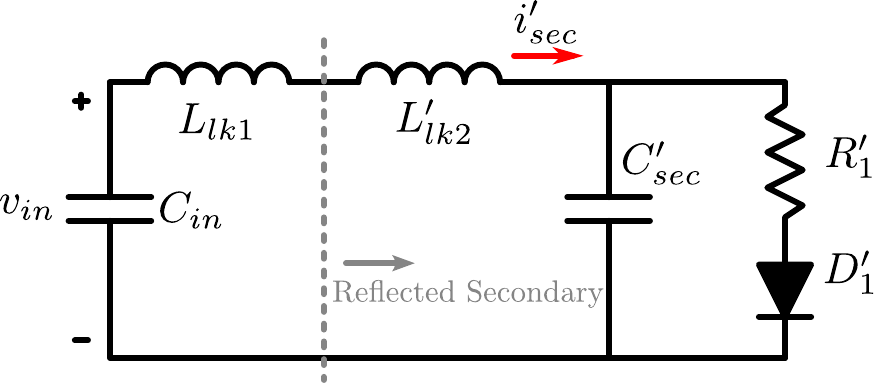}
    \caption{Equivalent circuit reflected on the primary side.}
    \label{fig:VS_fly_equi}
    \vspace{-10pt}
\end{figure}

One limitation of the RD and RDD diode branches is that they must block the secondary voltage, which is in the kV range, depending on the DBD applications. While there are many commercial HV diodes available with low current ratings, the corresponding losses in the pulse-shaping branches can be minimized by selecting the maximum allowable resistance value for the shaping resistor. However, the pulse-shaping resistance also affects the resonant frequency (\( \omega_o \)) of the circuit, which consists of the equivalent secondary capacitance (including the transformer's capacitances and the DBD load), as well as the leakage inductances, as depicted in Fig.~\ref{fig:VS_fly_equi}. The relationship is described by \eqref{eqn:fly_equi}, derived by solving the total impedance of the equivalent circuit. 
\begin{align}\label{eqn:fly_equi}
    \omega_o = \frac{1}{R'_{shape}C'_{sec}}\sqrt{\frac{R'^2_{shape}C'_{sec}-L_{lk}}{L_{lk}}}
\end{align}
Here, the apostrophe (') represents quantities reflected from the secondary to the primary side, and \( L_{lk} = L_{pri.lk} + L'_{sec.lk} \).

To suppress resonance, the numerator, \( R'^2_{shape}C'_{sec} - L_{lk} \), must be negative, resulting in purely imaginary resonant frequencies. This condition ensures that no oscillatory resonance modes are present in the system. For example, in larger DBD reactor setups, this condition limits the maximum resistance value that can be used effectively. As a result, the size of the DBD reactor directly influences the design and performance of the pulse-shaping branches.

A resistor-capacitor-diode (RCD) branch at the primary side can also be implemented for flyback-like solutions, similar to its application in conventional flyback DC-DC converters~\cite{Fly_boost_RDD}. In traditional RCD operation, the diode conducts to transfer energy from the transformer's leakage inductance to the capacitor, which subsequently dissipates the energy through the resistor, functioning as a passive voltage clamp. However, in a flyback-like pulse generator, the RCD branch operates in two distinct modes: leakage energy transfer mode and voltage limiting mode. In leakage energy transfer mode, the RCD branch rapidly transfers the energy stored in the transformer's leakage inductance to the capacitor. This process not only directs leakage current into the RCD branch but also slows the rate of increase in the drain-source voltage of the switching device, thereby reducing switching losses. In voltage limiting mode, the RCD branch acts as a voltage clamp, controlling the pulse voltage amplitude by managing energy feedback from the secondary side and ensuring it remains within safe limits. In summary, compared to the RCD snubber in conventional flyback DC-DC converters, the RCD branch in flyback-like pulse generators delivers dual advantages: reduced heat loss through optimized energy handling and enhanced output voltage control, specifically adapted to the HV requirements of pulse generation systems.

However, since the energy transfer mechanism in flyback converters relies on magnetizing inductance as the inductive energy storage element coupled to the DBD load, the achievable output pulse width is inherently determined by the duration of the magnetizing current discharge. This limitation generally confines output pulses to $\mu s$-scale durations, thereby restricting achievable PRF (\textless100kHz). Achieving higher PRF with narrower pulse widths requires reducing the magnetizing inductance ($L_{mag}$). Using topological variations, such as the inclusion of intermediate stages of inductive energy storage, along with reductions in the effective inductance of the pulse transformer, MHz range operation has been demonstrated for resistive loads~\cite{recent_flyback3}. Nevertheless, in typical flyback-based PPS configurations, decreasing $L_{mag}$ increases the required primary-side current to maintain constant pulse energy or peak voltage. This trade-off imposes high-power requirements on the switch rating and becomes a dominant design constraint in flyback-based pulse generators.

Additionally, flyback-based topologies still exhibit similar limitations to those commonly encountered in conventional DC/DC flyback converter. Resonant sub-oscillations arising among the leakage inductance, magnetizing inductance, and the DBD load can be suppressed by employing passive damping circuits; nevertheless, the primary-side switching device still experiences significant voltage reflected from the secondary side, especially at HV gains.

While flyback converters offer magnetic superposition and simple designs, their practical power level for single module typically remains limited to a few hundred watts due to core losses, passive branch losses, and snubber energy dissipation at high frequencies.

\subsubsection{Resonant Topologies}

\begin{figure}[!t]
    \centering
    \includegraphics[width=0.8\linewidth]{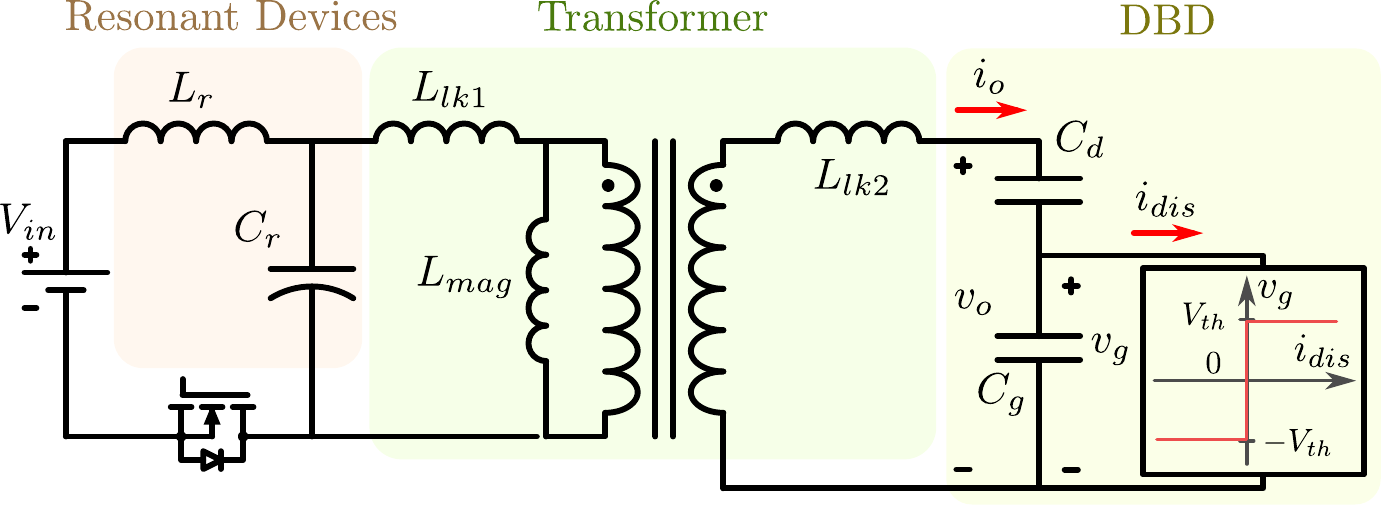}
    \caption{Forward-like PPS with ZCS capability.}
    \label{fig:Vs_forward}
    \vspace{-10pt}
\end{figure}
\begin{figure}[!t]
    \centering
    \includegraphics[width=0.5\linewidth]{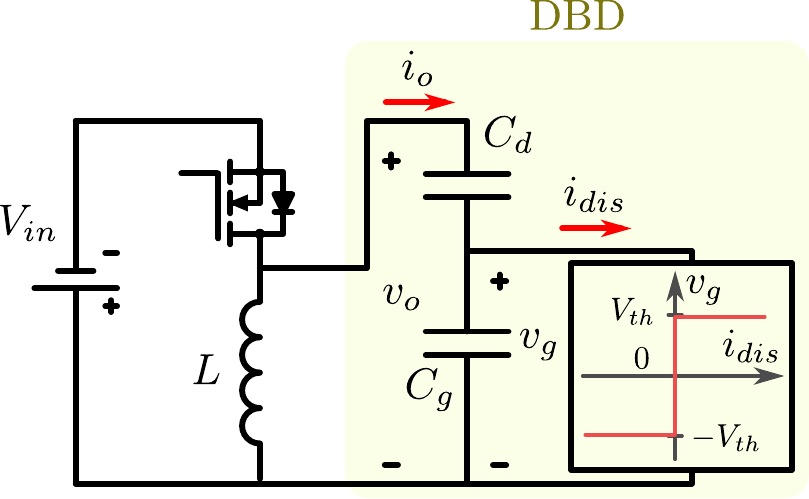}
    \caption{Transformerless single switch resonant PPS.}
    \label{fig:XFMR_single}
    \vspace{-10pt}
\end{figure}

Resonant circuit behavior can also be advantageous in PPS topologies, as it facilitates soft-switching operation. While flyback-like solutions have been explored as potential PPS structures, forward-like configurations can also be implemented, as shown in Fig.~\ref{fig:Vs_forward}. For example, a forward-like unipolar PPS utilizing a resonant inductor and capacitor was demonstrated in~\cite{forward1}, achieving zero-current turn-on and turn-off. However, a key limitation of this topology is the need to minimize leakage inductance and distributed capacitance, as an ideal transformer is assumed in the design.

In addition, resonant circuit implementation can enhance compactness and voltage gain. As shown in Fig.~\ref{fig:XFMR_single}, the circuit in~\cite{XFMR_single_switch} minimizes component count by incorporating a series inductor with the switching device. The inductor forms a resonant network with the parallel capacitive DBD load, enabling the generation of unipolar voltage pulses while ensuring zero-voltage turn-on for the switching device.

This approach highlights the benefits of resonant networks in improving efficiency. Moreover, transformerless designs, as seen in this topology, address a significant limitation of transformer-based PPS: the inherent constraint on achievable minimum pulse duration or oscillations due to transformer leakage inductance. By eliminating the transformer, smaller pulse widths and improved performance can be achieved. In~\cite{recent_res1}, a 1-MHz PRF is achieved by using LC resonance to reduce the pulse width to 7 ns while a voltage gain of 10. To further increase PRF, diode-based pulse superposition is introduced, enabling operation at up to 3 MHz.

\begin{table*}[!ht]
\centering
\caption{Specifications of Exemplary PPSs}
\label{table:PPS}
\begin{threeparttable}
\scalebox{0.8}{
\begin{tabular}{c|l|c|c|c|c|c|c|c|c}
\toprule
\multirow{2}{*}{\textbf{Topology}}&\multirow{2}{*}{\textbf{Features}} & \makecell{\textbf{Input} \\ \textbf{Voltage}} & \makecell{\textbf{Peak Output Voltage} \\ \textbf{(Positive / Negative)}} & \makecell{\textbf{Peak Current} \\ \textbf{(Positive / Negative)}} & \makecell{\textbf{Average} \\ \textbf{ Output Power}} & \makecell{\textbf{Efficiency}} & \makecell{\textbf{Rise} \\ \textbf{Time}} & \makecell{\textbf{Fall} \\ \textbf{Time}}& \makecell{\textbf{PRF}} \\ 
&& $V_{in}$ (V) & $v_{o.peak}$ (kV) & $i_{o.peak}$ (A) & $P_{o.avg}$ (W) & $\mu$ ($\%$)& ($\mu$s) & ($\mu$s) & (kHz) \\ 
\midrule
%%%%%%%%%%%%%%%%%%%%%%%%%%%%%%%%%%%%
\multirow{2}{*}{Bridge-Based}&\makecell[l]{\cite{HB_inverter} Half-bridge, bipolar output, \\ZCS turn on, ZVS turn off} 
& 120 & 6.68 / -6.68    & 7.18 / -7.18  & 116.13            & NR\tnote{1}             & 0.8   & 0.7   &100 \\ \cline{2-10}
&\makecell*[l]{\cite{HB_BLT} Half-bridge, bipolar output, \\pulse-forming line} 
& 850 & 5 / -5          & NR             & NR                  & NR              & 0.1   & 0.1   &2 $\sim$ 20  \\ \hline

Single-Switch&\makecell*[l]{\cite{XFMR_single_switch} ZCS switching
} 
& 48    & 2.5 / 0   & 4.15 / -4.15  & 5.8\tnote{2}            & NR             & 0.06   & 0.06   &63 \\ \hline
\multirow{3}{*}[-1em]{Forward-like}&\makecell*[l]{\cite{forward1} ZCS turn on} 
& 150   & 11.7 / 0   & 0.06 / -0.06   & 92.91 / 199.92\tnote{3}   & 84.61 / 88.9\tnote{3} & 10    & 10    &18 \\ \cline{2-10}
&\makecell*[l]{\cite{fly_RD} With RD branch.} & 310 & 3 / 0 & 1.7 / -0.4 & 190.5 & NR & 0.7 & 1.75 & 50 \\ \cline{2-10}
&\makecell*[l]{\cite{Fly_D} With feedback diode \\ across primary and secondary}                & 250   & 5 / 0     & 1.05 / -0.3   & 45    & 89.95\tnote{3}  & 0.25  & 0.75  & 10 \\ \hline

\multirow{5}{*}[-3em]{Flyback-like}&\makecell*[l]{\cite{fly_solar} No pulse-forming branch, \\ solar panel powered} & 12 & 17 / -3 & 0.9 / -0.4 & 150 & NR & 2.5 & 5 & 2 \\ \cline{2-10}
&\makecell*[l]{\cite{fly_modular} Modular structure, \\ series diode (single module results)}   & 30    & 7 / 0     & 0.4 / 0       & 18.2  &  40.4  & 10    & 25    & 1.3 \\ \cline{2-10}
&\makecell*[l]{\cite{Fly_lc_stacked} With RD branch. \\ Stacked secondary.} & 25 & 5.5 / -4 & 0.4 / -0.1 & 18.2 & 39 & 1.5 & 4 & 3 \\ \cline{2-10}
&\makecell*[l]{\cite{Fly_RDD} With RDD branch} & 80 & 30 / -5 & 0.4 / -0.28 & 986\tnote{4} &   NR  & 35 & 50 & 1 \\ \cline{2-10}
&\makecell*[l]{\cite{Fly_boost_RDD} Front-end boost converter \\ with RCD (pri.) and RDD (sec.) branches} & 38 & 7 / -2 & 1 / -0.2 & 7 & NR & 10 & 12 & 1.4 \\

\bottomrule
\end{tabular}}

\begin{tablenotes}
\scriptsize

\addtolength{\itemindent}{1cm}
\hspace*{0.5cm}
\begin{minipage}{\linewidth}
\item[1] Not Reported.
\item[2] Input Power. 
\item[3] Tested with resistive load. 
\item[4] Calculated with $P_{o.avg}=\frac{1}{2}C_{load}(\Delta v_o)^2f_{PRF}$.
\end{minipage}
\end{tablenotes}
\end{threeparttable}
\vspace{8pt}
\centering
\caption{Generalized Performance Comparison of PPS Topologies for DBD Applications}
\label{table:PPS2}
\renewcommand{\arraystretch}{1.1}
\begin{threeparttable}
\scalebox{0.8}{
\begin{tabular}{c|c|c|c|c|c|c|c|c|c|l}
\toprule
\textbf{Topology} & \makecell{\textbf{Output}\\\textbf{PRF}} & \makecell{\textbf{Pulse}\\\textbf{Width}\tnote{1}} & \makecell{\textbf{Voltage}\\\textbf{Gain}} & \makecell{\textbf{Power}\\\textbf{Density}} & \makecell{\textbf{Short-circuit}\\\textbf{Tolerance}} & \makecell{\textbf{Fast}\\\textbf{Transient}\tnote{2}} & \makecell{\textbf{Discharge}\\\textbf{Control}} & 
\textbf{Cost\tnote{3}} & 
\makecell{\textbf{Simplicity}} & \textbf{Key Advantages and Disadvantages} \\
\midrule
\makecell{Voltage-Fed\\Bridge-Based} & ++ & ++ & ++ & +++ & ++ & ++++ & ++ & ++ & ++ &
\makecell[l]{+ Bipolar capability.\\
+ MHz or ns pulse width achievable with PFL,\\
-- but waveform depends on impedance matching.\\
-- Additional damping is required for undershoot suppression.} \\\hline

\makecell{Voltage-Fed\\Forward-like} & ++ & ++ & ++ & +++ & ++ & +++ & ++ & +++ & +++ &
\makecell[l]{+ Fewer components.\\
+ Unipolar design possible.\\
-- Undershoot damping often needed.} \\\hline

\makecell{Voltage-Fed\\Flyback-like} & + & + & +++ & ++ & +++ & + & +++ & +++ & +++ &
\makecell[l]{+ High voltage gain with simple design.\\
+ Cost-effective implementation.\\
-- Lossy damping required to suppress oscillations.} \\\hline

\makecell{Marx-Based} & +++ & +++ & ++ & +++ & + & +++++ & + & +++ & ++++ &
\makecell[l]{+ Modular, transformerless architecture.\\
-- Uncontrollable surge current.\\
-- Achieving high gain requires many stages.} \\\hline

\makecell{Current-Fed\\Bridge-Based} & ++ & ++ & + & + & ++++ & ++ & +++++ & + & + &
\makecell[l]{+ Allows design without active front-end,\\
-- but with possible reduction in discharge waveform quality.\\
-- Requires balanced transformer design and switching.} \\\hline

\makecell{Current-Fed\\Push-Pull} & ++ & ++ & + & + & ++++ & ++ & +++++ & ++ & ++ &
\makecell[l]{
+ Current-fed operation with fewer components.\\
+ Allows design without active front-end,\\
-- but with possible reduction in discharge waveform quality.\\
-- Requires balanced transformer design and switching.} \\
\bottomrule
\end{tabular}}
\begin{tablenotes}
\scriptsize
\item[1] More plus signs (\(+\)) indicate shorter achievable minimum pulse width.
\item[2] Fast rising and falling time.
\item[3] More plus signs (\(+\)) indicate lower implementation cost.
\end{tablenotes}
\end{threeparttable}
\vspace{-10pt}
\end{table*}

%%%%%%%%%%%%%%%%%%%%%%%%%%%%%%%%%%%%%%%%%%%%%%%%%%%%%
\vspace{-10pt}
\subsection{Current-fed Systems}

\begin{figure}[!t]
    \centering
    \includegraphics[width=0.9\linewidth]{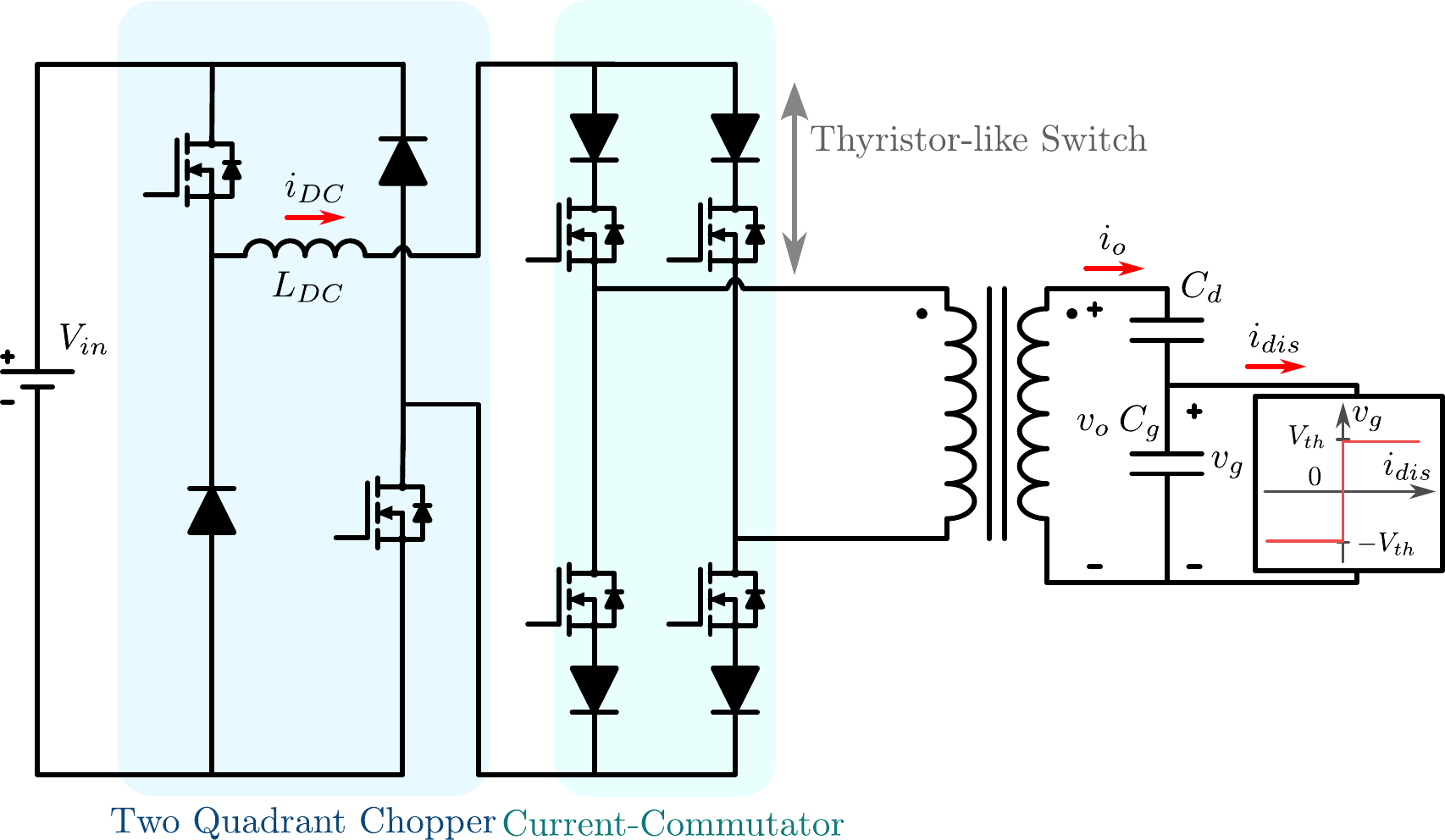}
    \caption{Current-fed square-wave bipolar current commutator with a front-end two-quadrant chopper.}
    \label{fig:CS_square}
    \vspace{-10pt}
\end{figure}
\begin{figure}[!t]
    \centering
    \includegraphics[width=0.7\linewidth]{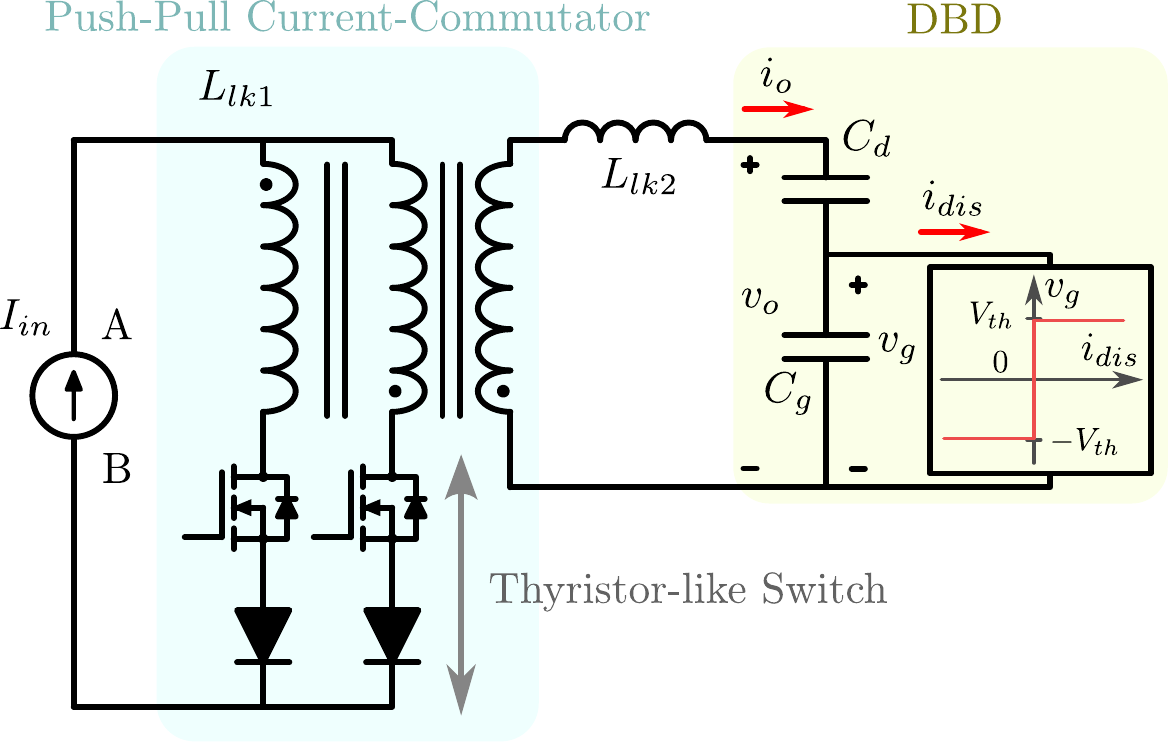}
    \caption{Current-fed push-pull topology.}
    \label{fig:CS_push_pull}
    \vspace{-10pt}
\end{figure}

\subsubsection{Bridge-Based}
Similar to the voltage-fed non-resonant bridge-based PPS, a square-wave bipolar current output can be achieved without a resonant tank in configurations like Fig.~\ref{fig:CS_square}. By assuming a small leakage inductance in the transformer, the output current waveform can ideally approximate a square-wave bipolar current~\cite{CS_square}. This design enables flexibility in operating the DBD load at various frequencies and pulse widths by adjusting the switching frequency and duty ratio.

However, the presence of parasitic components in the transformer distorts the injected current waveform to the DBD load, causing it to deviate from an ideal square wave. Additionally, HF oscillations are introduced, which adversely affect signal integrity and overall efficiency.

Given that DBD capacitances are in the range of tens of pF for prototype scale, it is essential to ensure that the parasitic capacitance remains smaller than the DBD setup. This improves efficiency and preserves optimal system performance—though achieving such low parasitic values in practice remains challenging.

Furthermore, because the voltage level and polarity across the switching devices depend on the voltage across the DBD load, reverse voltage-blocking series diodes are typically required. This adds up to four additional components, each with voltage ratings matching those of the MOSFETs, increasing the overall complexity of the design.

\subsubsection{Push-Pull Topologies}

Diez \textit{et al.} provide a comprehensive review of the push-pull current commutator designs for DBD excilamps with various front-end current converter configurations~\cite{CS_review}. Fig.~\ref{fig:CS_push_pull} shows the commutator consisting of two opposite primary windings in a push-pull arrangement, enabling bipolar current excitation for the DBD load~\cite{CS_fly_boost, CS_fly_buck_boost}. 

\addedthree{A key distinction from the two-quadrant chopper topology described in~\cite{CS_square}, which employs hysteresis control and assumes a stiff input current, is that the input current ($I_{in}$) for the push-pull commutator operates in a pulsed manner, corresponding to discontinuous conduction mode (DCM). Since current is only required when a pulse is generated, a pseudo-constant input current is sufficient for PPS structures.}

The front-end converter utilizes boost and buck-boost topologies operating in DCM. A significant advantage of DCM operation is the zero-current turn-on achieved by the main switch of the front-end current regulator, which reduces switching losses. Additionally, one of the commutator switches inherently achieves ZCS during turn-off when the primary winding current naturally falls to zero, further improving efficiency through partial soft-switching. A simplified front-end current source is also implemented in~\cite{CS_push_pull}, where the AC grid, rectified directly, feeds the commutator while preserving the push-pull structure. This design allows the peak current to scale with the AC grid voltage, simplifying the circuit while maintaining reliable performance.

However, the absence of a stable input current complicates the prediction of the current injected into the DBD load. As a result, it becomes difficult to accurately estimate the voltage across the DBD using \eqref{eqn:Cs_voltage}, particularly during the discharge event when the voltage exceeds the threshold value. This undermines one of the key advantages of current-fed systems—namely, the ability to precisely control and synchronize the discharge timing.

% Additionally, the discontinuous nature of the input current increases THD in the power supply, potentially challenging compliance with power quality standards. Conversely, using a front-end buck converter, for example, can produce continuous input current, potentially reducing THD, though this comes at the cost of losing the soft-switching benefits. 

%%%%%%%%%%%%%%%%%%%%%%%%%%%%%%%%%%%%%%%%%%%%%%%%%%%%%
%%%%%%%%%%%%%%%%%%%%%%%%%%%%%%%%%%%%%%%%%%%%%%%%%%%%%
%%%%%%%%%%%%%%%%%%%%%%%%%%%%%%%%%%%%%%%%%%%%%%%%%%%%%

\vspace{-10pt}
\subsection{Overview of PPS Topologies}

Table~\ref{table:PPS} summarizes representative cases of the reviewed PPS topologies, including input/output voltage levels, average output power, efficiency, PRF, and rise/fall times, where fast transient capability is a critical performance metric. Transformer-based solutions can be applied either for energy storage or energy transfer. Transfer-type approaches (e.g., bridge-based and forward-like) generally offer faster transients, higher PRF, and greater power density compared to storage-type approaches that utilize $L_{mag}$ (e.g., flyback-like), which typically achieve higher voltage gain with simpler designs. Both require additional secondary-side damping (undershoot suppression or oscillation damping) to improve pulse quality, which is often lossy and bulky.

Modular switched-capacitor methods, such as Marx generators, remain attractive due to their transformerless design, fast transient capability, and simple design. However, they suffer from uncontrollable output surge currents and reduced cost-effectiveness and power density as output power increases. Consequently, inductive storage methods (e.g., flyback, PFL, or resonant LC) are still preferred for their short-circuit tolerance and, in some cases, the ability to achieve higher voltage gain or PRF with nanosecond-scale pulse widths. However, their resonance interaction with the capacitive characteristic of the DBD load must be carefully addressed during the design process to prevent instability and maintain pulse quality. Furthermore, compared to the energy density of capacitors, inductive storage generally imposes a limitation on achievable power density.

Although current-fed PPS designs are relatively underexplored, they show strong potential by combining short-circuit tolerance with precise discharge control via output current regulation.
\addedthree{While current-fed PPSs typically share the same front-end architecture and control strategies used in current-fed resonant inverters (e.g., average-current regulation and hysteresis control), a key distinction lies in the discharge behavior, which further simplifies the overall design. Unlike current-fed resonant inverters, which require a continuous and nearly constant input current, PPS applications inherently benefit from discontinuous discharge events, requiring controlled input current only during the discharge pulse interval. This characteristic allows the front-end converter to be greatly simplified, or even omitted, thereby improving power density and system compactness.}

Table~\ref{table:PPS2} provides a generalized performance comparison of PPS topologies for DBD applications. Several PPS designs reported in the literature have yet to be evaluated with actual DBD loads~\cite{recent_pulse_review}. To meet the growing demand for MHz-range PRF, shorter pulse widths, faster transients, and higher power capability, future work should verify their compatibility with DBD operation.

\section{Identified Challenges}\label{sec:7}
%%%%%%%%%%%%%%%%%%%%%%%%%%%%%%%%%%%%%%%%%%%%%%%%%%%%%%%%%%%%%%%%%%%%%%%%%%%%%%%%%%%%%%%%%%%%%%%%%%
\subsection{Current-Fed Non-Isolated Systems}

One notable challenge is the limited exploration of current-fed non-isolated converters in DBD applications, despite their distinct advantages and potential for innovation. These topologies, which are already widely applied in DC–DC and DC–AC systems, offer a variety of configurations that could be adapted to meet the unique demands of DBD operation. At present, for example, sinusoidal power supplies for DBD applications are predominantly implemented using resonant inverters, where the output frequency is maintained within a narrow range to preserve a sinusoidal waveform.

However, to enable variable-frequency operation in DBD applications, conventional voltage-step-based VSIs are unsuitable due to the capacitive characteristics of DBD loads. In contrast, current-source-based inverters (CSIs) can support variable-frequency drive while inherently providing short-circuit tolerance and higher voltage gain—advantages that are particularly relevant to the capacitive nature of DBD loads. Moreover, unlike VSIs, whose voltage gain cannot exceed the input voltage, CSIs can achieve output voltage levels well above the input, making them a promising option for non-isolated designs~\cite{recent_current_Fed1, recent_current_Fed2}. Although limited in maximum operating frequency (below 100~kHz) and power density due to the lack of inherent soft-switching capability, their suitability for variable‑frequency operation, robust handling of capacitive loads, and precise discharge‑current control make CSIs a compelling target for further research and optimization in DBD power supplies.

\vspace{-10pt}
%%%%%%%%%%%%%%%%%%%%%%%%%%%%%%%%%%%%%%%%%%%%%%%%%%%%%%%%%%%%%%%%
%%%%%%%%%%%%%%%%%%%%%%%%%%%%%%%%%%%%%%%%%%%%%%%%%%%%%%%%%%%%%%%%
\subsection{\addedthree{Front-End Converter and Control}}

\addedthree{It is often assumed that achieving a higher voltage gain is inherently beneficial for DBD applications. However, intermediate diffuse stages such as TDBD and GDBD, which offer unique advantages, can transition into filamentary discharge under excessive voltage input, resulting in non-uniform plasma generation. To maintain operation within the range that suppresses microdischarges and leverages the benefits of diffuse modes, it is crucial to ensure that the voltage input remains adjustable to suit these conditions.}

\addedthree{To mitigate this issue, front-end conventional PWM-controlled voltage boost stages can be utilized to not only achieve adjustable voltage gain, but also alleviate the voltage gain requirement of the sinusoidal or pulse-generating stage. Usually, the second stage's frequency-dependent voltage gain is known at the operating frequency, the required gain for the downstream converter can be closed-loop controlled~\cite{rev3_voltage}. The simplicity and minimal component count of boost converters and their variations make designs cost-effective and compact~\cite{Fly_boost_RDD}.}

\addedthree{However, these front-end boost converters face control challenges due to various factors, such as mode transitions between CCM and DCM, non-minimum phase behavior, and limited control bandwidth caused by the presence of a right-half-plane (RHP) zero. To avoid these issues, proper modeling of the entire system is required. Since many advanced equivalent models have been reported~\cite{mis_equi}, the control logic of the front-end converter should incorporate not only the modeling of sinusoidal or pulse-generation stage but also the complex behavior of DBD. This ensures a comprehensive analysis of system operation, enabling the expected operation mode (e.g., filamentary and homogeneous DBD), proper front-end converter control, and preventing unexpected impedance mismatches, such as those occurring in resonant operation.}

\addedthree{Furthermore, in cascaded architectures where the load-stage converter is expected to produce a sinusoidal or pulsed output while behaving linearly with respect to its input voltage, which is the output of the front-end converter, resonant inverters often violate this assumption. In many Class-E and Class-F resonant stages, the voltage gain is not only frequency-dependent but also inherently nonlinear with respect to the input voltage, making the design of feedback loops with adjustable gain significantly more challenging.}

%%%%%%%%%%%%%%%%%%%%%%%%%%%%%%%%%%%%%%%%%%%%%%%%%%%%%%%%%%%%%%%%
%%%%%%%%%%%%%%%%%%%%%%%%%%%%%%%%%%%%%%%%%%%%%%%%%%%%%%%%%%%%%%%%
\subsection{Metric Gaps in Existing Literature}

\begin{figure}[tp!]
    \centering
    \includegraphics[width=0.9\linewidth]{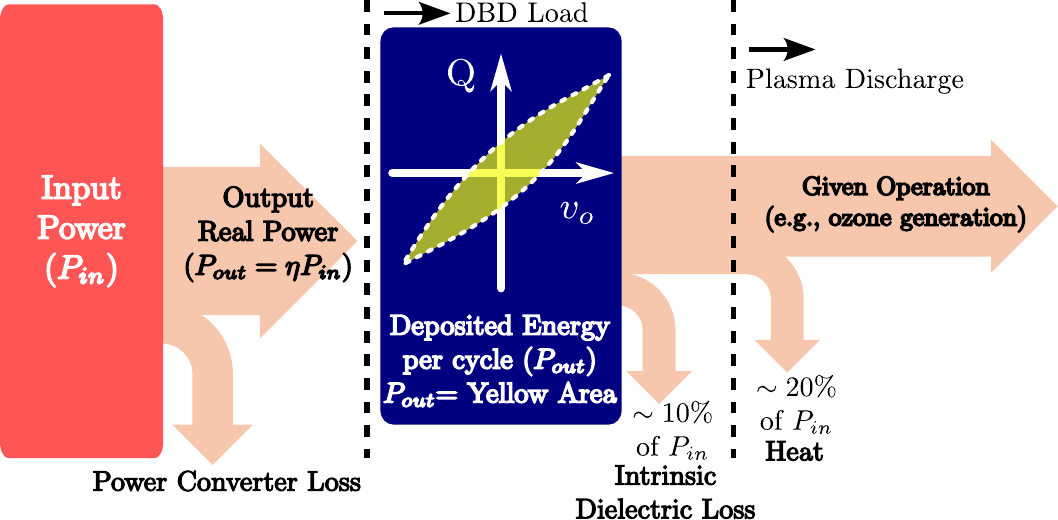}
    \caption{Power distribution in DBD system.}
    \label{fig:power_loss}
    \vspace{-10pt}
\end{figure}

A critical gap identified in existing DBD-related power electronics literature is the absence of a standardized, converter-focused performance metric. Many studies emphasize application-level outcomes, such as ozone yield or sterilization efficiency, which depend heavily on system-level parameters including DBD geometry, electrode material, gas composition, and environmental conditions. Even, as shown in Fig.~\ref{fig:power_loss}, the dissipated power of DBD load include intrinsic dielectric loss and heat generation (i.e., radiation and convection). Such metrics, although valuable in assessing end-use performance, are inadequate for benchmarking the electrical performance or efficiency of power converters, as they do not directly reflect converter-level power handling capabilities~\cite{challenges_heat}.

In addition, the reporting of electrical metrics across the literature is inconsistent. The literature lacks uniformity in specifying how output power is defined or measured, employing methods such as direct voltage-current integration, assumed input power, or charge-voltage (Q–V, Lissajous) plots, often without explicit clarification. Consequently, essential metrics for power electronics—such as power density ($\mathrm{W/cm^2}$), efficiency (\%), or energy per pulse (J)—are inconsistently applied or frequently omitted, significantly preventing meaningful comparisons among different converter topologies and excitation methods.

A practical approach to converter characterization is to use Q–V plots to determine the real power, or to employ capacitive or RC dummy loads that approximate the reactive behavior of DBD reactors, particularly under HF excitation~\cite{XFMR_class_E}. Such methods enable realistic evaluation of converter performance and efficiency without the need for active plasma discharges.

\section{Future Trends and Research Opportunities}\label{sec:8}
%%%%%%%%%%%%%%%%%%%%%%%%%%%%%%%%%%%%%%%%%%%%%%%%%%%%%%%%%%%%%%%%%%%%%%%%%%%%%%%%%%%%%%%%%%%%%%%%%%
This section highlights technological gaps and presents future research opportunities for power converter designs aimed at improving the performance of DBD systems.

\vspace{-10pt}
\subsection{Integration of New Materials}

Power supplies for DBD applications can greatly benefit from the integration of wide-bandgap (WBG) devices such as gallium nitride (GaN) high-electron-mobility transistors (HEMTs) and silicon carbide (SiC) \textsc{mosfet}s. These devices offer high thermal conductivity, high blocking voltage, and reduced conduction and switching losses. Their fast transient response—up to ten times faster than Si insulated-gate bipolar transistors (IGBTs)—enables high-power, high-PRF power supplies for DBD applications. Among WBG devices, GaN stands out due to its high switching speed, with 650-V GaN HEMTs achieving rise times between 2.5 and 12.5 ns~\cite{WBG1}. However, its adoption in high-power applications is constrained by the limited availability of commercially viable HV, high-current GaN devices compared to widely commercialized SiC \textsc{mosfet}s (e.g., 1.2 kV). The future commercialization of high-performance GaN devices will facilitate HF operation in DBD power supplies, enhancing plasma efficiency and enabling new applications.

For sinusoidal excitation of DBD, achieving MHz-range operation, WBG becomes an attractive choice. 
In pulsed-excitation systems, the ability of WBG devices to handle fast transients is particularly advantageous. Faster switching speeds enable shorter rise and fall times, improving ionization rates, plasma uniformity, and energy efficiency. Considering PPS often operates with nanosecond pulsewidths, minimizing delays caused by switching transients is essential for achieving high precision and performance.
However, in both HF sinusoidal and pulsed-excitation systems, careful implementation is required due to potential EMI issues caused by large $dv/dt$.

The integration of piezoelectric resonators can further enhance multiple designs, thanks to their superior quality factor in resonant tanks or discrete inductors with compact volume. Aligning with the implemenation of fast WBG devices, piezoelectric resonators function effectively at high frequencies, making them well-suited for high-power-density applications. Furthermore, they do not generate external magnetic fields, mitigating EMI concerns. Given that DBD power supplies often rely on resonant behavior, exploring various topologies incorporating piezoelectric resonators is a promising research direction.
A key design challenge, however, lies in the limitations of the simplified BVD model, which fails to predict spurious resonant modes, resulting in unexpected resonance~\cite{piezo}. Additionally, fabrication complexities and temperature sensitivity of the piezoelectric resonator remain challenges in practical implementations.

\vspace{-10pt}
\subsection{\addedthree{Slew-Rate Control}}
\addedthree{
As the rising and falling times of the PPS waveform influence the operating mode of DBD, including filamentary and diffuse modes, slew-rate control of PPS topologies will enhance the DBD performance. For example, while a high slew rate of the voltage transition enhances uniform plasma distribution, it can lead to excessive peak current, necessitating controlled slew-rate adjustment.
Active gate drivers, as discussed in~\cite{slew_review}, offer an effective solution for achieving adjustable slew rates, which can be further integrated into PPS, thereby enhancing the operational capability of PPS under different DBD conditions. By dynamically tuning the gate control, these drivers help mitigate excessive current spikes while ensuring optimal excitation conditions for various DBD loads.}
\vspace{-10pt}
\subsection{Separate Excitation and DC Biasing}
Besides slew-rate control, separate excitation and DC biasing strategies offer another promising approach to optimizing DBD operation. Combining AC excitation with a controlled DC bias enhances plasma tuning flexibility and thrust efficiency~\cite{bias1,bias2,bias4}. By independently applying DC bias, charge deposition on the dielectric surface can be managed, reducing performance degradation and enabling precise control over plasma formation and ion momentum transfer—critical for aerodynamic flow control.

Expanding on this approach, a hybrid waveform can be realized. A hybrid waveform of nanosecond pulses and low-frequency sinusoidal bias further prevents charge accumulation, improving efficiency and ensuring stable long-term operation~\cite{DBD_tendency4}. This approach maintains a uniform discharge while minimizing energy losses and mode transitions. 
Future research should focus on optimizing hybrid waveform configurations, developing active charge management strategies, and designing power supplies capable of supporting these advanced excitation methods.

\vspace{-10pt}
\subsection{Emerging Future Applications}
As research on DBD advances, multiple applications have emerged, each presenting unique power supply constraints.

\subsubsection{Aerospace and Space Applications}
SDBD has been researched for plasma thrusters and aerodynamic actuators, enabling flush-mounted, drag-free integration and conformability to curved airfoils, with its asymmetric geometry, generating wall-parallel ionic wind for momentum transfer~\cite{under_aero1,under_aero2,under_aero3, under_aero6}.

Despite these benefits, SDBD thrusters demand lightweight power converters capable of operating reliably across extreme environmental conditions including variable atmospheric pressure, temperature fluctuations, and EMI. Battery power constraints impose strict efficiency requirements and high-power-density converter designs. Additionally, the need for real-time thrust modulation demands rapid voltage and frequency control capabilities, pushing converter bandwidth requirements beyond conventional limits. EMI shielding and isolation become critical, necessitating robust component selection and thermal management. Further exploration of modulation strategies is also needed, for instance, burst-mode modulation has been shown to effectively control vortices and flow separation, enhancing boundary-layer reattachment or inducing turbulence as needed~\cite{mode_burst1,mode_burst5,mode_burst2,mode_burst4}. \addedthree{Moving beyond open-loop airflow control, recent studies have demonstrated closed-loop plasma-assisted aerodynamic feedback based on application-level measurements such as ionic wind speed or acoustic response~\cite{aero_feedback1,aero_feedback2}.}

\subsubsection{Advanced Manufacturing and Surface Treatment}  
DBD is applied to additive manufacturing for in-situ surface treatment and multi-material bonding, including plasma-assisted layer welding and composite–metal adhesion~\cite{under_3d1,under_3d2,under_3d3}.

However, the dynamic nature of manufacturing and surface treating processes demands rapid power adaptation, requiring converters with fast transient response and wide bandwidth control systems. Miniaturization for integration into print heads and robotic systems requires compact packaging and thermal management solutions. In addition, programmable waveform generation with precise amplitude, frequency, and pulse width control is essential—illustrated by thin-film deposition studies where frequency-shift keying-modulated sinusoidal waveforms enable control of HF/low-frequency ratios to balance nanoparticle deposition and matrix growth~\cite{DBD_tendency3}.

\subsubsection{Nanotechnology and Quantum Materials}
DBD also enables controlled synthesis of nanomaterials such as carbon quantum dots~\cite{under_carbon2}, silver oxide~\cite{under_silver1}, and gold nanoparticles~\cite{under_gold1}, offering energy efficiency, environmental benefits, and morphology control. It also facilitates ordered quantum dot array fabrication with high uniformity in a defect-free bottom-up approach~\cite{under_quantum}. It is noteworthy that precise control of plasma parameters demands ultra-stable power supplies with minimal voltage and current ripple, as small variations can significantly affect nanoparticle size distribution and morphology.

\section{Conclusion}\label{sec:9}

This review has highlighted the critical trade-offs among various excitation methods and power supply topologies, emphasizing their respective strengths and limitations. The requirements for power, voltage, current, frequency, waveform shape, and slew rate vary widely depending on DBD geometry and application constraints. Advances in materials, control strategies, and topology optimization present opportunities to improve the efficiency, scalability, and reliability of DBD systems. Future advancements in DBD power supply design will depend on addressing interdisciplinary gaps, particularly the lack of standardized performance metrics and the underexplored design space for certain topologies. This comprehensive review is intended to serve as a foundational resource for researchers and engineers, inspiring further progress in the field and supporting the power electronics-driven development of next-generation plasma technologies.

\ifCLASSOPTIONcaptionsoff
  \newpage
\fi

% References
% \bibliographystyle{IEEEtran}
% \bibliography{bib_power,bib_DBD.bib,bib_mis,bib_DBD_intro,bib_DBD_review, bib_rev1}

%,

% \end{IEEEbiography}


% % insert where needed to balance the two columns on the last page with
% %\newpage


%%%%%%%%%%%%%%%%%%%%%%%%%%%%%%%%%%%%%%%%%%%%%%%%%%%%%%%%%%
%%%%%%%%%%%%%%%%%%%%%%%%%%%%%%%%%%%%%%%%%%%%%%%%%%%%%%%%%%

% \newpage
% \section*{References}
% \nocite{Siepe2024}
% \printbibliography[heading=none]

  % you can also add \setlength{\bibitemsep}{0.5ex} here
\bibliographystyle{IEEEtran}
% \bibliography{bib_power,bib_DBD.bib,bib_mis,bib_DBD_intro,bib_DBD_review, bib_rev2_reviewer1, bib_rev2_reviewer3, bib_rev3}

\begin{thebibliography}{100}
\providecommand{\url}[1]{#1}
\csname url@samestyle\endcsname
\providecommand{\newblock}{\relax}
\providecommand{\bibinfo}[2]{#2}
\providecommand{\BIBentrySTDinterwordspacing}{\spaceskip=0pt\relax}
\providecommand{\BIBentryALTinterwordstretchfactor}{4}
\providecommand{\BIBentryALTinterwordspacing}{\spaceskip=\fontdimen2\font plus
\BIBentryALTinterwordstretchfactor\fontdimen3\font minus \fontdimen4\font\relax}
\providecommand{\BIBforeignlanguage}[2]{{%
\expandafter\ifx\csname l@#1\endcsname\relax
\typeout{** WARNING: IEEEtran.bst: No hyphenation pattern has been}%
\typeout{** loaded for the language `#1'. Using the pattern for}%
\typeout{** the default language instead.}%
\else
\language=\csname l@#1\endcsname
\fi
#2}}
\providecommand{\BIBdecl}{\relax}
\BIBdecl

\bibitem{DBD_basics1}
B.~Eliasson and U.~Kogelschatz, ``Modeling and applications of silent discharge plasmas,'' \emph{IEEE Transactions on Plasma Science}, vol.~19, no.~2, pp. 309--323, 1991.

\bibitem{DBD_basic2}
J.~Park, I.~Henins, H.~Herrmann, G.~Selwyn, J.~Jeong, R.~Hicks, D.~Shim, and C.~Chang, ``An atmospheric pressure plasma source,'' \emph{Applied Physics Letters}, vol.~76, no.~3, pp. 288--290, 2000.

\bibitem{DBD_basic3}
U.~Kogelschatz, B.~Eliasson, and W.~Egli, ``Dielectric-barrier discharges. principle and applications,'' \emph{Le Journal de Physique IV}, vol.~7, no.~C4, pp. C4--47, 1997.

\bibitem{DBD_material1}
J.~Xie, Q.~Chen, P.~Suresh, S.~Roy, J.~F. White, and A.~D. Mazzeo, ``based plasma sanitizers,'' \emph{Proceedings of the National Academy of Sciences}, vol. 114, no.~20, pp. 5119--5124, 2017.

\bibitem{DBD_material2}
R.~Li, Q.~Tang, S.~Yin, and T.~Sato, ``Investigation of dielectric barrier discharge dependence on permittivity of barrier materials,'' \emph{Applied physics letters}, vol.~90, no.~13, 2007.

\bibitem{DBD_material3}
J.~Kim, S.-j. Kim, Y.-N. Lee, I.-T. Kim, and G.~Cho, ``Discharge characteristics and plasma erosion of various dielectric materials in the dielectric barrier discharges,'' \emph{Applied Sciences}, vol.~8, no.~8, p. 1294, 2018.

\bibitem{DBD_temp1}
K.~Zhang, G.~Zhang, X.~Liu, A.~N. Phan, and K.~Luo, ``A study on co2 decomposition to co and o2 by the combination of catalysis and dielectric-barrier discharges at low temperatures and ambient pressure,'' \emph{Industrial \& Engineering Chemistry Research}, vol.~56, no.~12, pp. 3204--3216, 2017.

\bibitem{DBD_temp3}
A.~Bogaerts and E.~C. Neyts, ``Plasma technology: an emerging technology for energy storage,'' \emph{ACS Energy Letters}, vol.~3, no.~4, pp. 1013--1027, 2018.

\bibitem{DBD_temp4}
B.~Lee, D.-W. Kim, and D.-W. Park, ``Dielectric barrier discharge reactor with the segmented electrodes for decomposition of toluene adsorbed on bare-zeolite,'' \emph{Chemical Engineering Journal}, vol. 357, pp. 188--197, 2019.

\bibitem{DBD_polymer2}
P.~Cools, S.~Van~Vrekhem, N.~De~Geyter, and R.~Morent, ``The use of dbd plasma treatment and polymerization for the enhancement of biomedical uhmwpe,'' \emph{Thin Solid Films}, vol. 572, pp. 251--259, 2014.

\bibitem{DBD_wound1}
G.~Busco, E.~Robert, N.~Chettouh-Hammas, J.-M. Pouvesle, and C.~Grillon, ``The emerging potential of cold atmospheric plasma in skin biology,'' \emph{Free Radical Biology and Medicine}, vol. 161, pp. 290--304, 2020.

\bibitem{DBD_wound2}
S.~Arndt, A.~Schmidt, S.~Karrer, and T.~von Woedtke, ``Comparing two different plasma devices kinpen and adtec steriplas regarding their molecular and cellular effects on wound healing,'' \emph{Clinical Plasma Medicine}, vol.~9, pp. 24--33, 2018.

\bibitem{DBD_wound3}
B.~Boekema, M.~Vlig, D.~Guijt, K.~Hijnen, S.~Hofmann, P.~Smits, A.~Sobota, E.~Van~Veldhuizen, P.~Bruggeman, and E.~Middelkoop, ``A new flexible dbd device for treating infected wounds: in vitro and ex vivo evaluation and comparison with a rf argon plasma jet,'' \emph{Journal of Physics D: Applied Physics}, vol.~49, no.~4, p. 044001, 2015.

\bibitem{DBD_food_review}
B.~Zhou, H.~Zhao, X.~Yang, and J.-H. Cheng, ``Versatile dielectric barrier discharge cold plasma for safety and quality control in fruits and vegetables products: Principles, configurations and applications,'' \emph{Food Research International}, p. 115117, 2024.

\bibitem{DBD_food_review2}
E.~Feizollahi, N.~Misra, and M.~Roopesh, ``Factors influencing the antimicrobial efficacy of dielectric barrier discharge (dbd) atmospheric cold plasma (acp) in food processing applications,'' \emph{Critical Reviews in Food Science and Nutrition}, vol.~61, no.~4, pp. 666--689, 2021.

\bibitem{DBD_food1}
I.~Albertos, A.~Mart{\'\i}n-Diana, P.~Cullen, B.~K. Tiwari, S.~Ojha, P.~Bourke, C.~{\'A}lvarez, and D.~Rico, ``Effects of dielectric barrier discharge (dbd) generated plasma on microbial reduction and quality parameters of fresh mackerel (scomber scombrus) fillets,'' \emph{Innovative Food Science \& Emerging Technologies}, vol.~44, pp. 117--122, 2017.

\bibitem{DBD_catalyst_review}
F.~Th{\'e}venet, L.~Sivachandiran, O.~Guaitella, C.~Barakat, and A.~Rousseau, ``Plasma--catalyst coupling for volatile organic compound removal and indoor air treatment: a review,'' \emph{Journal of Physics D: Applied Physics}, vol.~47, no.~22, p. 224011, 2014.

\bibitem{DBD_catalyst_review2}
T.~Chang, Y.~Wang, Y.~Wang, Z.~Zhao, Z.~Shen, Y.~Huang, S.~K. Veerapandian, N.~De~Geyter, C.~Wang, Q.~Chen \emph{et~al.}, ``A critical review on plasma-catalytic removal of vocs: Catalyst development, process parameters and synergetic reaction mechanism,'' \emph{Science of the total Environment}, vol. 828, p. 154290, 2022.

\bibitem{DBD_catalyst_review3}
S.~Li, X.~Dang, X.~Yu, G.~Abbas, Q.~Zhang, and L.~Cao, ``The application of dielectric barrier discharge non-thermal plasma in vocs abatement: A review,'' \emph{Chemical Engineering Journal}, vol. 388, p. 124275, 2020.

\bibitem{mis_equi}
A.~V. Pipa and R.~Brandenburg, ``The equivalent circuit approach for the electrical diagnostics of dielectric barrier discharges: The classical theory and recent developments,'' \emph{Atoms}, vol.~7, no.~1, p.~14, 2019.

\bibitem{DBD_impedance}
R.~Diez, J.-P. Salanne, H.~Piquet, S.~Bhosle, and G.~Zissis, ``Predictive model of a dbd lamp for power supply design and method for the automatic identification of its parameters,'' \emph{The European Physical Journal-Applied Physics}, vol.~37, no.~3, pp. 307--313, 2007.

\bibitem{dbd_equi_non_ideal1}
U.~Pal, M.~Kumar, M.~Tyagi, B.~Meena, H.~Khatun, and A.~Sharma, ``Discharge analysis and electrical modeling for the development of efficient dielectric barrier discharge,'' in \emph{Journal of physics: Conference series}, vol. 208, no.~1.\hskip 1em plus 0.5em minus 0.4em\relax IOP Publishing, 2010, p. 012142.

\bibitem{dbd_equi_non_ideal2}
A.~Yehia, ``The electrical characteristics of the dielectric barrier discharges,'' \emph{Physics of Plasmas}, vol.~23, no.~6, 2016.

\bibitem{dbd_equi_non_ideal3}
Z.~Fang, S.~Ji, J.~Pan, T.~Shao, and C.~Zhang, ``Electrical model and experimental analysis of the atmospheric-pressure homogeneous dielectric barrier discharge in he,'' \emph{IEEE transactions on plasma science}, vol.~40, no.~3, pp. 883--891, 2012.

\bibitem{DBD_review1}
R.~Brandenburg, ``Dielectric barrier discharges: progress on plasma sources and on the understanding of regimes and single filaments,'' \emph{Plasma Sources Science and Technology}, vol.~26, no.~5, p. 053001, 2017.

\bibitem{DBD_catalyst}
K.~Ollegott, P.~Wirth, C.~Oberste-Beulmann, P.~Awakowicz, and M.~Muhler, ``Fundamental properties and applications of dielectric barrier discharges in plasma-catalytic processes at atmospheric pressure,'' \emph{Chemie Ingenieur Technik}, vol.~92, no.~10, pp. 1542--1558, 2020.

\bibitem{DBD_PB2}
D.~Mei and X.~Tu, ``Atmospheric pressure non-thermal plasma activation of co2 in a packed-bed dielectric barrier discharge reactor,'' \emph{ChemPhysChem}, vol.~18, no.~22, pp. 3253--3259, 2017.

\bibitem{DBD_FB1}
S.~Zhang, S.~Zhang, Z.~Liu, and K.~Yan, ``Remediation of 3, 4, 3', 4'-tetrachlorobiphenyl (pcb77) contaminated soil via a fluidized bed dielectric barrier discharge,'' \emph{Science of The Total Environment}, vol. 933, p. 173208, 2024.

\bibitem{DBD_FB2}
T.~Nozaki, X.~Chen, D.-Y. Kim, and H.-H. Kim, ``Plasma fluidized beds and their scalability,'' \emph{Current Opinion in Green and Sustainable Chemistry}, p. 100984, 2024.

\bibitem{DBD_FB3}
C.~Du, R.~Qiu, and J.~Ruan, \emph{Plasma fluidized bed}.\hskip 1em plus 0.5em minus 0.4em\relax Springer, 2018.

\bibitem{DBD_PB_lissa1}
L.~Jin, Z.~Xinbo, H.~Xueli, and T.~Xin, ``Plasma-assisted ammonia synthesis in a packed-bed dielectric barrier discharge reactor: roles of dielectric constant and thermal conductivity of packing materials,'' \emph{Plasma Science and Technology}, vol.~24, no.~2, p. 025503, 2022.

\bibitem{DBD_equivalent1}
S.~Liu and M.~Neiger, ``Electrical modelling of homogeneous dielectric barrier discharges under an arbitrary excitation voltage,'' \emph{Journal of Physics D: Applied Physics}, vol.~36, no.~24, p. 3144, 2003.

\bibitem{Food7}
J.~Kim, S.~Park, and W.~Choe, ``Surface plasma with an inkjet-printed patterned electrode for low-temperature applications,'' \emph{Scientific Reports}, vol.~11, no.~1, p. 12206, 2021.

\bibitem{DBD_SDBD_lissa1}
I.~Biganzoli, R.~Barni, A.~Gurioli, R.~Pertile, and C.~Riccardi, ``Experimental investigation of lissajous figure shapes in planar and surface dielectric barrier discharges,'' in \emph{Journal of Physics: Conference Series}, vol. 550, no.~1.\hskip 1em plus 0.5em minus 0.4em\relax IOP Publishing, 2014, p. 012039.

\bibitem{DBD_SDBD_lissa2}
H.~Jakob and M.~Kim, ``Electrical model for complex surface dbd plasma sources,'' \emph{IEEE Transactions on Plasma Science}, vol.~49, no.~10, pp. 3051--3058, 2021.

\bibitem{FE_DBD_human2}
L.-X. Zhao, H.-X. Zhao, H.~Chen, C.~Hu, Y.~Zhang, and H.-P. Li, ``Characteristics of portable air floating-electrode dielectric-barrier-discharge plasmas used for biomedicine,'' \emph{Plasma Chemistry and Plasma Processing}, vol.~43, no.~6, pp. 1567--1585, 2023.

\bibitem{FE_DBD_human}
L.~Kyu-Hang, K.~Sichan, J.~Hyun, S.~Byung-Koo, S.~Myung-Sun, and C.~Guangsup, ``Plasma skincare device based on floating electrode dielectric barrier discharge,'' \emph{Plasma Science and Technology}, vol.~21, no.~12, p. 125403, 2019.

\bibitem{DBD_tendency1}
N.~Mericam-Bourdet, M.~Kirkpatrick, F.~Tuvache, D.~Frochot, and E.~Odic, ``Effect of voltage waveform on dielectric barrier discharge ozone production efficiency,'' \emph{The European Physical Journal-Applied Physics}, vol.~57, no.~3, p. 30801, 2012.

\bibitem{DBD_tendency6}
J.~M. Williamson, D.~D. Trump, P.~Bletzinger, and B.~N. Ganguly, ``Comparison of high-voltage ac and pulsed operation of a surface dielectric barrier discharge,'' \emph{Journal of Physics D: Applied Physics}, vol.~39, no.~20, p. 4400, 2006.

\bibitem{DBD_tendency11}
U.~Kogelschatz, ``Dielectric-barrier discharges: their history, discharge physics, and industrial applications,'' \emph{Plasma chemistry and plasma processing}, vol.~23, no.~1, pp. 1--46, 2003.

\bibitem{DBD_tendency9}
A.~Nakano and H.~Nishida, ``The effect of the voltage waveform on performance of dielectric barrier discharge plasma actuator,'' \emph{Journal of Applied Physics}, vol. 126, no.~17, 2019.

\bibitem{DBD_tendency8}
N.~Benard and E.~Moreau, ``Role of the electric waveform supplying a dielectric barrier discharge plasma actuator,'' \emph{Applied Physics Letters}, vol. 100, no.~19, 2012.

\bibitem{DBD_tendency5}
M.~Forte, J.~Jolibois, J.~Pons, E.~Moreau, G.~Touchard, and M.~Cazalens, ``Optimization of a dielectric barrier discharge actuator by stationary and non-stationary measurements of the induced flow velocity: application to airflow control,'' \emph{Experiments in fluids}, vol.~43, pp. 917--928, 2007.

\bibitem{DBD_tendency10}
M.~C. Dumlao, D.~Xiao, D.~Zhang, J.~Fletcher, and W.~A. Donald, ``Effects of different waveforms on the performance of active capillary dielectric barrier discharge ionization mass spectrometry,'' \emph{Journal of The American Society for Mass Spectrometry}, vol.~28, no.~4, pp. 575--578, 2016.

\bibitem{DBD_bi_uni}
E.~Panousis, N.~Merbahi, F.~Clement, M.~Yousfi, J.-f. Loiseau, O.~Eichwald, and B.~Held, ``Analysis of dielectric barrier discharges under unipolar and bipolar pulsed excitation,'' \emph{IEEE Transactions on Dielectrics and Electrical Insulation}, vol.~16, no.~3, pp. 734--741, 2009.

\bibitem{DBD_diffuse}
F.~Massines, N.~Gh{\'e}rardi, N.~Naud{\'e}, and P.~S{\'e}gur, ``Recent advances in the understanding of homogeneous dielectric barrier discharges,'' \emph{The European Physical Journal-Applied Physics}, vol.~47, no.~2, p. 22805, 2009.

\bibitem{DBD_tendency_freq1}
W.~Jiang, J.~Tang, Y.~Wang, W.~Zhao, and Y.~Duan, ``Influence of driving frequency on discharge modes in a dielectric-barrier discharge with multiple current pulses,'' \emph{Physics of Plasmas}, vol.~20, no.~7, 2013.

\bibitem{DBD_tendency12}
R.~Bazinette, R.~Subileau, J.~Paillol, and F.~Massines, ``Identification of the different diffuse dielectric barrier discharges obtained between 50 khz to 9 mhz in ar/nh3 at atmospheric pressure,'' \emph{Plasma Sources Science and Technology}, vol.~23, no.~3, p. 035008, 2014.

\bibitem{mis_gas_breakdown2}
W.~Jiang, W.~Hao, W.~Zhijiang, Y.~Lin, and Y.~Zhang, ``Gas breakdown in radio-frequency field within mhz range: a review of the state of the art,'' \emph{Plasma Science and Technology}, vol.~24, no.~12, p. 124018, 2022.

\bibitem{mis_gas_breakdown}
J.~L. Walsh, Y.~T. Zhang, F.~Iza, and M.~G. Kong, ``Atmospheric-pressure gas breakdown from 2 to 100 mhz,'' \emph{Applied Physics Letters}, vol.~93, no.~22, 2008.

\bibitem{DBD_tendency_RF1}
R.~Bazinette, J.~Paillol, and F.~Massines, ``Optical emission spectroscopy of glow, townsend-like and radiofrequency dbds in an ar/nh3 mixture,'' \emph{Plasma Sources Science and Technology}, vol.~24, no.~5, p. 055021, 2015.

\bibitem{Food5}
X.~Y. Dong and Y.~L. Yang, ``A novel approach to enhance blueberry quality during storage using cold plasma at atmospheric air pressure,'' \emph{Food and Bioprocess Technology}, vol.~12, no.~8, pp. 1409--1421, 2019.

\bibitem{Food6}
D.~Zhou, R.~Sun, W.~Zhu, Y.~Shi, S.~Ni, C.~Wu, and T.~Li, ``Impact of dielectric barrier discharge cold plasma on the quality and phenolic metabolism in blueberries based on metabonomic analysis,'' \emph{Postharvest Biology and Technology}, vol. 197, p. 112208, 2023.

\bibitem{Food11}
S.~Rana, D.~Mehta, V.~Bansal, U.~Shivhare, and S.~K. Yadav, ``Atmospheric cold plasma (acp) treatment improved in-package shelf-life of strawberry fruit,'' \emph{Journal of food science and technology}, vol.~57, pp. 102--112, 2020.

\bibitem{Food13}
J.~Cheng, T.~Li, K.~Cong, C.~Wu, X.~Ge, G.~Fan, X.~Li, D.~Zhou, Z.~Yan, and Y.~Li, ``Effects of dielectric barrier discharge plasma and plasma-activated water on the surface microbial diversity of fresh goji berries during storage,'' \emph{Scientia Horticulturae}, vol. 313, p. 111920, 2023.

\bibitem{Food15}
Q.~Wu, C.~Shen, J.~Li, D.~Wu, and K.~Chen, ``Application of indirect plasma-processed air on microbial inactivation and quality of yellow peaches during storage,'' \emph{Innovative Food Science \& Emerging Technologies}, vol.~79, p. 103044, 2022.

\bibitem{Food16}
Y.~Wu, J.-H. Cheng, and D.-W. Sun, ``Subcellular damages of colletotrichum asianum and inhibition of mango anthracnose by dielectric barrier discharge plasma,'' \emph{Food Chemistry}, vol. 381, p. 132197, 2022.

\bibitem{Food34}
A.~Sudarsan and K.~M. Keener, ``Inactivation of salmonella enterica serovars and escherichia coli o157: H7 surrogate from baby spinach leaves using high voltage atmospheric cold plasma (hvacp),'' \emph{LWT}, vol. 155, p. 112903, 2022.

\bibitem{Food43}
T.~K. Ranjitha~Gracy, V.~Gupta, and R.~Mahendran, ``Influence of low-pressure nonthermal dielectric barrier discharge plasma on chlorpyrifos reduction in tomatoes,'' \emph{Journal of food process engineering}, vol.~42, no.~6, p. e13242, 2019.

\bibitem{Food14}
Y.~Du, S.~Mi, H.~Wang, S.~Yuan, F.~Yang, H.~Yu, Y.~Xie, Y.~Guo, Y.~Cheng, and W.~Yao, ``Intervention mechanisms of cold plasma pretreatment on the quality, antioxidants and reactive oxygen metabolism of fresh wolfberries during storage,'' \emph{Food Chemistry}, vol. 431, p. 137106, 2024.

\bibitem{Food30}
F.~Pasquali, A.~C. Stratakos, A.~Koidis, A.~Berardinelli, C.~Cevoli, L.~Ragni, R.~Mancusi, G.~Manfreda, and M.~Trevisani, ``Atmospheric cold plasma process for vegetable leaf decontamination: A feasibility study on radicchio (red chicory, cichorium intybus l.),'' \emph{Food control}, vol.~60, pp. 552--559, 2016.

\bibitem{Food40}
H.~Liu, D.~Guo, and X.~Feng, ``Plasma degradation of pesticides on the surface of corn and evaluation of its quality changes,'' \emph{Sustainability}, vol.~13, no.~16, p. 8830, 2021.

\bibitem{Food41}
X.~Feng, X.~Ma, H.~Liu, J.~Xie, C.~He, and R.~Fan, ``Argon plasma effects on maize: pesticide degradation and quality changes,'' \emph{Journal of the Science of Food and Agriculture}, vol.~99, no.~12, pp. 5491--5498, 2019.

\bibitem{Food10}
D.~Ziuzina, N.~Misra, L.~Han, P.~Cullen, T.~Moiseev, J.-P. Mosnier, K.~Keener, E.~Gaston, I.~Vilar{\'o}, and P.~Bourke, ``Investigation of a large gap cold plasma reactor for continuous in-package decontamination of fresh strawberries and spinach,'' \emph{Innovative Food Science \& Emerging Technologies}, vol.~59, p. 102229, 2020.

\bibitem{Food17}
E.~S. Lee, S.~Im, and S.~C. Min, ``Effects of in-package cold plasma treatment on the physicochemical and oral toxicological properties of raw grape tomatoes and the probability of salmonellosis from consumption of the tomatoes,'' \emph{Food Control}, vol. 152, p. 109809, 2023.

\bibitem{Food20}
Y.~E. Kim and S.~C. Min, ``Inactivation of salmonella in ready-to-eat cabbage slices packaged in a plastic container using an integrated in-package treatment of hydrogen peroxide and cold plasma,'' \emph{Food Control}, vol. 130, p. 108392, 2021.

\bibitem{Food24}
A.~Patange, D.~Boehm, D.~Ziuzina, P.~Cullen, B.~Gilmore, and P.~Bourke, ``High voltage atmospheric cold air plasma control of bacterial biofilms on fresh produce,'' \emph{International Journal of Food Microbiology}, vol. 293, pp. 137--145, 2019.

\bibitem{Food25}
D.~Ziuzina, L.~Han, P.~J. Cullen, and P.~Bourke, ``Cold plasma inactivation of internalised bacteria and biofilms for salmonella enterica serovar typhimurium, listeria monocytogenes and escherichia coli,'' \emph{International journal of food microbiology}, vol. 210, pp. 53--61, 2015.

\bibitem{Food4}
J.~Wang and Z.~Wu, ``Combined use of ultrasound-assisted washing with in-package atmospheric cold plasma processing as a novel non-thermal hurdle technology for ready-to-eat blueberry disinfection,'' \emph{Ultrasonics Sonochemistry}, vol.~84, p. 105960, 2022.

\bibitem{Food22}
S.~M. Hertrich, G.~Boyd, J.~Sites, and B.~A. Niemira, ``Cold plasma inactivation of salmonella in prepackaged, mixed salads is influenced by cross-contamination sequence,'' \emph{Journal of food protection}, vol.~80, no.~12, pp. 2132--2136, 2017.

\bibitem{Food27}
S.~C. Min, S.~H. Roh, B.~A. Niemira, G.~Boyd, J.~E. Sites, J.~Uknalis, and X.~Fan, ``In-package inhibition of e. coli o157: H7 on bulk romaine lettuce using cold plasma,'' \emph{Food microbiology}, vol.~65, pp. 1--6, 2017.

\bibitem{Food28}
S.~C. Min, S.~H. Roh, B.~A. Niemira, J.~E. Sites, G.~Boyd, and A.~Lacombe, ``Dielectric barrier discharge atmospheric cold plasma inhibits escherichia coli o157: H7, salmonella, listeria monocytogenes, and tulane virus in romaine lettuce,'' \emph{International Journal of Food Microbiology}, vol. 237, pp. 114--120, 2016.

\bibitem{Food45}
A.~D. Ngo, K.~Pai, C.~Timmons, L.~M. Ma, and J.~Jacob, ``Evaluation of cylindrical asymmetric surface dielectric barrier discharge actuators for surface decontamination and mixing,'' \emph{Plasma}, vol.~4, no.~4, pp. 755--763, 2021.

\bibitem{Food46}
K.~Lotfy, S.~M. Al-Qahtani, N.~A. Al-Harbi, K.~M. El-Absy, F.~A. Bu~Shulaybi, S.~A. Alali, and T.~A. Mashtoly, ``Influence of non-thermal plasma on the quality and nutritional content of palm dates,'' \emph{Applied Sciences}, vol.~12, no.~17, p. 8587, 2022.

\bibitem{Food21}
C.~Timmons, K.~Pai, J.~Jacob, G.~Zhang, and L.~M. Ma, ``Inactivation of salmonella enterica, shiga toxin-producing escherichia coli, and listeria monocytogenes by a novel surface discharge cold plasma design,'' \emph{Food Control}, vol.~84, pp. 455--462, 2018.

\bibitem{Food8}
P.~Dimitrakellis, M.~Giannoglou, A.~Zeniou, E.~Gogolides, and G.~Katsaros, ``Food container employing a cold atmospheric plasma source for prolonged preservation of plant and animal origin food products,'' \emph{MethodsX}, vol.~8, p. 101177, 2021.

\bibitem{Food9}
M.~Giannoglou, Z.-M. Xanthou, S.~Chanioti, P.~Stergiou, M.~Christopoulos, P.~Dimitrakellis, A.~Efthimiadou, E.~Gogolides, and G.~Katsaros, ``Effect of cold atmospheric plasma and pulsed electromagnetic fields on strawberry quality and shelf-life,'' \emph{Innovative Food Science \& Emerging Technologies}, vol.~68, p. 102631, 2021.

\bibitem{Food44}
C.~Maccaferri, A.~Sainz-Garc{\'\i}a, F.~Capelli, M.~Gherardi, F.~Alba-El{\'\i}as, and R.~Laurita, ``Evaluation of the antimicrobial efficacy of a large-area surface dielectric barrier discharge on food contact surfaces,'' \emph{Plasma Chemistry and Plasma Processing}, vol.~43, no.~6, pp. 1773--1790, 2023.

\bibitem{Food37}
Q.~Wang, R.~K. Pal, H.-W. Yen, S.~P. Naik, M.~K. Orzeszko, A.~Mazzeo, and D.~Salvi, ``Cold plasma from flexible and conformable paper-based electrodes for fresh produce sanitation: Evaluation of microbial inactivation and quality changes,'' \emph{Food Control}, vol. 137, p. 108915, 2022.

\bibitem{Food2}
Y.~Zhao, Y.~Xia, T.~Xi, D.~Zhu, Q.~Zhang, Z.~Qi, D.~Liu, and W.~Wang, ``Control of pathogenic bacteria on the surface of rolling fruits by an atmospheric pressure air dielectric barrier discharge system,'' \emph{Journal of Physics D: Applied Physics}, vol.~53, no.~16, p. 164005, 2020.

\bibitem{Food19}
Y.~Toyokawa, Y.~Yagyu, T.~Misawa, and A.~Sakudo, ``A new roller conveyer system of non-thermal gas plasma as a potential control measure of plant pathogenic bacteria in primary food production,'' \emph{Food Control}, vol.~72, pp. 62--72, 2017.

\bibitem{CH441}
H.~Wang, J.~Han, Z.~Bo, L.~Qin, Y.~Wang, and F.~Yu, ``Non-thermal plasma enhanced dry reforming of ch4 with co2 over activated carbon supported ni catalysts,'' \emph{Molecular Catalysis}, vol. 475, p. 110486, 2019.

\bibitem{CH442}
D.~Li, V.~Rohani, F.~Fabry, A.~P. Ramaswamy, M.~Sennour, and L.~Fulcheri, ``Direct conversion of co2 and ch4 into liquid chemicals by plasma-catalysis,'' \emph{Applied Catalysis B: Environmental}, vol. 261, p. 118228, 2020.

\bibitem{CO243}
D.~Mei, X.~Zhu, C.~Wu, B.~Ashford, P.~T. Williams, and X.~Tu, ``Plasma-photocatalytic conversion of co2 at low temperatures: Understanding the synergistic effect of plasma-catalysis,'' \emph{Applied Catalysis B: Environmental}, vol. 182, pp. 525--532, 2016.

\bibitem{CO246}
D.~Mei and X.~Tu, ``Conversion of co2 in a cylindrical dielectric barrier discharge reactor: Effects of plasma processing parameters and reactor design,'' \emph{Journal of CO2 Utilization}, vol.~19, pp. 68--78, 2017.

\bibitem{CO248}
R.~Benrabbah, C.~Cavaniol, H.~Liu, S.~Ognier, S.~Cavadias, M.~E. G{\'a}lvez, and P.~Da~Costa, ``Plasma dbd activated ceria-zirconia-promoted ni-catalysts for plasma catalytic co2 hydrogenation at low temperature,'' \emph{Catalysis Communications}, vol.~89, pp. 73--76, 2017.

\bibitem{CH439}
N.~Bouchoul, E.~Fourr{\'e}, J.-M. Tatibou{\"e}t, and C.~Batiot-Dupeyrat, ``Plasma-catalytic dry reforming of ch 4 over calcium oxide: catalyst structural and textural modifications,'' \emph{Plasma Chemistry and Plasma Processing}, vol.~39, pp. 713--727, 2019.

\bibitem{CO244}
I.~Michielsen, Y.~Uytdenhouwen, J.~Pype, B.~Michielsen, J.~Mertens, F.~Reniers, V.~Meynen, and A.~Bogaerts, ``Co2 dissociation in a packed bed dbd reactor: First steps towards a better understanding of plasma catalysis,'' \emph{Chemical Engineering Journal}, vol. 326, pp. 477--488, 2017.

\bibitem{CO249}
M.~Nizio, A.~Albarazi, S.~Cavadias, J.~Amouroux, M.~E. Galvez, and P.~Da~Costa, ``Hybrid plasma-catalytic methanation of co2 at low temperature over ceria zirconia supported ni catalysts,'' \emph{International Journal of Hydrogen Energy}, vol.~41, no.~27, pp. 11\,584--11\,592, 2016.

\bibitem{CO252}
R.~H. Rad, V.~Br{\"u}ser, M.~Schiorlin, J.~Sch{\"a}fer, and R.~Brandenburg, ``Enhancement of co2 splitting in a coaxial dielectric barrier discharge by pressure increase, packed bed and catalyst addition,'' \emph{Chemical Engineering Journal}, vol. 456, p. 141072, 2023.

\bibitem{CO253}
A.~M. Banerjee, J.~Billinger, K.~J. Nordheden, and F.~J. Peeters, ``Conversion of co2 in a packed-bed dielectric barrier discharge reactor,'' \emph{Journal of Vacuum Science \& Technology A}, vol.~36, no.~4, 2018.

\bibitem{CO254}
Q.~Wang, Y.~Cheng, and Y.~Jin, ``Dry reforming of methane in an atmospheric pressure plasma fluidized bed with ni/$\gamma$-al2o3 catalyst,'' \emph{Catalysis Today}, vol. 148, no. 3-4, pp. 275--282, 2009.

\bibitem{CO255}
N.~Bouchoul, H.~Touati, E.~Fourr{\'e}, J.-M. Clacens, and C.~Batiot-Dupeyrat, ``Efficient plasma-catalysis coupling for ch4 and co2 transformation in a fluidized bed reactor: Comparison with a fixed bed reactor,'' \emph{Fuel}, vol. 288, p. 119575, 2021.

\bibitem{CO256}
X.~Chen, Z.~Sheng, S.~Murata, S.~Zen, H.-H. Kim, and T.~Nozaki, ``Ch4 dry reforming in fluidized-bed plasma reactor enabling enhanced plasma-catalyst coupling,'' \emph{Journal of CO2 Utilization}, vol.~54, p. 101771, 2021.

\bibitem{VOC107}
O.~Karatum and M.~A. Deshusses, ``A comparative study of dilute vocs treatment in a non-thermal plasma reactor,'' \emph{Chemical engineering journal}, vol. 294, pp. 308--315, 2016.

\bibitem{VOC108}
A.~A. Assadi, A.~Bouzaza, and D.~Wolbert, ``Comparative study between laboratory and large pilot scales for voc's removal from gas streams in continuous flow surface discharge plasma,'' \emph{Chemical Engineering Research and Design}, vol. 106, pp. 308--314, 2016.

\bibitem{VOC117}
P.~Liang, W.~Jiang, L.~Zhang, J.~Wu, J.~Zhang, and D.~Yang, ``Experimental studies of removing typical vocs by dielectric barrier discharge reactor of different sizes,'' \emph{Process Safety and Environmental Protection}, vol.~94, pp. 380--384, 2015.

\bibitem{VOC114}
Z.~Ye, Y.~Zhang, P.~Li, L.~Yang, R.~Zhang, and H.~Hou, ``Feasibility of destruction of gaseous benzene with dielectric barrier discharge,'' \emph{Journal of Hazardous materials}, vol. 156, no. 1-3, pp. 356--364, 2008.

\bibitem{VOC110}
N.~Jiang, L.~Guo, C.~Qiu, Y.~Zhang, K.~Shang, N.~Lu, J.~Li, and Y.~Wu, ``Reactive species distribution characteristics and toluene destruction in the three-electrode dbd reactor energized by different pulsed modes,'' \emph{Chemical Engineering Journal}, vol. 350, pp. 12--19, 2018.

\bibitem{VOC106}
R.~Zhu, Y.~Mao, L.~Jiang, and J.~Chen, ``Performance of chlorobenzene removal in a nonthermal plasma catalysis reactor and evaluation of its byproducts,'' \emph{Chemical Engineering Journal}, vol. 279, pp. 463--471, 2015.

\bibitem{VOC109}
B.~Wang, C.~Chi, M.~Xu, C.~Wang, and D.~Meng, ``Plasma-catalytic removal of toluene over ceo2-mnox catalysts in an atmosphere dielectric barrier discharge,'' \emph{Chemical Engineering Journal}, vol. 322, pp. 679--692, 2017.

\bibitem{VOC112}
D.~Dobslaw, O.~Ortlinghaus, and C.~Dobslaw, ``A combined process of non-thermal plasma and a low-cost mineral adsorber for voc removal and odor abatement in emissions of organic waste treatment plants,'' \emph{Journal of Environmental Chemical Engineering}, vol.~6, no.~2, pp. 2281--2289, 2018.

\bibitem{VOC113}
Y.~Cai, X.~Zhu, W.~Hu, C.~Zheng, Y.~Yang, M.~Chen, and X.~Gao, ``Plasma-catalytic decomposition of ethyl acetate over lamo3 (m= mn, fe, and co) perovskite catalysts,'' \emph{Journal of Industrial and Engineering Chemistry}, vol.~70, pp. 447--452, 2019.

\bibitem{surface1}
G.~Borcia, C.~Anderson, and N.~Brown, ``Surface treatment of natural and synthetic textiles using a dielectric barrier discharge,'' \emph{Surface and Coatings Technology}, vol. 201, no.~6, pp. 3074--3081, 2006.

\bibitem{surface2}
------, ``Dielectric barrier discharge for surface treatment: application to selected polymers in film and fibre form,'' \emph{Plasma Sources Science and Technology}, vol.~12, no.~3, p. 335, 2003.

\bibitem{surface3}
H.-E. Wagner, R.~Brandenburg, K.~Kozlov, A.~Sonnenfeld, P.~Michel, and J.~Behnke, ``The barrier discharge: basic properties and applications to surface treatment,'' \emph{Vacuum}, vol.~71, no.~3, pp. 417--436, 2003.

\bibitem{surface4}
C.~Jia, P.~Chen, W.~Liu, B.~Li, and Q.~Wang, ``Surface treatment of aramid fiber by air dielectric barrier discharge plasma at atmospheric pressure,'' \emph{Applied Surface Science}, vol. 257, no.~9, pp. 4165--4170, 2011.

\bibitem{surface6}
Z.~Fang, X.~Qiu, Y.~Qiu, and E.~Kuffel, ``Dielectric barrier discharge in atmospheric air for glass-surface treatment to enhance hydrophobicity,'' \emph{IEEE transactions on plasma science}, vol.~34, no.~4, pp. 1216--1222, 2006.

\bibitem{surface8}
C.~Zhang, T.~Shao, K.~Long, Y.~Yu, J.~Wang, D.~Zhang, P.~Yan, and Y.~Zhou, ``Surface treatment of polyethylene terephthalate films using dbd excited by repetitive unipolar nanosecond pulses in air at atmospheric pressure,'' \emph{IEEE transactions on plasma science}, vol.~38, no.~6, pp. 1517--1526, 2010.

\bibitem{surface9}
J.~Huang, Y.~Zhu, S.~Guo, L.~Guo, W.~Yu, S.~Akram, X.~Zhu, X.~Cui, and Z.~Fang, ``Surface treatment of large-area epoxy resin by water-perforated metal plate electrodes dielectric barrier discharge: Hydrophobic modification and uniformity improvement,'' \emph{Applied Surface Science}, vol. 639, p. 158166, 2023.

\bibitem{surface10}
Z.~Fang, H.~Yang, and Y.~Qiu, ``Surface treatment of polyethylene terephthalate films using a microsecond pulse homogeneous dielectric barrier discharges in atmospheric air,'' \emph{IEEE transactions on plasma science}, vol.~38, no.~7, pp. 1615--1623, 2010.

\bibitem{surface11}
Y.~Liu, C.~Su, X.~Ren, C.~Fan, W.~Zhou, F.~Wang, and W.~Ding, ``Experimental study on surface modification of pet films under bipolar nanosecond-pulse dielectric barrier discharge in atmospheric air,'' \emph{Applied surface science}, vol. 313, pp. 53--59, 2014.

\bibitem{DBD_rise_Xie_2019}
Q.~Xie, W.~Gan, C.~Zhang, X.~Che, P.~Yan, and T.~Shao, ``Effect of rise time on nanosecond pulsed surface dielectric barrier discharge actuator,'' \emph{IEEE Transactions on Dielectrics and Electrical Insulation}, vol.~26, no.~2, pp. 346--352, 2019.

\bibitem{DBD_rise_Liu_2001}
S.~Liu and M.~Neiger, ``Excitation of dielectric barrier discharges by unipolar submicrosecond square pulses,'' \emph{Journal of Physics D: Applied Physics}, vol.~34, no.~11, p. 1632, 2001.

\bibitem{DBD_rise_benard_2012}
N.~Benard, N.~Zouzou, A.~Claverie, J.~Sotton, and E.~Moreau, ``Optical visualization and electrical characterization of fast-rising pulsed dielectric barrier discharge for airflow control applications,'' \emph{Journal of Applied Physics}, vol. 111, no.~3, 2012.

\bibitem{DBD_rise_Zhang_2019}
S.~Zhang, Z.~Chen, B.~Zhang, and Y.~Chen, ``Numerical investigation on the effects of dielectric barrier on a nanosecond pulsed surface dielectric barrier discharge,'' \emph{Molecules}, vol.~24, no.~21, p. 3933, 2019.

\bibitem{mod_excimer1}
S.~Beleznai, G.~Mihajlik, I.~Maros, L.~Bal{\'a}zs, and P.~Richter, ``High frequency excitation waveform for efficient operation of a xenon excimer dielectric barrier discharge lamp,'' \emph{Journal of Physics D: Applied Physics}, vol.~43, no.~1, p. 015203, 2009.

\bibitem{mod_ozone1}
Y.~Zhang, L.~Wei, X.~Liang, H.~Deng, and M.~{\v{S}}imek, ``Characteristics of the discharge and ozone generation in oxygen-fed coaxial dbd using an amplitude-modulated ac power supply,'' \emph{Plasma Chemistry and Plasma Processing}, vol.~38, no.~6, pp. 1199--1208, 2018.

\bibitem{mod_ozone2}
Y.~Zhang, L.~Wei, X.~Liang, and M.~{\v{S}}imek, ``Ozone production in coaxial dbd using an amplitude-modulated ac power supply in air,'' \emph{Ozone: Science \& Engineering}, vol.~41, no.~5, pp. 437--447, 2019.

\bibitem{mod_ozone3}
M.~{\v{S}}imek, S.~Pek{\'a}rek, and V.~Prukner, ``Influence of power modulation on ozone production using an ac surface dielectric barrier discharge in oxygen,'' \emph{Plasma Chemistry and Plasma Processing}, vol.~30, no.~5, pp. 607--617, 2010.

\bibitem{clarification_food_ozone}
A.~J. Brodowska, A.~Nowak, and K.~{\'S}migielski, ``Ozone in the food industry: Principles of ozone treatment, mechanisms of action, and applications: An overview,'' \emph{Critical reviews in food science and nutrition}, vol.~58, no.~13, pp. 2176--2201, 2018.

\bibitem{clarification_food_excimer}
P.~Ning, Y.~Han, Y.~Liu, S.~Liu, Z.~Sun, X.~Wang, B.~Wang, F.~Gao, Y.~Wang, Y.~Wang \emph{et~al.}, ``Study on disinfection effect of a 222-nm uvc excimer lamp on object surface,'' \emph{AMB Express}, vol.~13, no.~1, p. 102, 2023.

\bibitem{ozone_kw_1}
E.~J. Van~Heesch, T.~Huiskamp, K.~Yan, F.~Beckers, H.~Smulders, G.~Winands, R.~Lemmens, P.~Blom, S.~D. Segura, W.~Hoeben \emph{et~al.}, ``Pulsed processing by cold plasma, applied to industrial emission control,'' \emph{Frontiers in Chemistry}, vol.~12, p. 1386055, 2024.

\bibitem{ozone_kw_2}
T.~Krishna, P.~Himasree, S.~Srinivasa~Rao, N.~B. Kundakarla, H.-J. Kim \emph{et~al.}, ``Design and development of a digital controlled dielectric barrier discharge (dbd) ac power supply for ozone generation,'' \emph{Journal of Scientific \& Industrial Research}, vol.~79, no.~12, pp. 1057--1068, 2020.

\bibitem{ozone_kw_3}
N.~Lisi, U.~Pasqual~Laverdura, R.~Chierchia, I.~Luisetto, and S.~Stendardo, ``A water cooled, high power, dielectric barrier discharge reactor for co2 plasma dissociation and valorization studies,'' \emph{Scientific Reports}, vol.~13, no.~1, p. 7394, 2023.

\bibitem{ozone_kw_4}
M.~Nur, A.~I. Susan, Z.~Muhlisin, F.~Arianto, A.~W. Kinandana, I.~Nurhasanah, S.~Sumariyah, P.~J. Wibawa, G.~Gunawan, and A.~Usman, ``Evaluation of novel integrated dielectric barrier discharge plasma as ozone generator,'' \emph{Bulletin of Chemical Reaction Engineering \& Catalysis}, vol.~12, no.~1, pp. 24--31, 2017.

\bibitem{DBD_application_review}
U.~Kogelschatz and B.~Eliasson, ``Fundamentals and applications of dielectric-barrier discharges,'' in \emph{HAKONE VII Int. Symp. On High Pressure Low Temperature Plasma Chemistry, Greifswald}, 2000.

\bibitem{ozone_ozcon}
{Ozcon Environmental}, ``Grv2-ep ozone generator,'' [Online]. Available: \url{https://www.ozcon.co.uk/products/ozone-generators/industrial-ozone-generators/grzv2-ozone-generator/}, 2025, accessed: Aug. 1, 2025.

\bibitem{ozone_zenith}
{Pinnacle Ozone Solutions}, ``The zenith ozone generators: High-output \& energy-efficient,'' [Online]. Available: \url{https://pinnacleozone.com/products/ozone-generators/the-zenith/}, 2025, accessed: Aug. 1, 2025.

\bibitem{ozone_zonosistem}
{ZonoSistem}, ``Gr5000-ep industrial ozone generator v4,'' [Online]. Available: \url{https://www.zonosistem.com/en_US/shop/product/fabgr-0047-gr5000-ep-industrial-ozone-generator-v4-8674}, 2025, accessed: Aug. 1, 2025.

\bibitem{HB_inverter}
X.~Tang, W.~Fang, and T.~Jiang, ``Analysis and design of a compact bipolar high-voltage short-pulse power supply for dbd application,'' \emph{IEEE Transactions on Plasma Science}, vol.~50, no.~11, pp. 4650--4659, 2022.

\bibitem{recent_current_fed}
M.~Ponce-Silva, J.~A. Aqui, V.~H. Olivares-Peregrino, and M.~A. Oliver-Salazar, ``Assessment of the current-source, full-bridge inverter as power supply for ozone generators with high power factor in a single stage,'' \emph{IEEE Transactions on Power Electronics}, vol.~31, no.~12, pp. 8195--8204, 2016.

\bibitem{current_fed_pf1}
I.~Castro, K.~Martin, A.~Vazquez, M.~Arias, D.~G. Lamar, and J.~Sebastian, ``An ac--dc pfc single-stage dual inductor current-fed push--pull for hb-led lighting applications,'' \emph{IEEE Journal of Emerging and Selected Topics in Power Electronics}, vol.~6, no.~1, pp. 255--266, 2017.

\bibitem{current_fed_pf2}
C.~Li, Y.~Zhang, Z.~Cao, and D.~Xu, ``Single-phase single-stage isolated zcs current-fed full-bridge converter for high-power ac/dc applications,'' \emph{IEEE Transactions on Power Electronics}, vol.~32, no.~9, pp. 6800--6812, 2016.

\bibitem{current_fed_pf3}
W.~Uddin, T.~A. Wagaye, and M.~Kim, ``Quasi-single-stage current-fed resonant ac--dc converter having improved heat distribution,'' \emph{IEEE Transactions on Power Electronics}, vol.~37, no.~11, pp. 13\,499--13\,512, 2022.

\bibitem{current_fed_pf4}
J.~Hassan, J.-W. Lim, T.~A. Wagaye, B.~Han, S.-W. Lee, M.-J. Kim, and M.~Kim, ``Highly efficient quasi-single-stage ac-dc converter employing bidirectional switch,'' \emph{IEEE Transactions on Transportation Electrification}, vol.~8, no.~2, pp. 2746--2757, 2022.

\bibitem{current_fed_pf5}
B.~Singh, S.~Singh, A.~Chandra, and K.~Al-Haddad, ``Comprehensive study of single-phase ac-dc power factor corrected converters with high-frequency isolation,'' \emph{IEEE transactions on Industrial Informatics}, vol.~7, no.~4, pp. 540--556, 2011.

\bibitem{current_fed_pf6}
M.~Narimani and G.~Moschopoulos, ``A three-level integrated ac--dc converter,'' \emph{IEEE Transactions on Power Electronics}, vol.~29, no.~4, pp. 1813--1820, 2013.

\bibitem{current_fed_pf7}
C.~Li and D.~Xu, ``Family of enhanced zcs single-stage single-phase isolated ac--dc converter for high-power high-voltage dc supply,'' \emph{IEEE Transactions on Industrial Electronics}, vol.~64, no.~5, pp. 3629--3639, 2017.

\bibitem{CS_push_pull}
X.~Tang, Z.~Lin, Z.~Zhou, Y.~Chen, and M.~Zhang, ``Analysis and experimental validation of an integrated current-source power supply with high power factor for dbd applications,'' \emph{IEEE Transactions on Plasma Science}, vol.~51, no.~5, pp. 1290--1301, 2023.

\bibitem{dbd_function_gen}
N.~Balcon, N.~Benard, Y.~Lagmich, J.-P. Boeuf, G.~Touchard, and E.~Moreau, ``Positive and negative sawtooth signals applied to a dbd plasma actuator--influence on the electric wind,'' \emph{Journal of electrostatics}, vol.~67, no. 2-3, pp. 140--145, 2009.

\bibitem{XFMR_FB}
X.~Bonnin, J.~Brandelero, N.~Videau, H.~Piquet, and T.~Meynard, ``A high voltage high frequency resonant inverter for supplying dbd devices with short discharge current pulses,'' \emph{IEEE Transactions on Power Electronics}, vol.~29, no.~8, pp. 4261--4269, 2014.

\bibitem{XFMR_FB_LCL}
M.~Amjad, Z.~Salam, M.~Facta, and S.~Mekhilef, ``Analysis and implementation of transformerless lcl resonant power supply for ozone generation,'' \emph{IEEE transactions on power electronics}, vol.~28, no.~2, pp. 650--660, 2012.

\bibitem{XFMR_single_switch}
D.~Florez, H.~Piquet, E.~Bru, and R.~Diez, ``Single-switch transformer-less power supply for low temperature plasma jet – 3.3 kv sic mosfet opportunities,'' \emph{IEEE Transactions on Power Electronics}, vol.~39, no.~6, pp. 7230--7237, 2024.

\bibitem{XFMR_thy}
M.~A. Diop, A.~Belinger, and H.~Piquet, ``10 kv sic mosfet evaluation for dielectric barrier discharge transformerless power supply,'' \emph{plasma}, vol.~3, no.~3, pp. 103--116, 2020.

\bibitem{mis_low_pf}
A.~Schonknecht and R.~De~Doncker, ``Novel topology for parallel connection of soft-switching high-power high-frequency inverters,'' \emph{IEEE Transactions on Industry Applications}, vol.~39, no.~2, pp. 550--555, 2003.

\bibitem{DBD_power_review}
U.~Kogelschatz, ``Dielectric-barrier discharges: Their history, discharge physics, and industrial applications,'' \emph{Plasma Chemistry and Plasma Processing}, vol.~23, pp. 1--46, 2003.

\bibitem{DBD_cap}
X.~Bonnin, H.~Piquet, N.~Naud{\'e}, M.~C. Bouzidi, N.~Gherardi, and J.-M. Blaqui{\`e}re, ``Design of a current converter to maximize the power into homogeneous dielectric barrier discharge (dbd) devices,'' \emph{The European Physical Journal-Applied Physics}, vol.~64, no.~1, p. 10901, 2013.

\bibitem{modular1}
S.~Jiang, L.~Qiu, Z.~Li, L.~Zhang, and J.~Rao, ``A new all-solid-state bipolar high-voltage multilevel generator for dielectric barrier discharge,'' \emph{IEEE transactions on plasma science}, vol.~48, no.~4, pp. 1076--1081, 2020.

\bibitem{modular2}
F.~Dragonas, G.~Grandi, and G.~Neretti, ``High-voltage high-frequency arbitrary waveform multilevel generator for dielectric barrier discharge,'' in \emph{2014 International Symposium on Power Electronics, Electrical Drives, Automation and Motion}.\hskip 1em plus 0.5em minus 0.4em\relax IEEE, 2014, pp. 57--61.

\bibitem{modular3}
S.~A. Saleh, B.~Allen, E.~Ozkop, and B.~G. Colpitts, ``Multistage and multilevel power electronic converter-based power supply for plasma dbd devices,'' \emph{IEEE Transactions on Industrial Electronics}, vol.~65, no.~7, pp. 5466--5475, 2017.

\bibitem{recent_modular}
F.~A. Dragonas, G.~Neretti, P.~Sanjeevikumar, and G.~Grandi, ``High-voltage high-frequency arbitrary waveform multilevel generator for dbd plasma actuators,'' \emph{IEEE Transactions on Industry Applications}, vol.~51, no.~4, pp. 3334--3342, 2015.

\bibitem{recent_FB3}
M.~Amjad and Z.~Salam, ``Analysis, design, and implementation of multiple parallel ozone chambers for high flow rate,'' \emph{IEEE Transactions on Industrial Electronics}, vol.~61, no.~2, pp. 753--765, 2014.

\bibitem{mis_lcc_zero}
V.~M. López, A.~Navarro-Crespin, R.~Schnell, C.~Brañas, F.~J. Azcondo, and R.~Zane, ``Current phase surveillance in resonant converters for electric discharge applications to assure operation in zero-voltage-switching mode,'' \emph{IEEE Transactions on Power Electronics}, vol.~27, no.~6, pp. 2925--2935, 2012.

\bibitem{XFMR_class_E}
J.~Xu, K.~Surakitbovorn, B.~Wang, M.~A. Cappelli, and J.~Rivas-Davila, ``Frequency-selective mhz power amplifier for dielectric barrier discharge plasma generation,'' \emph{IEEE Open Journal of Power Electronics}, vol.~3, pp. 846--855, 2022.

\bibitem{XFMR_piezo}
M.~Yang, E.~A. Stolt, Z.~Ye, and J.~M. Rivas-Davila, ``Piezoelectric based class-e resonant inverter for driving surface dielectric barrier discharge plasma,'' in \emph{2024 IEEE 10th International Power Electronics and Motion Control Conference (IPEMC2024-ECCE Asia)}, 2024, pp. 367--372.

\bibitem{phase_shift_FB1}
V.~Kinnares and P.~Hothongkham, ``Circuit analysis and modeling of a phase-shifted pulsewidth modulation full-bridge-inverter-fed ozone generator with constant applied electrode voltage,'' \emph{IEEE Transactions on Power Electronics}, vol.~25, no.~7, pp. 1739--1752, 2010.

\bibitem{phase_shift_FB2}
O.~Koudriavtsev, S.~Wang, Y.~Konishi, and M.~Nakaoka, ``A novel pulse-density-modulated high-frequency inverter for silent-discharge-type ozonizer,'' \emph{IEEE Transactions on industry applications}, vol.~38, no.~2, pp. 369--378, 2002.

\bibitem{phase_shift_FB3}
P.~Hothongkham and V.~Kinnares, ``Measurement of an ozone generator using a phase-shifted pwm full bridge inverter,'' in \emph{The 2010 International Power Electronics Conference-ECCE ASIA-}.\hskip 1em plus 0.5em minus 0.4em\relax IEEE, 2010, pp. 1552--1559.

\bibitem{recent_FB2}
J.~A. L{\'o}pez-Fern{\'a}ndez, R.~Pe{\~n}a-Eguiluz, R.~L{\'o}pez-Callejas, A.~Mercado-Cabrera, B.~Jaramillo-Sierra, B.~Rodr{\'\i}guez-Mendez, R.~Valencia-Alvarado, and A.~E. Mu{\~n}oz-Castro, ``A 10-to 30-khz adjustable frequency resonant full-bridge multicell power converter,'' \emph{IEEE Transactions on Industrial Electronics}, vol.~62, no.~4, pp. 2215--2223, 2014.

\bibitem{CS_dcdc_1}
K.~R. Sree and A.~K. Rathore, ``Hybrid modulated extended secondary universal current-fed zvs converter for wide voltage range: Analysis, design, and experimental results,'' \emph{IEEE Transactions on Industrial Electronics}, vol.~62, no.~7, pp. 4471--4480, 2014.

\bibitem{CS_dcdc_2}
S.-J. Jang, C.-Y. Won, B.-K. Lee, and J.~Hur, ``Fuel cell generation system with a new active clamping current-fed half-bridge converter,'' \emph{IEEE Transactions on Energy Conversion}, vol.~22, no.~2, pp. 332--340, 2007.

\bibitem{CS_dcdc_3}
X.~Kong and A.~M. Khambadkone, ``Analysis and implementation of a high efficiency, interleaved current-fed full bridge converter for fuel cell system,'' \emph{IEEE Transactions on Power Electronics}, vol.~22, no.~2, pp. 543--550, 2007.

\bibitem{CS_dcdc_4}
U.~Prasanna and A.~K. Rathore, ``Extended range zvs active-clamped current-fed full-bridge isolated dc/dc converter for fuel cell applications: analysis, design, and experimental results,'' \emph{IEEE Transactions on Industrial Electronics}, vol.~60, no.~7, pp. 2661--2672, 2012.

\bibitem{CS_third}
S.~Hao, X.~Liu, W.~Li, Y.~Deng, and X.~He, ``Energy compression of dielectric barrier discharge with third harmonic circulating current in current-fed parallel-series resonant converter,'' \emph{IEEE Transactions on Power Electronics}, vol.~31, no.~12, pp. 8528--8540, 2016.

\bibitem{CS_square}
D.~Florez, R.~Diez, H.~Piquet, and A.~K. Hay~Harb, ``Square-shape current-mode supply for parametric control of the dbd excilamp power,'' \emph{IEEE Transactions on Industrial Electronics}, vol.~62, no.~3, pp. 1451--1460, 2015.

\bibitem{mis_math}
A.~J. Gilbert, C.~M. Bingham, D.~A. Stone, and M.~P. Foster, ``Normalized analysis and design of lcc resonant converters,'' \emph{IEEE Transactions on power Electronics}, vol.~22, no.~6, pp. 2386--2402, 2007.

\bibitem{CS_CLCC}
L.~Chang, T.~Guo, J.~Liu, C.~Zhang, Y.~Deng, and X.~He, ``Analysis and design of a current-source clcc resonant converter for dbd applications,'' \emph{IEEE Transactions on Power Electronics}, vol.~29, no.~4, pp. 1610--1621, 2013.

\bibitem{CS_LLCC}
S.~Hao, C.~Zhang, T.~Guo, X.~Liu, S.~Hu, J.~Liu, and X.~He, ``A current-fed asymmetric llcc resonant converter for dbd applications,'' in \emph{2014 IEEE Applied Power Electronics Conference and Exposition - APEC 2014}, 2014, pp. 873--878.

\bibitem{current_fed_buck}
C.~Ordiz, J.~Alonso, M.~Dalla~Costa, J.~Ribas, and A.~Calleja, ``Development of a high-voltage closed-loop power supply for ozone generation,'' in \emph{2008 Twenty-Third Annual IEEE Applied Power Electronics Conference and Exposition}.\hskip 1em plus 0.5em minus 0.4em\relax IEEE, 2008, pp. 1861--1867.

\bibitem{CS_push_pull_inverter_reverse}
J.~Alonso, J.~Cardesin, J.~Martin-Ramos, J.~Garcia, and M.~Rico-Secades, ``Using current-fed parallel-resonant inverters for electrodischarge applications: a case of study,'' in \emph{Nineteenth Annual IEEE Applied Power Electronics Conference and Exposition, 2004. APEC'04.}, vol.~1.\hskip 1em plus 0.5em minus 0.4em\relax IEEE, 2004, pp. 109--115.

\bibitem{CS_push_pull_inverter}
T.-J. Liang, R.-Y. Chen, and J.-F. Chen, ``Current-fed parallel-resonant dc--ac inverter for cold-cathodefluorescent lamps with zero-current switching,'' \emph{IEEE transactions on power electronics}, vol.~23, no.~4, pp. 2206--2210, 2008.

\bibitem{recent_res_EF}
S.~Aldhaher, D.~C. Yates, and P.~D. Mitcheson, ``Modeling and analysis of class ef and class e/f inverters with series-tuned resonant networks,'' \emph{IEEE transactions on power electronics}, vol.~31, no.~5, pp. 3415--3430, 2015.

\bibitem{recent_res_load_independent}
------, ``Load-independent class e/ef inverters and rectifiers for mhz-switching applications,'' \emph{IEEE Transactions on Power Electronics}, vol.~33, no.~10, pp. 8270--8287, 2018.

\bibitem{new_excimer1}
S.~K. Ram, B.~K. Verma, V.~K. Saini, R.~P. Lamba, P.~K. Das, S.~Mishra, S.~Devassy, and U.~N. Pal, ``Short-pulse width high-voltage bipolar impulse generator for dbd-based 222 nm excimer source,'' \emph{IEEE Transactions on Plasma Science}, 2024.

\bibitem{new_excimer2}
B.~K. Verma, S.~K. Ram, V.~K. Saini, U.~N. Pal, and R.~P. Lamba, ``Implementation and analysis of unipolar high-voltage pulse modulator for 172-nm vuv excilamp,'' \emph{IEEE Transactions on Plasma Science}, 2024.

\bibitem{recent_FB1}
Y.~Chen, Q.~Wang, and Y.~Wang, ``Analysis and comparison of dbd power supplies with different control methods,'' \emph{IEEE Transactions on Plasma Science}, vol.~52, no.~5, pp. 1747--1757, 2024.

\bibitem{recent_forward}
Y.~Lv, X.~Liu, Y.~Wang, S.~Lu, J.~Zhang, and H.~Zhang, ``High frequency bipolar pulse transformer with an undershoot damping circuit for tumor application,'' \emph{IEEE Transactions on Power Electronics}, 2024.

\bibitem{recent_pfl2}
Y.~Mi, C.~Bian, P.~Li, C.~Yao, and C.~Li, ``A modular generator of nanosecond pulses with adjustable polarity and high repetition rate,'' \emph{IEEE Transactions on Power Electronics}, vol.~33, no.~12, pp. 10\,654--10\,662, 2018.

\bibitem{recent_pfl1}
J.~Ma, L.~Yu, L.~Ren, C.~Yao, S.~Dong, J.~Ma, and W.~Xu, ``Nanosecond pulse generator based on inductive energy storage forming line with impedance matching modulation capability,'' \emph{IEEE Transactions on Industrial Electronics}, vol.~71, no.~12, pp. 15\,643--15\,653, 2024.

\bibitem{recent_pfl4}
J.~Ma, C.~Yao, L.~Yu, C.~Li, C.~Liu, L.~Zhao, and S.~Dong, ``Compact nanosecond pulse generator based on distributed inductive energy storage of twisted pair wire,'' \emph{IEEE Transactions on Power Electronics}, 2024.

\bibitem{recent_pfl3}
J.~Ma, L.~Yu, S.~Dong, C.~Yao, L.~Gao, W.~Sun, and Y.~He, ``Mhz nanosecond rectangular pulse generator with high voltage gain and multimode,'' \emph{IEEE Transactions on Power Electronics}, vol.~36, no.~8, pp. 8978--8987, 2021.

\bibitem{HB_BLT}
H.~Gui, Z.~Zhao, Q.~Shi, X.~Liu, and C.~Yao, ``All-solid-state nanosecond pulse power supply based on blts and pulse transformer for dbd application,'' \emph{IEEE Transactions on Power Electronics}, vol.~38, no.~8, pp. 10\,085--10\,092, 2023.

\bibitem{recent_pulse_review}
Y.~Zhuge, J.~Liang, M.~Fu, T.~Long, and H.~Wang, ``Comprehensive overview of power electronics intensive solutions for high-voltage pulse generators,'' \emph{IEEE Open Journal of Power Electronics}, vol.~5, pp. 1--20, 2023.

\bibitem{recent_Marx2}
S.~Shen, J.~Yan, G.~Sun, and W.~Ding, ``Improved auxiliary triggering topology for high-power nanosecond pulse generators based on avalanche transistors,'' \emph{IEEE Transactions on Power Electronics}, vol.~36, no.~12, pp. 13\,634--13\,644, 2021.

\bibitem{recent_Marx4}
C.~Yu, H.~Kim, S.~Jang, T.~Kim, S.~Son, C.~Kwon, and H.~Cha, ``Solid state pulsed power modulator with high repetition rate and short pulse width for high-speed pulsed lasers,'' \emph{IEEE Transactions on Industrial Electronics}, vol.~71, no.~1, pp. 388--397, 2023.

\bibitem{Marx_review}
Z.~Zhong, J.~Rao, H.~Liu, and L.~M. Redondo, ``Review on solid-state-based marx generators,'' \emph{IEEE Transactions on Plasma Science}, vol.~49, no.~11, pp. 3625--3643, 2021.

\bibitem{Marx_boost}
L.~Yu, W.~Sun, C.~Yao, S.~Dong, K.~Li, D.~He, Y.~Jin, and Z.~Bo, ``A novel boost marx pulse generator based on single-driver series-connected sic mosfets,'' \emph{IEEE Transactions on Industrial Electronics}, vol.~71, no.~2, pp. 2070--2079, 2024.

\bibitem{Marx_ind_excimer}
Y.~Wang, L.~Tong, Q.~Han, and K.~Liu, ``Repetitive high-voltage all-solid-state marx generator for excimer dbd uv sources,'' \emph{IEEE Transactions on Plasma Science}, vol.~44, no.~10, pp. 1933--1940, 2016.

\bibitem{recent_Marx3}
F.~Wu, C.~Yao, Y.~Chen, L.~Yu, S.~Dong, and H.~Wang, ``All-solid-state ultrashort pulse generator by capacitive chopping circuit,'' \emph{IEEE Transactions on Power Electronics}, vol.~38, no.~8, pp. 9897--9906, 2023.

\bibitem{Saket}
S.~Kapse and J.~Roy, ``High voltage flyback inspired converter driving dielectric barrier discharge for ozone generation,'' in \emph{2025 IEEE International Communications Energy Conference (INTELEC)}, 2025, pp. 173--178.

\bibitem{Fly_lc_stacked}
Y.~Feng, C.~Zhang, L.~Dou, C.~Zhang, and T.~Shao, ``Repetitive high boost ratio pulsed power generator for dielectric barrier discharge applications,'' \emph{IEEE Transactions on Industrial Electronics}, vol.~71, no.~7, pp. 7053--7062, 2024.

\bibitem{recent_flyback2}
Z.~Zhang, H.~He, H.~Yu, K.~Li, K.~Bian, and W.~Chen, ``A modular push--pull--flyback high-voltage pulse generator for electric field emulation during a lightning strike,'' \emph{IEEE Transactions on Power Electronics}, vol.~38, no.~6, pp. 7322--7335, 2023.

\bibitem{Fly_D}
S.~Rai, A.~Dhakar, and U.~Pal, ``A compact nanosecond pulse generator for dbd tube characterization,'' \emph{Review of Scientific Instruments}, vol.~89, no.~3, 2018.

\bibitem{Fly_RDD}
S.~Jin, J.~Chen, Z.~Li, C.~Zhang, Y.~Zhao, and Z.~Fang, ``Novel rdd pulse shaping method for high-power high-voltage pulse current power supply in dbd application,'' \emph{IEEE Transactions on Industrial Electronics}, vol.~69, no.~12, pp. 12\,653--12\,664, 2022.

\bibitem{Fly_boost_RDD}
S.~Jin, S.~Wang, K.~Zhao, Y.~Jiang, S.~Zhao, Y.~Zhao, and Z.~Fang, ``A high-drive-performance microsecond pulse power module for portable dbd plasma source device,'' \emph{IEEE Transactions on Power Electronics}, vol.~38, no.~12, pp. 15\,072--15\,085, 2023.

\bibitem{recent_flyback3}
Q.~Yang, C.~Yao, S.~Dong, K.~Qian, R.~Liang, and H.~Liu, ``High-voltage ultrashort pulse power module with mhz repetition rate based on ssst,'' \emph{IEEE Transactions on Power Electronics}, 2024.

\bibitem{forward1}
Y.~Wang, S.~Yang, G.~Wu, Z.~Lu, S.~Jiang, Z.~Li, and J.~Rao, ``A novel repetitive high-voltage resonant pulse generator for plasma-assisted milling,'' \emph{IEEE Transactions on Plasma Science}, vol.~49, no.~8, pp. 2350--2358, 2021.

\bibitem{recent_res1}
S.~Chen, X.~Cheng, S.~Wang, Y.~Wang, and J.~Liang, ``3-mhz high-voltage pulse generator with novel multilevel lc resonant network for adjustable pulse width,'' \emph{IEEE Transactions on Power Electronics}, 2024.

\bibitem{fly_RD}
A.~K. Dhakar, S.~K. Rai, V.~K. Saini, S.~K. Sharma, and U.~N. Pal, ``Simplified high-voltage short-pulse power modulator for dbd plasma application,'' \emph{IEEE Transactions on Plasma Science}, vol.~49, no.~4, pp. 1422--1427, 2021.

\bibitem{fly_solar}
Z.~Fang, Y.~Shi, F.~Liu, and R.~Zhou, ``Compact microsecond pulsed power generator driven by solar energy for dielectric barrier discharge applications,'' \emph{IEEE Transactions on Dielectrics and Electrical Insulation}, vol.~26, no.~2, pp. 390--396, 2019.

\bibitem{fly_modular}
S.~Jin, C.~Zhang, Y.~Peng, and Z.~Fang, ``Novel ipox architecture for high-voltage microsecond pulse power supply using energy efficiency and stability model design method,'' \emph{IEEE Transactions on Power Electronics}, vol.~36, no.~9, pp. 10\,852--10\,865, 2021.

\bibitem{CS_review}
R.~Diez, H.~Piquet, D.~Florez, and X.~Bonnin, ``Current-mode approach in power supplies for dbd excilamps: Review of 4 topologies,'' \emph{IEEE Transactions on Plasma Science}, vol.~43, no.~1, pp. 452--460, 2015.

\bibitem{CS_fly_boost}
R.~Diez, H.~Piquet, S.~Bhosle, J.-M. Blaqui{\`e}re, and N.~Roux, ``Design of a current converter for the study of the uv emission in dbd excilamps,'' in \emph{2008 IEEE International Symposium on Industrial Electronics}.\hskip 1em plus 0.5em minus 0.4em\relax IEEE, 2008, pp. 62--67.

\bibitem{CS_fly_buck_boost}
R.~Diez, H.~Piquet, M.~Cousineau, and S.~Bhosle, ``Current-mode power converter for radiation control in dbd excimer lamps,'' \emph{IEEE Transactions on Industrial Electronics}, vol.~59, no.~4, pp. 1912--1919, 2011.

\bibitem{recent_current_Fed1}
S.~S. Nag and S.~Mishra, ``Current-fed switched inverter,'' \emph{IEEE Transactions on Industrial Electronics}, vol.~61, no.~9, pp. 4680--4690, 2013.

\bibitem{recent_current_Fed2}
S.~Lee, F.~Chen, T.~M. Jahns, and B.~Sarlioglu, ``Review on switching device fault, protection, and fault-tolerant topologies of current source inverter,'' in \emph{2021 IEEE 13th International Symposium on Diagnostics for Electrical Machines, Power Electronics and Drives (SDEMPED)}, vol.~1, 2021, pp. 489--495.

\bibitem{rev3_voltage}
Z.~Ding, C.~Ren, M.~Xu, J.~Sun, J.~Sun, W.~Yang, and X.~Li, ``A linear irradiance control power supply for dielectric barrier discharge excimer ultraviolet lamps,'' in \emph{2024 IEEE Applied Power Electronics Conference and Exposition (APEC)}.\hskip 1em plus 0.5em minus 0.4em\relax IEEE, 2024, pp. 1689--1693.

\bibitem{challenges_heat}
A.~A. Abdelaziz, T.~Ishijima, T.~Seto, N.~Osawa, H.~Wedaa, and Y.~Otani, ``Characterization of surface dielectric barrier discharge influenced by intermediate frequency for ozone production,'' \emph{Plasma Sources Science and Technology}, vol.~25, no.~3, p. 035012, 2016.

\bibitem{WBG1}
M.~Buffolo, D.~Favero, A.~Marcuzzi, C.~De~Santi, G.~Meneghesso, E.~Zanoni, and M.~Meneghini, ``Review and outlook on gan and sic power devices: industrial state-of-the-art, applications, and perspectives,'' \emph{IEEE Transactions on Electron Devices}, 2024.

\bibitem{piezo}
W.~D. Braun, E.~A. Stolt, L.~Gu, J.~Segovia-Fernandez, S.~Chakraborty, R.~Lu, and J.~M. Rivas-Davila, ``Optimized resonators for piezoelectric power conversion,'' \emph{IEEE Open Journal of Power Electronics}, vol.~2, pp. 212--224, 2021.

\bibitem{slew_review}
S.~Zhao, X.~Zhao, Y.~Wei, Y.~Zhao, and H.~A. Mantooth, ``A review of switching slew rate control for silicon carbide devices using active gate drivers,'' \emph{IEEE Journal of Emerging and Selected Topics in Power Electronics}, vol.~9, no.~4, pp. 4096--4114, 2020.

\bibitem{bias1}
H.~Yan, L.~Yang, X.~Qi, and C.~Ren, ``Effect of a direct current bias on the electrohydrodynamic performance of a surface dielectric barrier discharge actuator for airflow control,'' \emph{Journal of Applied Physics}, vol. 117, no.~6, 2015.

\bibitem{bias2}
P.~Zhao and S.~Roy, ``Study of spectrum analysis and signal biasing for dielectric barrier discharge actuator,'' in \emph{50th AIAA Aerospace Sciences Meeting including the New Horizons Forum and Aerospace Exposition}, 2012, p. 408.

\bibitem{bias4}
H.~Mahdavi and F.~Sohbatzadeh, ``The effects of applying different bias voltages and phase differences on performance of an asymmetric surface dielectric barrier discharge; an experimental investigation,'' \emph{Journal of Theoretical and Applied Physics}, vol.~13, pp. 165--177, 2019.

\bibitem{DBD_tendency4}
D.~F. Opaits, A.~V. Likhanskii, G.~Neretti, S.~Zaidi, M.~N. Shneider, R.~B. Miles, and S.~O. Macheret, ``Experimental investigation of dielectric barrier discharge plasma actuators driven by repetitive high-voltage nanosecond pulses with dc or low frequency sinusoidal bias,'' \emph{Journal of applied physics}, vol. 104, no.~4, 2008.

\bibitem{under_aero1}
Z.~Wu, J.~Xu, P.~Chen, K.~Xie, and N.~Wang, ``Maximum thrust of single dielectric barrier discharge thruster at low pressure,'' \emph{AIAA Journal}, vol.~56, no.~6, pp. 2235--2241, 2018.

\bibitem{under_aero2}
H.~Xu, Y.~He, and S.~R. Barrett, ``A dielectric barrier discharge ion source increases thrust and efficiency of electroaerodynamic propulsion,'' \emph{Applied Physics Letters}, vol. 114, no.~25, 2019.

\bibitem{under_aero3}
N.~Gomez-Vega, H.~Xu, J.~M. Abel, and S.~R. Barrett, ``Performance of decoupled electroaerodynamic thrusters,'' \emph{Applied Physics Letters}, vol. 118, no.~7, 2021.

\bibitem{under_aero6}
M.~Ahangar, N.~Alebrahim, and S.~Ahmadian, ``Performance characterization and comparison of two-and tri-electrode configurations in a dbd plasma channel thruster utilizing volumetric and surface discharge mechanisms,'' \emph{Physics of Plasmas}, vol.~32, no.~2, 2025.

\bibitem{mode_burst1}
S.~Sekimoto, T.~Nonomura, and K.~Fujii, ``Burst-mode frequency effects of dielectric barrier discharge plasma actuator for separation control,'' \emph{AIAA journal}, vol.~55, no.~4, pp. 1385--1392, 2017.

\bibitem{mode_burst5}
T.~Matsunuma, ``Effects of burst ratio and frequency on the passage vortex reduction of a linear turbine cascade using a dielectric barrier discharge plasma actuator,'' in \emph{Actuators}, vol.~11, no.~8.\hskip 1em plus 0.5em minus 0.4em\relax MDPI, 2022, p. 210.

\bibitem{mode_burst2}
M.~Xue, C.~Gao, H.-D. Xi, and F.~Liu, ``Vortices induced by a dielectric barrier discharge plasma actuator under burst-mode actuation,'' \emph{AIAA Journal}, vol.~58, no.~6, pp. 2428--2441, 2020.

\bibitem{mode_burst4}
Y.~Wang, C.~Gao, and Y.~Wang, ``Flow characteristics of dielectric barrier discharge plasma actuator array in burst mode,'' \emph{AIAA Journal}, vol.~59, no.~11, pp. 4581--4597, 2021.

\bibitem{aero_feedback1}
P.~W. Kwan and X.~Huang, ``A pedagogical study of aerodynamic feedback control by dielectric barrier discharge plasma,'' \emph{IEEE Transactions on Industrial Electronics}, vol.~67, no.~1, pp. 451--460, 2020.

\bibitem{aero_feedback2}
Y.~Wang, S.~Wu, Z.~Zhu, Z.~Li, and D.~Hong, ``Closed-loop control of ionic wind speed generated by nanosecond pulsed surface dielectric barrier discharge,'' \emph{IEEE Transactions on Plasma Science}, vol.~51, no.~10, pp. 3018--3026, 2023.

\bibitem{under_3d1}
C.~B. Sweeney, M.~L. Burnette, M.~J. Pospisil, S.~A. Shah, M.~Anas, B.~R. Teipel, B.~S. Zahner, D.~Staack, and M.~J. Green, ``Dielectric barrier discharge applicator for heating carbon nanotube-loaded interfaces and enhancing 3d-printed bond strength,'' \emph{Nano letters}, vol.~20, no.~4, pp. 2310--2315, 2020.

\bibitem{under_3d2}
S.~Ma, T.~Ma, S.~Tsuchikawa, T.~Inagaki, H.~Wang, and H.~Jiang, ``Effect of dielectric barrier discharge (dbd) plasma treatment on physicochemical and 3d printing properties of wheat starch,'' \emph{International Journal of Biological Macromolecules}, vol. 269, p. 132159, 2024.

\bibitem{under_3d3}
M.~Kari{\v{z}}, D.~K. Tomec, S.~Dahle, M.~K. Kuzman, M.~{\v{S}}ernek, and J.~{\v{Z}}igon, ``Effect of sanding and plasma treatment of 3d-printed parts on bonding to wood with pvac adhesive,'' \emph{Polymers}, vol.~13, no.~8, p. 1211, 2021.

\bibitem{DBD_tendency3}
P.~Brunet, R.~Rinc{\'o}n, Z.~Matouk, M.~Chaker, and F.~Massines, ``Tailored waveform of dielectric barrier discharge to control composite thin film morphology,'' \emph{Langmuir}, vol.~34, no.~5, pp. 1865--1872, 2018.

\bibitem{under_carbon2}
M.~Yang, C.~Li, Y.~Tian, L.~Wu, J.~Hu, and X.~Hou, ``Dielectric barrier discharge-accelerated one-pot synthesis of sulfur quantum dots for fluorescent sensing of lead ions and l-cysteine,'' \emph{Chemical Communications}, vol.~58, no.~62, pp. 8614--8617, 2022.

\bibitem{under_silver1}
A.~Ananth and Y.~S. Mok, ``Dielectric barrier discharge (dbd) plasma assisted synthesis of ag2o nanomaterials and ag2o/ruo2 nanocomposites,'' \emph{Nanomaterials}, vol.~6, no.~3, p.~42, 2016.

\bibitem{under_gold1}
A.~Bjelajac, A.-M. Phillipe, J.~Guillot, Y.~Fleming, J.-B. Chemin, P.~Choquet, and S.~Bulou, ``Gold nanoparticles synthesis and immobilization by atmospheric pressure dbd plasma torch method,'' \emph{Nanoscale Advances}, vol.~5, no.~9, pp. 2573--2582, 2023.

\bibitem{under_quantum}
L.~Chen, D.~Li, A.~Wang, W.~Guo, X.~Su, J.~Shang, W.~Du, S.~Liu, and Z.~Ma, ``Negative corona discharge strategy for efficient quantum dot light-emitting diodes,'' \emph{Optics Letters}, vol.~49, no.~12, pp. 3392--3395, 2024.

\end{thebibliography}

% Color coded review
  % \printbibliography

\end{document}